\documentclass[aps,pra,reprint,showpacs,floatfix,superscriptaddress]{revtex4-1}

\usepackage{amsmath}
\usepackage{slashed}
\usepackage{txfonts}
\usepackage{microtype}
\usepackage{graphicx}
\usepackage[breaklinks=true]{hyperref}
\usepackage{color}

\def\sout{\bgroup\markoverwith
{\textcolor{red}{\rule[0.5ex]{2pt}{0.5pt}}}\ULon}
\def\be{\begin{equation}}
\def\ee{\end{equation}}
\def\bes{\begin{equation*}}
\def\ees{\end{equation*}}
\def\bea{\begin{eqnarray}}
\def\eea{\end{eqnarray}}
\def\beas{\begin{eqnarray*}}
\def\eeas{\end{eqnarray*}}
\def\bal#1\eal{\begin{align}#1\end{align}}
\def\bals#1\eals{\begin{align*}#1\end{align*}}
\newcommand{\bra}[1]{\langle #1|}
\newcommand{\ket}[1]{|#1\rangle}
\newcommand{\braket}[2]{\langle #1|#2\rangle}
\newcommand{\bk}[1]{\langle #1\rangle}

\renewcommand{\vec}{\vectorsym}
\newcommand{\del}{\partial}
\renewcommand*{\vec}[1]{\boldsymbol{#1}}

\graphicspath{{figures/}}

\usepackage[normalem]{ulem}

\def\R {\mathbb{R}}
\def\Z {\mathbb{Z}}
\newcommand{\bx}{\vec{x}}
\newcommand{\bA}{\vec{A}}
\newcommand{\bF}{{\vec{F}}}
\newcommand{\cH}{\mathcal{H}}
\newcommand{\cN}{{\hat n}}
\newcommand{\ha}{{\hat a}}
\newcommand{\hS}{{\hat S}}
\newcommand{\hU}{{\hat U}}
\DeclareMathOperator{\re}{\mathrm{Re}}

\begin{document}

\title{A Quantum Impurity Model for Anyons}  

\author{E. Yakaboylu}
\email{enderalp.yakaboylu@ist.ac.at}
\affiliation{IST Austria (Institute of Science and Technology Austria), Am Campus 1, 3400 Klosterneuburg, Austria}

\author{A. Ghazaryan}
\affiliation{IST Austria (Institute of Science and Technology Austria), Am Campus 1, 3400 Klosterneuburg, Austria}

\author{D. Lundholm}
\affiliation{Uppsala University, Department of Mathematics - Box 480, SE-751 06 Uppsala, Sweden}

\author{N. Rougerie}
\affiliation{Universit\'{e} Grenoble- Alpes \& CNRS, LPMMC (UMR 5493) - B.P. 166, F-38042 Grenoble, France}

\author{M. Lemeshko}
\affiliation{IST Austria (Institute of Science and Technology Austria), Am Campus 1, 3400 Klosterneuburg, Austria}

\author{R. Seiringer}
\affiliation{IST Austria (Institute of Science and Technology Austria), Am Campus 1, 3400 Klosterneuburg, Austria}

\date{\today}

\begin{abstract}


One of the hallmarks of quantum statistics, tightly entwined with the concept of topological phases of matter, is the prediction of anyons. Although anyons are predicted to be realized in certain fractional quantum Hall systems, they have not yet been unambiguously detected in experiment. Here we introduce a simple quantum impurity model, where bosonic or fermionic impurities turn into anyons as a consequence of their interaction with the surrounding many-particle bath. A cloud of phonons dresses each impurity in such a way that it effectively attaches fluxes/vortices to it and thereby converts it into an Abelian anyon. The corresponding quantum impurity model, first, provides a new approach to the numerical solution of the many-anyon problem, along with a new concrete perspective of anyons as emergent quasiparticles built from composite bosons or fermions. More importantly, the model paves the way towards realizing anyons using impurities in crystal lattices as well as ultracold gases. In particular, we consider two heavy electrons interacting with a two-dimensional lattice crystal in a magnetic field, and show that when the impurity-bath system is rotated at the cyclotron frequency, impurities behave as anyons as a consequence of the angular momentum exchange between the impurities and the bath. A possible experimental realization is proposed by identifying the statistics parameter in terms of the mean square distance of the impurities and the magnetization of the impurity-bath system, both of which are accessible to experiment. Another proposed application are impurities immersed in a two-dimensional weakly interacting Bose gas.

\end{abstract}

\maketitle

\section{Introduction}

A topological classification of interacting quantum states is crucial in the context of current research on topological states of matter~\cite{PhysRevLett.49.405,PhysRevLett.95.146802,PhysRevLett.98.106803,PhysRevLett.61.2015}. The discovery of such states in the fractional quantum Hall effect (FQHE)~\cite{Tsui_82} has revolutionized our understanding of the quantum properties of matter, and hence they are becoming landmarks for the current as well as future research directions in physics. One of the most important characterizations of topological states of matter is in terms of the underlying fractionalized excitations. As proposed by Laughlin~\cite{Laughlin_83}, the excitations in the FQHE are fractionally-charged quasiparticles, which were later demonstrated to be anyons with fractional statistics~\cite{Arovas_84}. Since then, anyons have received a significant amount of attention, also because of their potential role in quantum computation~\cite{lloyd2002quantum,kitaev2003fault,freedman2003topological,Nayak_08}.
Experimental evidence of anyons, on the other hand, is not yet conclusive and currently contested, despite that two recent works have reported encouraging results in that direction~\cite{bartolomei2020fractional,nakamura2020direct}.

Anyons are a type of quasiparticle whose quantum statistics interpolates 
between bosons and fermions. They occur only in lower-dimensional systems, i.e. mainly in two dimensions and to some extent in one dimension, although the latter will not be our focus here. The possibility of anyons in two dimensions may be traced to the algebraic triviality of the rotation group SO(2) and the topological non-triviality of a 2D configuration space with a point removed. Indeed, the symmetrization postulate which had been taken for granted during the first half of a century of quantum mechanics was called into question for such geometries in the 1960's-80's~\cite{Girardeau_1965,Souriau_1970,StrWil-70,laidlaw1971feynman,leinaas1977theory,goldin1981representations,Wilczek_82,Wilczek_82b,Wu_84,Froehlich_1990}. As elaborated for the first time by Leinaas and Myrheim~\cite{leinaas1977theory}, this leads to the possibility that when two identical particles are exchanged in two dimensions, the statistics parameter $\alpha$ can assume any intermediate value between $0$ (bosons) and $1$ (fermions):
\be
\label{spin_stat_1}
\psi_\mathrm{A}(r, \varphi + \pi) = e^{i \pi \alpha } \,  \psi_\mathrm{A}(r, \varphi) \, ,
\ee
where $\psi_\mathrm{A}(r, \varphi)$ is the two-body wave function in relative coordinates, $(r,\varphi)$. As a consequence, anyons have a peculiar property: when they are interchanged twice in the same way, the wave function does not return to the original. Namely, under a $2\pi$ rotation, the relative wave function is not single-valued, 
$\psi_\mathrm{A}(r, \varphi + 2 \pi) = e^{2 i \pi \alpha } \,  \psi_\mathrm{A}(r, \varphi)$. 
Nevertheless, if $\psi_\mathrm{A}(r, \varphi)$ is an eigenstate of some Hamiltonian $\hat H$, one may introduce a single-valued wave function, $\psi(r, \varphi) = \exp[-i \alpha \varphi ] \psi_\mathrm{A}(r, \varphi)$, which is governed by the Hamiltonian
\be
\label{anyon_ham}
 e^{-i \alpha \varphi} \, \hat H \left( \del/ \del \varphi \right) \, e^{i \alpha \varphi} = \hat H  \left( \del/ \del \varphi  + i \alpha \right) \, ,
\ee
where the statistics parameter now emerges as a gauge field. 
This establishes a connection between the statistics and gauge fields, and further implies that the orbital angular momentum of two particles in relative coordinates, which is given by $-i \del/ \del \varphi + \alpha$, is 
nonintegral~\cite{leinaas1977theory,Wilczek_82}. Such a configuration can be obtained with a magnetic field which substitutes the role of the statistics gauge field. This concept of anyons was discussed by Wilczek~\cite{Wilczek_82,Wilczek_82b}, who realized them as a flux-tube-charged-particle composite -- a charged particle `orbiting around' a magnetic flux $\alpha \pi$.

The picture of flux-tube-charged-particle composites provides a formal description of anyons. From the practical point of view, however, it does not give much insight concerning a physical realization. Indeed there has been a recent upsurge in interest concerning the realization of anyons as emergent quasiparticles in experimentally feasible systems,
in particular from the perspective of deriving robust, testable predictions such as density signatures \cite{CooSim-15,ZhaSreGemJai-14,ZhaSreJai-15,MorTurPolWil-17,UmuMacComCar-18,Yakaboylu_2018,CorDubLunRou-19}.
Also the emergence of anyons in a FQHE setting by means of attachment of flux via Laughlin quasiholes was recently revisited and elaborated on in Ref.~\cite{Lundholm_2016}. Here our main motivation is to define a physical Hamiltonian for a bipartite system such that the statistics gauge field emerges as a consequence of the interaction between the two subsystems~\cite{Moody_86,Yakaboylu17}. In particular, due to its experimental feasibility, we consider a quantum impurity model.

The presence of individual quantum particles, called impurities, is almost inevitable in many quantum settings. In several situations, ranging from crystals to helium nanodroplets to neutron stars, impurities are coupled to a complex many-body environment~\cite{landau1933uber,pekar1946local,frohlich1954electrons,Lemeshko_2015,duer2018probing}. 
Their interaction with a surrounding quantum-mechanical medium is the focus of quantum impurity problems. Impurities do not only appear to be good descriptions of experimental reality, but also provide intricate examples of quantum critical phenomena~\cite{kondo1964resistance,Anderson_67}. In general, quasiparticles formed by impurities are considered as an elementary building block of complex many-body systems. A well-known example is the polaron, which has been introduced as a quasiparticle consisting of an electron dressed by lattice excitations in a crystal~\cite{landau1933uber,pekar1946local,frohlich1954electrons}.  Over the years, with the help of recent advances in ultracold atomic physics, which enable a high degree of control over experimental parameters such as interactions and impurity concentration, quantum impurities have been investigated in several different experimental and theoretical studies~\cite{Schirotzek_09,Jorgensen_16,Cetina96,fukuhara2013quantum,koschorreck2012attractive,kohstall2012metastability,Camargo_18,Lemeshko_2015,PhysRevX.6.011012,Yakaboylu_2018_quantum,koepsell2019imaging}. In recent works~\cite{Yakaboylu17,Yakaboylu_2018}, it has been demonstrated that the many-particle environment manifests itself as an external gauge field with respect to the impurities. It has been also shown in Ref.~\cite{Yakaboylu_2018} that the angular momentum of a quantum planar rotor becomes fractional when it interacts with a 2D many-particle environment. This configuration resembles a two-anyon problem in relative coordinates when the relative distance between impurities is fixed.

In this manuscript, we consider identical impurities immersed in a many-particle bath and show that they turn into anyons in the introduced model as a consequence of their interaction with the surrounding bath. Particularly, we treat the impurities as a \textit{slow/heavy} and the surrounding bath as a \textit{fast/light} system, and demonstrate that the latter manifests itself as a statistics gauge field with respect to the impurities. Excitations of the surrounding bath -- a cloud of phonons -- attach vortices to each impurity so that the quasiparticle formed from the impurity dressed by phonons becomes an anyon. The main contributions of the manuscript can be summarized as follows:\\
(1) The introduced impurity model provides a new perspective of anyons as emergent quasiparticles built from composite bosons or fermions and a coherent state of quantized vortices. Such a perspective is not only conceptually helpful but also opens up new and simple numerical approaches to investigate the spectra of many anyons.
\\
(2) The model further paves the way towards realizing anyons in terms of impurities in standard condensed matter systems such as crystal lattices and ultracold gases. This offers a significant practical advantage over the strongly correlated materials which to date typically have been investigated to realize anyons.
\\
(3) The statistics parameter of the emerging anyons is identified as the phonon angular momentum, which allows us to measure the statistics parameter in terms of the mean square distance of the impurities and the magnetization of the impurity-bath system.

The paper is organized as follows. In Sec.~\ref{singular_free} we present the basic machinery for anyons and give a brief review of the regular ideal anyon Hamiltonian. Afterwards, in Sec.~\ref{impurity_model}, based on the emergent gauge picture, we derive the Hamiltonian of a quantum impurity model whose adiabatic limit corresponds to anyons. We present a transparent model where the impurities couple only to a single phonon mode. Then, we exemplify the model by investigating two- and three-impurity problems and their exact numerical spectra. In Sec.~\ref{new_perp} we present a new approach to the numerical solution of the many-anyon problem and provide a new perspective of anyons in terms of composite bosons or fermions. We discuss a possible experimental realization of the model in Sec.~\ref{exp_real} by considering heavy impurities interacting with collective excitations of a bath within the Fr\"{o}hlich-Bogoliubov model. We conclude the paper in Sec.~\ref{sec_conc} with a discussion of our results and questions for future work. The Appendix provides some further technical details. Throughout the manuscript we use natural units ($\hbar \equiv M \equiv 1$), unless otherwise stated.

\section{Anyon Hamiltonian} \label{singular_free}

In general, one can derive the statistics gauge field within Chern-Simons 
theory~\cite{Jackiw-90,iengo1992anyon,dunne1999aspects,de1993topological}. 
The action of a system of non-relativistic charged particles with mass $M$ coupled to the Abelian Chern-Simons gauge field $\mathcal{A}_\mu$ is given by
\be
S = \frac{M}{2} \int dt\, \sum_{q=1}^N \dot{\vec{x}}_q^2 + \int d^3 y \, \mathcal{A}_\mu j^\mu + \frac{\kappa}{2} \int d^3 y \, \epsilon^{\mu \nu \rho} \mathcal{A}_\mu \del_\nu \mathcal{A}_\rho \, .
\ee
Here $j^\mu (y) = \sum_{q=1}^N \dot{y}^\mu \delta^{(2)}(\vec{y} - \vec{x}_q)$ is a point-like source, $\kappa = 1/(2\pi \alpha)$ the level parameter, which assumes any number~\footnote{We note that in Chern-Simons theory, the flux is given by $\Phi= 1/\kappa$ so that the statistics parameter reads $\alpha = \Phi/(2\pi)$, whereas in the flux-tube-charged-particle composite picture, i.e., in Maxwell theory, the statistics parameter is given by $\alpha = 2 \xi/(2\pi)$ with $\xi$ being the flux of the each composite. In other words, the statistics phase is half of the flux in the former case, whereas in the latter case it is equivalent to the flux. This can be intuitively understood as follows. Interchanging two particles in Chern-Simons theory gives only the phase from the charges moving around the fluxes, but no contribution from the fluxes moving around the charges, whereas in Maxwell theory interchanging two composites gives the sum of these two phases~\cite{iengo1992anyon}.}, and $\epsilon^{\mu \nu \rho}$ the 2+1 dimensional Levi-Civita symbol. The letters $\mu, \nu, \rho$ indicate the 2+1 dimensional Lorentz indices, i.e., $\mu=0$ denotes the time component, whereas $\mu=1,2$ are the space components. By solving the zeroth component of the equations of motion, $\delta S/ \delta \mathcal{A}^0 = 0$ and defining $\mathcal{A} = \alpha \vec{A}$, one obtains the statistics gauge field:
\be
\label{anyon_gauge_field_0} 
A^i_q = \frac{\del}{\del x_q^i} \sum_{p'>p}^N \Theta_{p'p}  = \sum_{p(\neq q)=1}^N  \frac{\epsilon^{i j}\left( x^j_p - x^j_q \right)}{|\vec{x}_p - \vec{x}_q|^2} \, ,
\ee
where $\sum_{q>p}$ denotes the summation over both of the particle indices $q$ and $p$ with the condition $q>p$, $\Theta_{qp} = \tan^{-1}\left[x^2_{qp}/x^1_{qp}\right] $ is the relative polar angle between particles $q$ and $p$, $x^i_{qp} = x^i_q - x^i_p$, and we define $\epsilon^{ij} = \epsilon^{0ij}$ with $\epsilon^{12}=1$. The Hamiltonian for $N$ ideal non-interacting anyons with statistics parameter $\alpha$ can be written as
\be
\label{N_anyon}
 \hat{H}_\text{$N$-anyon} = - \frac{1}{2} \sum_{q=1}^N \left[\vec{\nabla}_q + i \alpha  \vec{A}_q\right]^2 \, .
\ee
Without loss of generality, we consider the Hamiltonian $\hat{H}_\text{$N$-anyon}$ to act on bosonic states,
i.e. the Hilbert space $\cH = L^2_{\textrm{sym}}(\R^{2N})$ of square-integrable functions on $\R^{2N}$ which are symmetric w.r.t. exchange. Later we also consider fermionic states by changing $\alpha$ to $\alpha-1$ and acting on antisymmetric functions $ L^2_{\textrm{asym}}(\R^{2N})$.

\subsection{Regular anyon Hamiltonian}
\label{sec_reg_anyon_ham}

The eigenvalue problem for the Hamiltonian~\eqref{N_anyon} has been solved analytically only for the two-anyon case \cite{leinaas1977theory,Wilczek_82b,AroSchWilZee-85} (see also Refs.~\cite{vercin1991two,myrheim1992two} for a system of two anyons in the presence of the Coulomb potential), while for three or more particles only part of the spectrum is known exactly.
The three- and four-anyon spectra have been investigated by means of numerical diagonalization techniques~\cite{Murthy_91,Sporre_91,SpoVerZah-92,SpoVerZah-93}, and a subspace of exact eigenstates is also known analytically for arbitrary $N$
\cite{Wu_84_anyon,Chou-91a,Chou-91b,Myrheim-99,Khare-05}.
Rigorous upper and lower bounds on the exact ground-state energy were established in Refs.~\cite{ChiSen-92,LunSol-13a,LunSol-13b,LunSol-14,Lundholm_2017,LunSei-17}.
Another approach has been to first regularize the Hamiltonian \eqref{N_anyon}
by making the fluxes extended \cite{Trugenberger-92,Trugenberger-92b,ChoLeeLee-92,LarLun-16},
and in this situation an exact average-field theory and a corresponding Thomas-Fermi theory may be derived in the almost-bosonic limit $\alpha \sim N^{-1} \to 0$ 
\cite{LunRou-15,CorLunRou-16,CorLunRou-proc-17,CorDubLunRou-19,Girardot-19}.
Also singular or point-interacting anyons may be considered \cite{Grundberg-etal-91,ManTar-91,BorSor-92,MurLawBhaDat_92,CorOdd-18},
as well as anyons regularized by a strong magnetic field \cite{Ouvry-07}. See Refs.~\cite{Froehlich_1990,iengo1992anyon,Myrheim-99,Khare-05} for reviews.

For ideal point-like anyons, however, the form of the Hamiltonian \eqref{N_anyon} leads to some singularity problems when the anyon spectrum is investigated from the bosonic end. 
For example, let us consider two anyons confined additionally in a harmonic-oscillator potential. 
The Hamiltonian in relative coordinates is given by
\be
\label{2imp_rel}
 \hat{H}_\text{2-anyon} = - \frac{1}{2 r^2} \left( \frac{\del}{\del \varphi} + i \alpha \right)^2 - \frac{1}{2 r}\frac{\del}{\del r}\left(r \frac{\del}{\del r} \right) + \frac{r^2}{2}  \, .
\ee
We observe that this form of the Hamiltonian allows neither a perturbative treatment of the problem nor the use of diagonalization techniques with respect to the free operator. Namely, the matrix element $ \bra{l,m} r^{-2} \ket{l,m} $, where the state $\ket{l,m}$ is the eigenstate of the harmonic oscillator in polar coordinates with the principal and magnetic quantum numbers $l$ and $m$, respectively, is logarithmic divergent for the $m = 0$ states. For the three-anyon case, on the other hand, the problem occurs for all the zeroth-order bosonic eigenstates~\cite{mccabe1991perturbative}. This is a well-known problem, which necessitates a certain (arguably somewhat ambiguous) scheme to select regular solutions corresponding to free anyons; see, for instance, Refs.~\cite{amelino1994perturbative,comtet1995magnetic,mccabe1991perturbative,Mashkevich_1996}
(indeed free anyons must be regular, which in technical terms corresponds to selecting the Friedrichs extension away from two-particle diagonals as the preferred self-adjoint realization of Eq.~\eqref{N_anyon}~\cite{LunSol-14}.)

This problem can be overcome with the similarity transformation 
$ \hat{\tilde{H}}_\text{2-anyon} = r^{-\alpha} \hat H_\text{2-anyon} r^{\alpha} $, 
which is a self-adjoint Hamiltonian in the $L^2$ space weighted by the measure $r^{2\alpha +1}dr$. In the transformed Hamiltonian the divergent term vanishes~\cite{de1993topological}. In fact, this transformation corresponds to a `real gauge transformation' leading to an imaginary vector potential. In other words, it leads to the replacement of $\del/\del r \to \del/\del r + \alpha /r$ in the Hamiltonian~\eqref{2imp_rel}. If we further combine this transformation with the one given in Eq.~\eqref{anyon_ham}, the regular (singular-free) two-anyon Hamiltonian can be obtained directly from the bosonic ($\alpha=0$) Hamiltonian via the transformation:
\be
\hat{\tilde{H}}_\text{2-anyon} = \exp\left[ -i \alpha(\varphi - i \ln r)\right] \hat H_\text{2-boson} \exp\left[i \alpha(\varphi - i \ln r)\right].
\ee
We can generalize this and obtain the regular $N$-anyon Hamiltonian as
\be
\label{N_anyon_free}
\hat{\tilde{H}}_\text{$N$-anyon} = \left( \prod_{q>p}^N z_{qp}^{-\alpha} \right) \hat{H}_\text{$N$-boson} \left( \prod_{q>p}^N z_{qp}^{\alpha} \right) = - \frac{1}{2} \sum_{q=1}^N \left[\vec{\nabla}_q + i \alpha  \tilde{\vec{A}}_q \right]^2 \,
\ee
with the complex gauge field
\be
\label{anyon_gauge_field}
\tilde{\vec{A}}_q = \vec{\nabla}_q \sum_{p'>p}^N \left( \Theta_{p'p} - i \ln r_{p'p} \right) \, ,
\ee
where we define $\hat H_\text{$N$-boson} = - \sum_{q=1}^N \vec{\nabla}_q^2 /2 $, $z_{qp} = x^1_{qp} + i x^2_{qp} = e^{i (\Theta_{qp} - i \, \ln r_{qp} )}$, 
and $r_{qp} = |\vec{x}_q - \vec{x}_p|$. 
In the gauge field~\eqref{anyon_gauge_field} the second term identifies the imaginary vector potential that eliminates the singularities arising due to the first term. This can be observed with the disappearance of the $\tilde{\vec{A}}_q^2$ term in the Hamiltonian, i.e., $\tilde{\vec{A}}_q^2 = \vec{A}_q^2 - \vec{A}_q^2 = 0$ because of the Cauchy-Riemann equations (note that the terms in \eqref{anyon_gauge_field} are each other's harmonic conjugates). The $N$-anyon Hamiltonian in this gauge can be written as
\be
\label{N_anyon_ham_2}
\hat{\tilde{H}}_\text{$N$-anyon} = - \frac{1}{2} \sum_{q=1}^N \left(\vec{\nabla}_q^2 + 2 i \alpha \sum_{p(\neq q)=1}^N \frac{\epsilon^{i j}x^j_{pq} + i x^i_{pq}}{r_{pq}^2} \frac{\del}{\del x^i_q} \right) \, .
\ee
The above Hamiltonian is again self-adjoint in a weighted space (with the weight $\prod_{q>p}r_{qp}^{2\alpha}$ multiplying the usual measure), and it may serve as a model Hamiltonian for the calculation of the corresponding anyon spectra. Therefore, in the rest of the paper we will also consider this computationally convenient regular Hamiltonian $\hat{\tilde{H}}_\text{$N$-anyon}$, even though the main emphasis will be placed on the physical one~$ \hat{H}_\text{$N$-anyon} $.

\section{Impurity Model} \label{impurity_model}

We start by considering impurities immersed in a weakly interacting bath. Within the Fr\"{o}hlich-Bogoliubov theory~\cite{frohlich1954electrons,bogolyubov1947theory,Pitaevskii2016}, a general Hamiltonian of an impurity problem is given by
\bal
\label{imp_ham}
\hat H_\text{qim} & =  -\sum_{q=1}^N \frac{\vec{\nabla}_q^2}{2}  + \sum_{l = 1}^\Lambda \omega_l  \, \hat b^\dagger_{l} \hat b_{l} \\
\nonumber & + \sum_{l = 1}^\Lambda \lambda_l (\vec{x}) \left( e^{-i \beta_l (\vec{x}) } \hat b^\dagger_{l} + e^{i \beta_l (\vec{x}) } \hat b_{l} \right) \, .
\eal
Here the first term corresponds to the kinetic energy of the identical impurities, which are considered to be either bosons or fermions. The second term is the kinetic energy of the many-particle bath, whose collective excitations are given with a gapped dispersion relation $\omega_l$. In what follows, we will refer these excitations simply as ``phonons'' regardless of its actual meaning. $\Lambda$ is the total number of phonon modes. The creation and annihilation operators associated to each phonon mode, $\hat b^\dagger_{l}$ and $\hat b_{l}$, obey the commutation relation  $ \left[ \hat b_{l}, \hat b^\dagger_{l'}\right]  = \delta_{l l'}$. The final term describes the interaction between the impurities and the many-particle bath. While we assume that $\lambda_l (\vec{x})$ is a real function, $\beta_l (\vec{x})$ will at this stage be allowed to be complex for a later purpose. Both functions may depend on all the variables $\vec{x} = \{ \bx_1,\ldots,\bx_N \}$.

As we discuss in more detail in Sec.~\ref{exp_real} on specific applications, we consider heavy impurities. This, together with a gapped dispersion relation, allows us to treat the bosonic or fermionic impurities as the slow system and the rest of the Hamiltonian as the fast one:
\be
\label{H_fast}
\hat H_\text{fast} (\vec{x} ) =\sum_{l} \omega_l  \, \hat b^\dagger_{l} \hat b_{l}  + \sum_{l} \lambda_l (\vec{x}) \left( e^{-i \beta_l (\vec{x} ) } \hat b^\dagger_{l} + e^{i \beta_l (\vec{x} ) } \hat b_{l} \right),
\ee
where the coordinates of impurities, $\vec{x}$, are regarded as parameters. 
We note that even if $\beta_l$ is complex, $\hat H_\text{fast} (\vec{x} )$ is self-adjoint in the $\hat w = \exp\left[ - 2 \sum_{l} \text{Im}[\beta_l ] \hat b^\dagger_{l} \hat b_{l} \right]$-weighted Fock space. The eigenstates and eigenvalues of the Hamiltonian 
\eqref{H_fast} 
can be found by applying the following two transformations:
\be
\label{s_trans}
\hat S = \exp\left[ - i  \sum_{l} \beta_l \, \hat b^\dagger_{l} \hat b_{l} \right] \, , \, \hat U  = \exp\left[ - \sum_{l} \frac{\lambda_l}{\omega_l}\left( \hat b^\dagger_{l} - \hat b_{l} \right) \right] \, ,
\ee
where the transformation $\hat S$ is, in general, a similarity transformation, as  $\beta_l $ might be complex, whereas $\hat U$ is unitary. Therefore, the eigenstates can be written as $\hat S \ket{\psi_n}$ with $\ket{\psi_n} = \hat U \ket{n}$. Here the states $\ket{n}$ symbolically represent  normalized phonon states with the collective index $n$. Namely, $\ket{0}$ is the vacuum state of the bath, $\ket{1} \equiv \hat b^\dagger_l \ket{0}$ a one-phonon state, and so on. (Later, in simplified models, where we consider a single phonon mode, the states $\ket{n}$ correspond to the usual harmonic-oscillator eigenstates). The eigenvalues, on the other hand, are given by $\varepsilon_n = \sum_{i=1}^n \omega_{l_i} + \varepsilon_0$ with the ground state energy $\varepsilon_0 = - \sum_{l} \lambda_l^2/\omega_l$. 

Let us assume that there exists a large energy gap between the vacuum state $\ket{0}$ and the excited states $\hat b^\dagger_l \ket{0}$, i.e., we consider the limit $\omega_l \to \infty$, which corresponds to a physical realization of heavy impurities interacting with gapped excitations of a bath, see Sec.~\ref{exp_real}. In this adiabatic limit the lowest energy spectrum of the Hamiltonian~\eqref{imp_ham} is given by the Schr\"{o}dinger equation
\be
\label{bos_anyon_Ham}
\left( - \frac{1}{2} \sum_{q=1}^N \left[ \vec{\nabla}_q + i \vec{G}_q \right]^2 + W(\vec{x}) \right)  \chi_0^E (\vec{x}) = E \, \chi_0^E (\vec{x})\, ,
\ee
where
\be
\label{emergent_gauge_field}
\vec{G}_q = -i \bra{\psi_0} \hat S^{-1} \vec{\nabla}_q \hat S \ket{\psi_0} = -\sum_{l}  \left( \lambda_l (\vec{x}) / \omega_l \right)^2 \vec{\nabla}_q \beta_l (\vec{x})
\ee
is the emergent gauge field and $W(\vec{x})$ the emergent scalar potential. The corresponding lowest energy eigenstates of the Hamiltonian~\eqref{imp_ham} are, then, given by 
\be 
\label{total_state}
\ket{\Psi^E (\vec{x})} = \chi_0^E \hat S \hat U \ket{0} \, ;
\ee
see Appendix~A for details. If we are able to match the emergent gauge field~\eqref{emergent_gauge_field} with the statistics gauge field, the aforementioned adiabatic solution of the full problem described by the Hamiltonian~\eqref{imp_ham} corresponds to the problem of $N$ anyons interacting with the potential $W(\vec{x})$. Although our essential focus is the statistics gauge field given by Eq.~\eqref{anyon_gauge_field_0}, i.e., the matching of $\vec{G}_q = \alpha \vec{A}_q$, we also consider the other choice~\eqref{anyon_gauge_field} within a toy model for computational purposes. We emphasize that the condition $\omega_l \to \infty$ does not necessarily cancel out the emergent gauge field $\vec{G}_q $, as we demonstrate in particular examples below.

At first sight, it looks like we have made the $N$-anyon problem more complicated, as we consider in Eq.~\eqref{imp_ham} bosonic or fermionic particles in a many-particle bath, instead of particles interacting with the statistics gauge field. However, the corresponding quantum impurity setup, first, lays the groundwork for realizing anyons in experiment (an experimental realization will be proposed in Sec.~\ref{exp_real}). Furthermore, as will be discussed next, this formulation reveals new insights into the structure of the anyon Hamiltonians \eqref{N_anyon} and \eqref{N_anyon_ham_2}, and on how anyons may emerge from composite bosons (fermions) -- the topological bound state of a boson (fermion) and an even number of quantized vortices (cf. e.g. \cite{jain2007composite}). Finally, it introduces new techniques for numerical investigation of the $N$-anyon problem.

\subsection{Emergent interacting anyons}

By using the definition of the statistics gauge field~\eqref{anyon_gauge_field_0} as well as imposing adiabaticity in the problem leading to Eqs.~\eqref{bos_anyon_Ham} and \eqref{emergent_gauge_field} we can identify the parameters of the many-particle bath that turn the impurities into anyons. In order to present features of the introduced model in a transparent way, we consider for simplicity the following $l$-independent expressions: 
\be
\label{beta_for_gauge}
\beta_l (\vec{x}) = \beta (\vec{x}) = -s \, \sum_{q>p}^N \, \Theta_{qp}  \, ,
\ee
constant $\lambda_l = \lambda$, and $\omega_l = \omega$. Here $s$ is an integer such that $\beta(\bx_q)$ has the correct periodicity under the continuous exchange of particles and the Hamiltonian~\eqref{imp_ham} commutes with the permutation operators. Without loss of generality it can be set to its lowest possible non-trivial value $s=2$. The minus sign in \eqref{beta_for_gauge} is chosen simply to eventually make our emergent anyons positively oriented.

We now introduce a unitary transformation $ \hat a^\dagger_{l} = \sum_{l'} v_{l l'} \hat b^\dagger_{l'} $, with $\sum_{l''} v_{l' l''} v_{l l''}^* = \delta_{l' l}$, such that $\hat a^\dagger_1 = \sum_{l=1}^{\Lambda} \hat b^\dagger_l / \sqrt{\Lambda}$ and consider the choice of $\omega = \sqrt{\Lambda}$, which simplifies the Hamiltonian. Then, the impurity Hamiltonian~\eqref{imp_ham} can be written as
\be
\label{imp_ham_anyon_int}
\hat H_\text{qim}   = -\frac{1}{2}\sum_{q=1}^N \vec{\nabla}_q^2 +  \omega \left( \hat a^\dagger \hat a + \lambda^2 \right) + \lambda\omega \left( F \, \hat a^\dagger + F^{-1} \, \hat a \right)  \, 
\ee
with
\be
\label{F}
F = \prod_{q>p}^N \left(\frac{z_{qp}}{|z_{qp}|} \right)^2 \, .
\ee
For later convenience we use the notation $F^{-1}$, instead of $F^{*}$, even though they are equivalent in this particular case. 
In Eq.~\eqref{imp_ham_anyon_int} we have neglected the term $\sum_{l \ge 2}^{\Lambda} \omega \, \hat a^\dagger_{l} \hat a_{l}$ as the impurities couple only to a single phonon mode, and omit the subindex, $\hat a^\dagger_1 \to \hat a^\dagger$. We further subtracted the ground state energy of the phonon part of the Hamiltonian, $ -\omega\lambda^2$. Then, \emph{defining}
\be
\label{alpha}
\alpha = 2 \lambda^2,
\ee
and keeping the coupling $\lambda$ fixed in the limit of $\omega \to \infty$, 
the lowest energy spectrum of the quantum impurity problem~\eqref{imp_ham_anyon_int} is governed by Eq.~\eqref{bos_anyon_Ham} with the gauge field $\vec{G}_q = \alpha \vec{A}_q$ and scalar potential $ W = \alpha \sum_{q=1}^N \vec{A}_q^2$. 

The emergence of the anyon Hamiltonian can also be obtained without going to the gauge picture, by direct diagonalization of the Hamiltonian~\eqref{imp_ham_anyon_int}. Namely, if we apply the corresponding $\hat S$ and $\hat U$ transformations,
\be
\hat S = F^{\hat a^\dagger \hat a} \, , \quad 
\hat U  = \exp\left[ - \sqrt{\alpha/2}\left( \hat a^\dagger - \hat a \right) \right] \, ,
\ee
the transformed Hamiltonian can be written as
\bal
\label{imp_ham_23}
& \hat H'_\text{qim} = \hat U^{-1} \hat S^{-1} \hat H_\text{qim} \hat S \hat U =  \omega \, \hat a^\dagger \hat a \\
\nonumber & - \frac{1}{2} \sum_{q=1}^N \left[ \vec{\nabla}_q + 2 i \vec{A}_q (\vec{x})\left(\hat a^\dagger - \sqrt{\frac{\alpha}{2}} \right) \left(\hat a - \sqrt{\frac{\alpha}{2}} \right) \right]^2 \, .
\eal
If we diagonalize the Hamiltonian in the basis of the eigenstates of the free operators, the transition 
between different phonon states for any bosonic (respectively fermionic) eigenstate $\ket{\Phi}$ of the free $N$-particle Hamiltonian is quantified by the matrix element $\left\langle \bra{n} \hat H'_\text{qim} \ket{m}\right\rangle_{\Phi} /\left\langle \bra{m} \hat H'_\text{qim} \ket{m}- \bra{n} \hat H'_\text{qim} \ket{n}\right\rangle_{\Phi}$ with $n \neq m$. In particular, the transition matrix element between the vacuum state and the one-phonon state is
\be
\label{transition} 
\sqrt{\alpha/2}\frac{\left\langle \sum_q \left(2i [\vec{\nabla}_q,\vec{A}_q]_{+} -(1+\alpha)\vec{A}_q^2 \right) \right\rangle_{\Phi}}{2 \omega- \left\langle\sum_q \left(2i [\vec{\nabla}_q,\vec{A}_q]_{+} -(1+2\alpha)\vec{A}_q^2 \right)\right\rangle_{\Phi}} \, ,
\ee
where $[,]_{+}$ is the anti-commutator. Then, in the limit of $\omega \to \infty$ the transition matrix element becomes negligible (we refer to Appendix~C for more details), and hence, the vacuum expectation value
\be
\label{int_anyons}
\bra{0} \hat H'_\text{qim} \ket{0} = - \frac{1}{2} \sum_{q=1}^N \left[ \vec{\nabla}_q + i\alpha \vec{A}_q (\vec{x}) \right]^2 + W(\vec{x}) \, ,
\ee
decouples from the rest of the spectrum. This is basically the statement of the adiabatic theorem. As a result, the energy levels belonging to the vacuum sector governed by Eq.~\eqref{int_anyons} describes interacting anyons in the potential $ W(\vec{x}) = \alpha \sum_{q=1}^N \vec{A}_q^2 (\vec{x}) $. We note, however, that for ideal point-like anyons the expectation values $\bk{\vec{A}_q^2}_{\Phi}$ in Eq.~\eqref{transition} are divergent for certain bosonic eigenstates of the free Hamiltonian, as we discussed in Sec.~\ref{sec_reg_anyon_ham}, and therefore, the adiabatic theorem breaks down. In order to avoid these singularities the corresponding spectrum can be calculated instead from the fermionic end by changing $\alpha$ to $\alpha-1$ and $ L^2_{\textrm{sym}}(\R^{2N})$ to $ L^2_{\textrm{asym}}(\R^{2N})$ for the impurities.

Let us investigate the form of the Hamiltonian~\eqref{imp_ham_anyon_int} and the emergence of anyons in more detail. In this simplified model the interaction between the impurities and phonon field is described by the factor $F$. The $\hat S$ transformation attaches two flux/vortex units to bosonic (fermionic) impurities to convert them into composite bosons (fermions). (In fact, in this particular case~\eqref{F} we call it a flux rather than a vortex, as it is just a phase factor.) Then, a cloud of phonons, which manifests itself through a coherent state, dresses each impurity and forms a quasiparticle governed in the limit of $\omega \to \infty$ by the anyon Hamiltonian \eqref{int_anyons}. The corresponding coherent state of phonons is given by
\be
\label{coherent_state}
	|\psi_0\rangle = \hat{U}|0\rangle = e^{-\alpha/4} \sum_{n=0}^\infty \frac{\left(-\sqrt{\alpha/2}\right)^n}{\sqrt{n!}} |n\rangle \, ,
\ee
which involves infinitely many phonons, weighted according to $\alpha$. In this adiabatic limit, the states in the Fock space decouple from each other, and each resulting energy level is filled by anyons with the statistics parameter $2n + \alpha$, where $n$ corresponds to the energy levels of the Fock space sector. Specifically, if the impurities are initially bosons, in the vacuum state, $n=0$, they turn into interacting anyons with statistics parameter $\alpha$ and spectrum described by Eq.~\eqref{int_anyons}. In the excited states, $n \neq 0$, on the other hand, the impurities become composite bosons with $2n$ units of flux and interacting with the statistics gauge field $\alpha \vec{A}_q$ as well as the scalar potential $\alpha(1+2n)\sum_q\bA_q^2$. See Appendix~C for more details concerning the diagonalization of Eq.~\eqref{imp_ham_23}.

Below we focus only on the vacuum state and present two simple examples. For an easier comparison with the results existing in the literature, we investigate impurities confined additionally in a harmonic-oscillator potential and allow for fermionic impurities with the coupling $\alpha -1 = s \lambda^2$ in order to deal with the singular interaction. Furthermore, since the interaction term depends only on relative coordinates, we factor out the center-of-mass problem, and make a transformation to relative coordinates. In general, the relative coordinates for an $N$-body problem are given by Jacobi coordinates: $ \vec{R} = \sum_{i=1}^N z_i / \sqrt{N} $ and $u_m = (\sum_{i=1}^m z_i - m z_{m+1})/\sqrt{m(m+1)}$, where $m$ runs from 1 to $N-1$. 

\subsubsection{Two anyons}

\begin{figure}
  \centering
        \includegraphics[width=\linewidth]{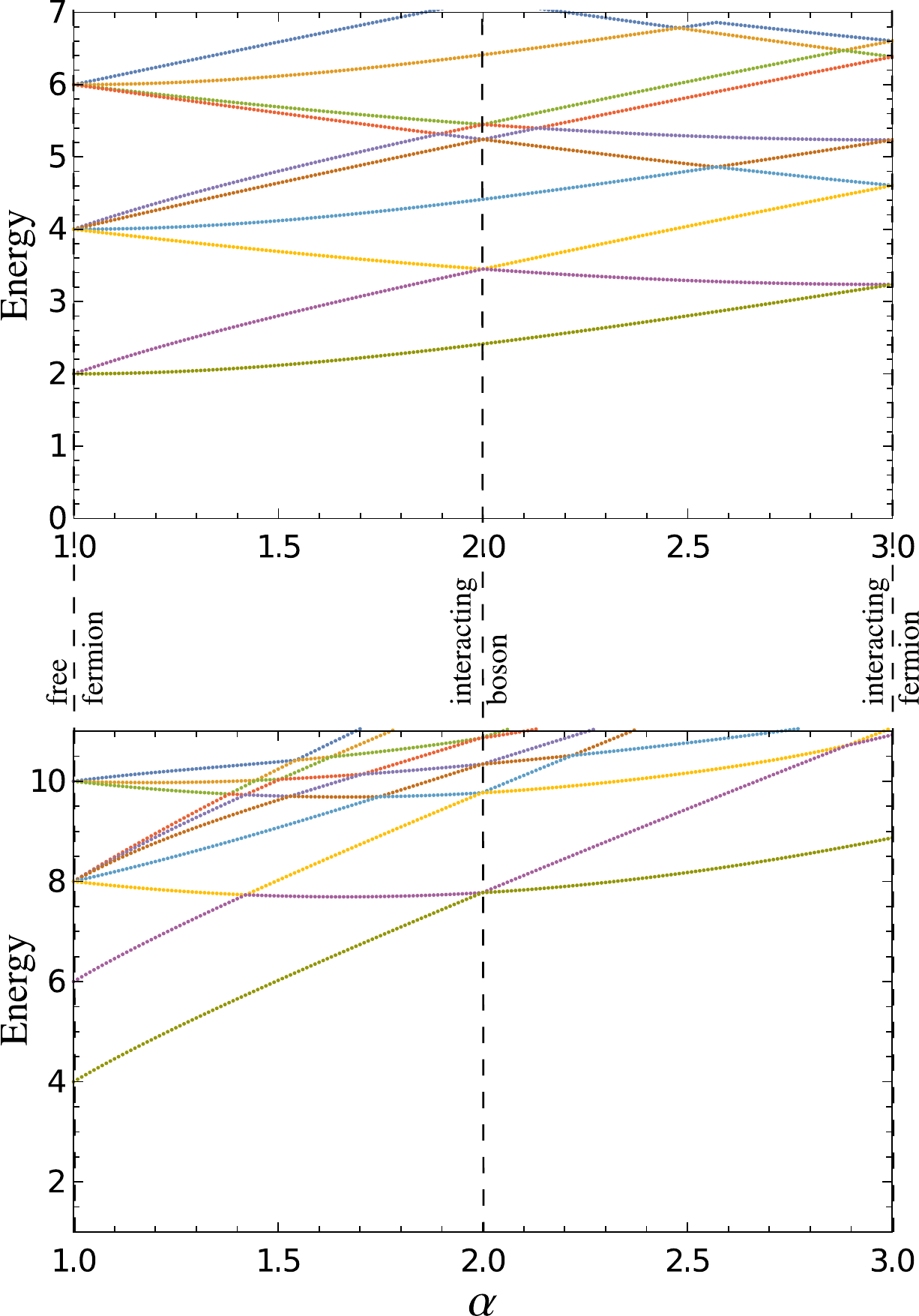} 
 \caption{Calculations of the two-anyon (Top) and three-anyon (Bottom) spectra for the interacting anyon model in an external harmonic-oscillator potential~\eqref{imp_ham_anyon_int}. The energies are given in units of the harmonic frequency. The spectra have been calculated from the fermionic end, i.e., the coupling is chosen as $\alpha -1 = s \lambda^2$ such that $\alpha=1$ corresponds to free fermions. The applied parameters are $l_\text{max}=10$, $m_\text{max} = 21$, and $\omega=23$ with up to 5 phonons for the two-anyon case. For the three-anyon case we consider all the antisymmetric impurity wave functions~\eqref{3_impt_wave_function} restricted by the condition $E_{n_\text{max} m_\text{max}} \le 26$ and limit the maximum number of phonons to 10 with $\omega=54$. For clarity, we do not display all the curves in the second plot.} 
 \label{fig_int}
\end{figure}

As a first example, we consider the simplest case -- the two-impurity problem. The Hamiltonian in Jacobi coordinates with the notation $u_1 = r \exp(i \varphi)$, which is simply usual relative coordinates for a 2-body problem, becomes
\bal
\label{two_imp_int_1}
\nonumber \hat H_\text{2-imp} =&  - \frac{\vec{\nabla}^2}{2}  + \frac{ r^2}{2}  + \omega \,\left( \hat a^\dagger \hat a + \frac{\alpha-1}{2}  \right)   \\ 
  &  + \omega\sqrt{\frac{\alpha-1}{2}} \left[ e^{2i \varphi}  \hat a^\dagger +  e^{-2i \varphi}  \hat a \right]  \, ,
\eal
where $\alpha \ge 1$. This two-impurity problem can be solved numerically by diagonalizing the Hamiltonian with the eigenstates of the free Hamiltonian -- the first line of Eq.~\eqref{two_imp_int_1}. These eigenstates are the phonon states~$\ket{n}$ times the antisymmetric impurity states. The latter are the usual harmonic oscillator wave functions:
\be
\label{2_har_osc}
\Phi_{lm} = \sqrt{\frac{2(l!)}{(l+|m|+1)!}} e^{-r^2/2}r^{m}L_{l}^{m}(r^2)\exp(i m \varphi)/\sqrt{2\pi} \, ,
\ee
where $L_{l}^{m}(r^2)$ are associated Laguerre polynomials. The corresponding eigenvalues are given by $2l + |m| + 1$. The antisymmetric impurity states in Jacobi coordinates follow from the parity of the angular quantum number of the impurities $m$: while even $m$ refers to bosons, the odd ones correspond to fermions. Due to the finite number of impurity states considered in numerics, there is an intricate relation between the number of impurity wave functions, maximum number of phonons, and actual value of $\omega$ in order to achieve convergence. The converged result of the anyonic spectra for the Hamiltonian~\eqref{two_imp_int_1} is obtained with $\omega \gtrsim 20$, up to 5 number of phonons, and several hundred impurity states in the fermionic basis, which is shown in In Fig.\ref{fig_int} (Top). In fact, this result can also be found analytically. Namely, it follows from Eq.~\eqref{int_anyons} that the lowest levels are described by the Hamiltonian
\bal
\label{2imp_rel_int}
\nonumber \bra{0} \hat H'_\text{2-imp} \ket{0} & = - \frac{1}{2 r^2}\left[ \left( \frac{\del}{\del \varphi} + i (\alpha-1) \right)^2 - 2 (\alpha-1) \right] \\
&- \frac{1}{2 r}\frac{\del}{\del r}\left(r \frac{\del}{\del r} \right)  + \frac{r^2}{2}  \, .
\eal
Then, the corresponding eigenvalues directly follow from the harmonic oscillator ones by replacing $|m|
$ with $\sqrt{(m+(\alpha-1))^2+2(\alpha-1)}$ to yield $2l + \sqrt{(m+(\alpha-1))^2+2(\alpha-1)} + 1$, which agrees with the numerical result shown in Fig.\ref{fig_int} (Top).

\subsubsection{Three anyons}

As a next example, we study the three-impurity problem. In Jacobi coordinates, with the notation $u_1 = \eta = \eta_x + i \eta_y$ and $u_2 = \xi = \xi_x + i \xi_y$, the Hamiltonian is given by 
\bal 
\label{two_imp_int_3}
\nonumber \hat{\tilde{H}}_\text{3-imp} & =  -\frac{1}{2}\left( \vec{\nabla}_\eta^2 + \vec{\nabla}_\xi^2 \right) + \frac{1}{2} \left( \vec{\eta}^2 + \vec{\xi}^2 \right)  + \omega \left( \hat a^\dagger \hat a + \frac{\alpha-1}{2} \right) \\
& + \omega\sqrt{\frac{\alpha-1}{2}}  \left( \frac{\eta^3 - 3 \eta \xi^2}{|\eta^3 - 3 \eta \xi^2|}\right)^2  \hat a^\dagger + \text{H.c.}  \, 
\eal
with $\alpha \ge 1$, $\vec{\eta}^2 = \eta_x^2 + \eta_y^2 $, and $\vec{\xi}^2 = \xi_x^2 + \xi_y^2 $. In contrast to the two-impurity problem, the implementation of the permutation symmetry for the impurity states in Jacobi coordinates is not trivial due to the relations $P_{23} \eta = (\eta +\sqrt{3} \xi)/2$ and $P_{23} \xi = ( \sqrt{3}\eta - \xi)/2$. Accordingly, by following the convention introduced by Kilpatrick and Larsen~\cite{kilpatrick1987set}, we use hyperspherical coordinates, $(\rho, \theta, \phi, \psi)$, 
\bals
\eta & = \rho e^{-i \psi}\left(\cos \theta \cos \phi + i \sin \theta \sin \phi \right) \, ,\\
\xi & = \rho e^{-i \psi} \left(\cos \theta \sin \phi - i \sin \theta \cos \phi \right)\, ,
\eals
where the symmetrization of the wave function is straightforward. Namely, the bosonic (+) or fermionic (-) wave functions are given by
\be
\label{3_impt_wave_function}
\Phi^{\pm}_{n m \nu \mu} = \sqrt{\frac{2 (n!)}{(m+n+1)!}} e^{-\rho^2/2}\rho^{m}L_{n}^{m+1}(\rho^2) Y^{\pm}_{m \nu \mu} (\theta,\phi,\psi) \, ,
\ee
where $Y^{\pm}_{m \nu \mu} (\theta,\phi,\psi)$ are the hyperspherical harmonics; see Appendix~B. The wave functions are normalized over the volume $d V = \rho^3  d \rho \, \cos(2\theta) d \theta \, d \phi \, d \psi$ for $\rho \in [0,\infty) \, , \theta \in [-\pi/4,\pi/4]\, , \phi \in [-\pi/2,\pi/2]\, , \psi \in [0,2\pi]$. Here, $n= 0,1,2, \cdots$ is the radial quantum number and $m= 0,1,2, \cdots$ is one of the angular momentum numbers such that the corresponding spectrum reads $ E_{n m} = (2n + m + 2) $. The other two angular quantum numbers $(\nu, \mu)$ have the same parity as $m$ and they are restricted by: $|\nu|, |\mu| \le m$ and $\nu = 3q$ with a non-negative integer $q$; see also Ref.~\cite{khare1991perturbative}. In Fig.~\ref{fig_int} (Bottom), we show the corresponding spectra. In the diagonalization procedure over one thousand fermionic impurity states are considered and the maximum number of phonons is limited to 10 with $\omega \approx 50$ resulting in converged spectra in the regime considered.

\subsection{A toy model for free anyons}

Eq.~\eqref{int_anyons}, which describes the lowest levels of the impurity problem~\eqref{imp_ham_anyon_int} in the limit of $\omega \to \infty$, can also serve as a new platform for studying the original $N$-anyon problem without the presence of the interaction potential $W$, i.e., free anyons. Namely, instead of Eq.~\eqref{beta_for_gauge}, if we define 
\be
\beta_l (\vec{x}) = \beta (\vec{x}) = - s \, \sum_{q>p}^N \left( \Theta_{qp} - i  \, \ln r_{qp} \right)
\ee
with the same $\omega$ and $\lambda$ as in the previous case, the emergent gauge field is given by Eq.~\eqref{anyon_gauge_field}. In this case the corresponding impurity Hamiltonian analogous to Eq.~\eqref{imp_ham_anyon_int} can be written as
\be
\label{imp_ham_anyon_free}
\hat{\tilde{H}}_\text{qim}   = -\frac{1}{2}\sum_{q=1}^N \vec{\nabla}_q^2 +  \omega \left( \hat a^\dagger \hat a + \lambda^2 \right) +  \lambda \omega \left( \tilde{F} \, \hat a^\dagger + \tilde{F}^{-1} \, \hat a \right)  \, 
\ee
with
\be
\label{F_tilde}
\tilde{F} = \prod_{q>p}^N\, z_{qp}^2 \, .
\ee
We note that as $\tilde{F}^{-1} \neq \tilde{F}^{*}$, the Hamiltonian~\eqref{imp_ham_anyon_free} is not Hermitian in this case. 
Nevertheless, its non-Hermiticity is harmless for our purposes, and indeed the fast part of the Hamiltonian is self-adjoint in a Fock space weighted by $|\tilde{F}|^{2\ha^\dagger \ha}$. 
This allows us to rewrite the Hamiltonian in the gauge picture so that the lowest energy levels of the Hamiltonian~\eqref{imp_ham_anyon_free} become real in the limit of $\omega \to \infty$. Namely, if we diagonalize the Hamiltonian, the energy levels belonging to the vacuum state in this limit are given by Eq.~\eqref{int_anyons} with the replacement $\vec{A}_q \to \tilde{\vec{A}}_q$. Moreover, as $\tilde{\vec{A}}_q^2 = 0$, the emergent scalar potential $W(\vec{x})$ vanishes and hence Eq.~\eqref{int_anyons} for this case simply corresponds to the free $N$-anyon Hamiltonian defined in the regular gauge~\eqref{N_anyon_ham_2}:
\be
\label{free_anyons}
\bra{0} \hat{\tilde{H}}'_\text{qim} \ket{0} = - \frac{1}{2} \sum_{q=1}^N \left[ \vec{\nabla}_q + i\alpha \tilde{\vec{A}}_q (\vec{x}) \right]^2  \, .
\ee
Therefore, the Hamiltonian~\eqref{imp_ham_anyon_free} describes free anyons in the limit of $\omega \to \infty$. The mechanism of how anyons emerge out of the impurity-bath coupling in this particular model is very similar to the previous interacting case. The only difference is the form of composite bosons/fermions. In contrast to the factor $F$, in the Hamiltonian~\eqref{imp_ham_anyon_free} the attachment of flux/vortex is performed by $\tilde{F}$. As the latter includes also a length, we will call the multiplication by $\tilde{F}$ vortex attachment. Therefore, in this toy model, impurities are dressed by vortices by means of phonons. 

Similar to the previous interacting anyon case, the corresponding two- and three-impurity problems in this toy model can be solved by diagonalizing the Hamiltonian~\eqref{imp_ham_anyon_free} with the eigenstates of the free Hamiltonian by considering the impurities confined additionally in a harmonic-oscillator potential. Instead of following this approach, below we present new computational techniques for the numerical solution, along with a new concrete perspective of anyons in relation to composite bosons or fermions.

\section{A new perspective on anyons} \label{new_perp}

The simplified impurity model given in Eq.~\eqref{imp_ham_anyon_free}  or~\eqref{imp_ham_anyon_int} also provides some useful analytical insights for the $N$-anyon problem. Here we focus solely on the Hamiltonian~\eqref{imp_ham_anyon_free} corresponding to free anyons. Nevertheless the interacting case, Eq.~\eqref{imp_ham_anyon_int}, follows straightforwardly. The Hamiltonian~\eqref{imp_ham_anyon_free} can be diagonalized in the Fock space with the displacement operator $ \hat T = \exp\left[ - \sqrt{\alpha} \left( \tilde{F} \, \hat a^\dagger -\tilde{F}^{-1} \, \hat a \right)/\sqrt{2} \right] $. Then, the $N$-anyon spectrum, which emerges in the limit of $\omega \to \infty$, is given by the Hamiltonian
\be
\label{N_anyon_energy}
\tilde{H}_\text{$N$-anyon} = \bra{0} \hat T^{-1} \hat{\tilde{H}}_\text{qim} \hat T \ket{0} = \sum_{n=0}^\infty \frac{e^{-\alpha/2}}{n!} \left(\frac{\alpha}{2} \right)^n \hat{\tilde{H}}_\text{$N$-comp}^{(2n)} \, ,
\ee
where we made use of the coherent state~\eqref{coherent_state} (see Appendix~C for details). Here the Hamiltonian 
\be
\label{N_comp}
\hat{\tilde{H}}_\text{$N$-comp}^{(2n)} = \tilde{F}^{-n} \hat{H}_\text{$N$-boson} \tilde{F}^n  = - \frac{1}{2}\sum_{q=1}^N \left[\vec{\nabla}_q + i 2 n  \tilde{\vec{A}}_q \right]^2
\ee
describes $N$ composite bosons (fermions) -- the topological bound state of a boson (fermion) and an even number of quantized vortices. This formula shows concretely how anyons emerge from composite bosons or, put it differently, it depicts how a fractional vortex manifests itself through integer vortices. A similar expression can also be obtained for the interacting anyon model~\eqref{imp_ham_anyon_int} by removing tildes from $\tilde{F}$ and $\tilde{F}^{-1}$ in the displacement operator. In this case the expression~\eqref{N_anyon_energy} describes the interacting anyon model.

First of all, Eq.\eqref{N_anyon_energy} naturally simplifies to $H_\text{$N$-boson}$ for $\alpha \to 0$, and, in general, it can be given by a power series 
\be
\label{pow_exp}
\hat{\tilde{H}}_\text{$N$-anyon} = \sum_{n=0}^\infty \frac{1}{n!}\left(\frac{\alpha}{2}\right)^n K_n
\ee
with
\be
K_n = \sum_{j=0}^n {n \choose j} (-1)^{n-j}\hat{\tilde{H}}_\text{$N$-comp}^{(2j)} \, .
\ee
One sees from Eq.~\eqref{int_anyons} or Eq.~\eqref{free_anyons} that $K_n = 0$ for all $n \ge 3$. Furthermore, for the free anyon model $K_2$ also vanishes as a result of the fact that $\tilde{\vec{A}}_q^2 = 0$. Therefore, the power series expansion~\eqref{pow_exp} simply yields
\be
\label{simple_expression_2}
\hat{\tilde{H}}_\text{$N$-anyon} = \hat{\tilde{H}}_\text{$N$-comp}^{(0)} + \frac{\alpha}{2} \left(\hat{\tilde{H}}_\text{$N$-comp}^{(2)} - \hat{\tilde{H}}_\text{$N$-comp}^{(0)}\right) \, .
\ee
It is straightforward to show that the above simple expression is equivalent to the $N$-anyon Hamiltonian~\eqref{N_anyon_ham_2}. In this form we are quite justified to term this process as ``statistics transmutation" from bosons at $\alpha=0$ and $\alpha=2$ to intermediate/fractional values of $\alpha$. As we discuss below, Eq.~\eqref{simple_expression_2} admits new numerical techniques for studying various many-anyon problems.

\subsection{New numerical techniques for free anyons}

Although the resulting formula~\eqref{simple_expression_2} looks almost trivial, we would like to emphasize that it is derived within the introduced quantum impurity model by the non-trivial algebraic properties of the coherent state~\eqref{coherent_state}. This, first, shows the consistency of our approach, and further provides a new approach to the numerical solution of the $N$-anyon problem. Specifically, the Hamiltonian~\eqref{simple_expression_2} also reads
\be
\hat{\tilde{H}}_\text{$N$-anyon} = H_\text{$N$-boson} + \frac{\alpha}{2} \left( \tilde{F}^{-1} H_\text{$N$-boson} \tilde{F} - H_\text{$N$-boson} \right) \, .
\ee
The above Hamiltonian can be further written in terms of the impurity basis states, $\ket{\Phi_m}$, which are the (anti-)symmetric eigenstates of the free $N$-particle Hamiltonian~$\hat H_\text{$N$-boson}$, as
\be
\label{simple_expression}
 \underline{E}_\text{$N$-anyon} = \underline{E}_\text{$N$-boson} + \frac{\alpha}{2} \left( \underline{\tilde{Z}}^{-1} \underline{E}_\text{$N$-boson} \underline{\tilde{Z}} - \underline{E}_\text{$N$-boson} \right) \, ,
\ee
where we define the elements of the matrices $\left(\underline{E}_\text{$N$-boson}\right)_{nm} = \bra{\Phi_n} \hat H_\text{$N$-boson} \ket{\Phi_m}$, and 
\be
\label{Z_from_integrals}
\left(\underline{\tilde{Z}}\right)_{nm} = \bra{\Phi_n} \tilde{F}  \ket{\Phi_m} \, , \quad \left(\underline{\tilde{Z}}^{-1}\right)_{nm} = \bra{\Phi_n} \tilde{F}^{-1}  \ket{\Phi_m} \, .
\ee 
We note that the inverse matrix $\underline{\tilde{Z}}^{-1}$ is deduced from the relation $(\underline{\tilde{Z}} \underline{\tilde{Z}}^{-1})_{nm} =  \sum_{k} \bra{\Phi_n} \tilde{F} \ket{\Phi_k} \bra{\Phi_k} \tilde{F}^{-1}  \ket{\Phi_m} = \delta_{nm}$, where we use the fact that the (anti-)symmetrizer, $\sum_{k} \ket{\Phi_k} \bra{\Phi_k} = \mathcal{S}$, commutes with the interaction term $\tilde{F}$, and $\mathcal{S}\ket{\Phi_k} = \ket{\Phi_k}$ for all (anti-)symmetric states.

We underline that all the diagonalization techniques that we are aware of in the literature are based on the diagonalization of the matrix whose entries are given by the interaction term $ \bra{\Phi_n} [\vec{\nabla}_q,\tilde{\vec{A}}_q]_{+} \ket{\Phi_m}$. However, Eq.~\eqref{simple_expression} is based on the matrix $\underline{Z}$, which is much easier to construct in comparison to the former. Thus, the matrix equation~\eqref{simple_expression} could be of use as a powerful technique in calculating the $N$-anyon spectrum. This will be illustrated below in particular examples.

\subsubsection{Examples}

\begin{figure}
  \centering
    \includegraphics[width=\linewidth]{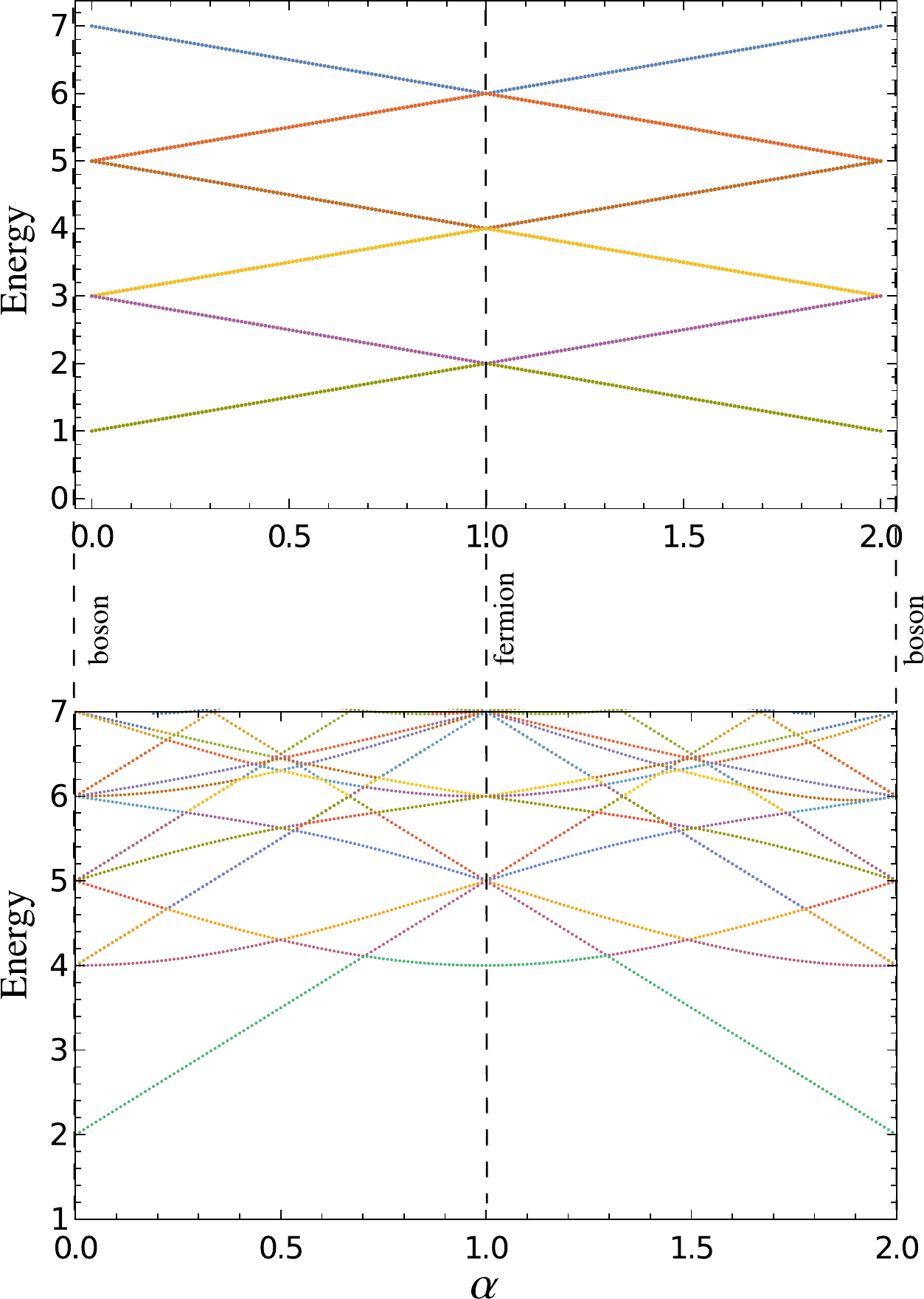}
 \caption{Calculations of the two-anyon (Top) and three-anyon (Bottom) spectra for the original harmonic-oscillator anyon problem without the scalar potential $W$~\eqref{imp_ham_anyon_free}. The energies are given in units of the harmonic frequency. The spectra are here calculated from the bosonic end by using Eq.~\eqref{simple_expression}. The applied parameters are $l_\text{max}=10$, $m_\text{max} = 20$ for the two-anyon case. For the three-anyon case we consider all the symmetric impurity wave functions~\eqref{3_impt_wave_function} restricted by the condition $E_{n_\text{max} m_\text{max}} \le 26$.}
 \label{fig_free}
\end{figure}

We again consider our impurities confined additionally in a harmonic-oscillator potential. Moreover, since in this toy model there are no singular terms in the corresponding transition matrix element~\eqref{transition}, i.e., $\tilde{\vec{A}}_q^2 = 0$, we consider bosonic impurities. We first consider the two-impurity problem. The Hamiltonian in relative coordinates is given by
\bal
\label{two_imp_1}
\nonumber \hat{\tilde{H}}_\text{2-imp} =&  - \frac{\vec{\nabla}^2}{2}  + \frac{ r^2}{2}  + \omega \,\left( \hat a^\dagger \hat a + \frac{\alpha}{2} \right)   \\ 
  &  + \omega\sqrt{\frac{\alpha}{2}} \left[ (\sqrt{2} r e^{i \varphi})^2  \hat a^\dagger + (\sqrt{2} r e^{i \varphi})^{-2}  \hat a \right]  \, .
\eal
The corresponding matrix $\underline{\tilde{Z}}$ can be constructed from the matrix elements $\bra{\Phi_{lm}} (\sqrt{2} r e^{i \varphi})^2 \ket{\Phi_{l'm'}}$ and the matrix $\underline{\tilde{Z}}^{-1}$ from the elements $\bra{\Phi_{lm}} (\sqrt{2} r e^{i \varphi})^{-2} \ket{\Phi_{l'm'}}$, where $\Phi_{lm}$ are the harmonic oscillator wave functions~\eqref{2_har_osc}. We note that as the phase factor excludes the $m=m'=0$ states leading to singularities, the latter matrix elements are finite. The corresponding spectra are presented in Fig~\ref{fig_free} (Top), which agrees with the well-known result $2l + |m + \alpha| +1$ of the two-anyon spectrum. In the numerics the Hilbert space dimension is limited to several hundreds ($\approx300$) impurity basis states. 

Next, we study the three-impurity problem, which reflects the full many-body character of anyons. In Jacobi coordinates the Hamiltonian for the three-impurity problem is given by 
\bal 
\label{two_imp_3}
\hat{\tilde{H}}_\text{3-imp} =&  \frac{1}{2}\left( P_{\vec{\eta}}^2 + P_{\vec{\xi}}^2 \right) + \frac{1}{2} \left( \vec{\eta}^2 + \vec{\xi}^2 \right)  + \omega \left( \hat a^\dagger \hat a + \frac{\alpha}{2} \right) \\
\nonumber & + \omega\sqrt{\frac{\alpha}{2}}  \left( \frac{1}{2}(\eta^3 - 3 \eta \xi^2)^2  \hat a^\dagger + 2 (\eta^3 - 3 \eta \xi^2)^{-2} \hat a \right)  \, .
\eal
In this particular example, instead of constructing the matrices $\underline{\tilde{Z}}$ and $\underline{\tilde{Z}}^{-1}$ separately, we first constructed the matrix $\underline{\tilde{Z}}^{-1}$ by using the symmetric impurity wave functions~\eqref{3_impt_wave_function}. Afterwards we take its pseudoinverse and define $\underline{\tilde{Z}}$. Although the other way around is also possible, we find the former way more convenient for numerical reasons as $\underline{\tilde{Z}}^{-1}$ is a more stable matrix. Similar to the two-impurity case, all the entries of $\underline{\tilde{Z}}^{-1}$ are finite due to the presence of the phase factor in the interaction term; see Appendix~B for the calculation of the matrix elements in hyperspherical coordinates. In Fig.~\ref{fig_free} (Bottom), we show the corresponding spectra, which agree with the known three-anyon spectra~\cite{Murthy_91,Sporre_91}. We found that one can access the spectrum for the interval $0 \le \alpha \le 1 $ by considering less than a thousand impurity wave functions. Moreover, in contrast to the diagonalization of the corresponding anyon Hamiltonian~\eqref{N_anyon_ham_2}, the calculation of the matrix $\underline{\tilde{Z}}^{-1}$ is quite straightforward; see Appendix~B. This appears to us as a very promising way for numerical investigation of various many-anyon problems.

\subsection{Emergence of anyons from composite bosons}

In addition to the new numerical approach discussed above, Eq.\eqref{N_anyon_energy} provides a natural geometric interpretation of this statistics transmutation of impurities in terms of vector bundles. 
Note that in general we may consider the $N$-anyon Hamiltonian~\eqref{N_anyon} resp. \eqref{N_anyon_ham_2} and its domain of functions as defining a complex line bundle (i.e. a rank one hermitian vector bundle) over the configuration space of $N$ identical particles in the plane. In the cases considered here there are no further magnetic interactions than the statistical one, and this then defines a locally flat line bundle which is characterized solely by the statistics parameter $\alpha \in \R$. Using the trivial bosonic bundle $\alpha=0$ as reference, the other possible bundles are geometrically defined by the holonomy of the connection along loops in the configuration space, i.e. the exchange phase which comes in units of $e^{i\pi\alpha}$. We note that $\alpha$ and $\alpha+2$ are unitarily equivalent by a gauge transformation when the free Hamiltonian is considered, however, upon introducing interactions or further regularity conditions into the domain of the kinetic energy operator it makes sense to consider these as different bundles. A family of such bundles $\alpha = 2n$, $n \in \Z$, may be characterized geometrically by the minimal winding number $n$ of the phase as two particles are simply exchanged, or the winding number $2n$ as one particle continuously encircles another one. This is the same as the number of unit fluxes attached to each particle, and the permutation symmetry enforces the same number to each particle. We may thus talk about an even-integer family of bosonic bundles having the physical interpretation as composite bosons with $2n$ quanta of flux attached to each boson, cf. Eq.~\eqref{N_comp}. Note that multiplication by $F$ of Eq.~\eqref{F} resp. $F^*$ changes these winding numbers, and indeed these are the gauge transformations that unitarily transform one such bundle into the other \footnote{In order to uniquely label the bundles and identify them as composite bosons one needs to select some sub-bundle where the exact winding comes out algebraically (and not just via the phase which is periodic over the even integers), such as some subspace of holomorphic functions or some finite range of allowed angular momenta. Note in particular that the space $\cH\tilde{F}^n$ contains also functions on the form $|\tilde{F}|^{2n}$ for which the winding is zero, however by further restricting $\cH$ it is possible to ensure zero overlap between such function spaces and thus the one-to-one labeling by $n$ via the holomorphic factor $\tilde{F}^n$.
This happens for example when one switches on a strong external magnetic field, and indeed the bosonic Laughlin factor $\tilde{F}$ represents the smallest symmetric attachment of an integer number of vortices to every particle.}.


Thus, our starting point in the statistics transmutation was the geometrically trivial bundle $\cH$ of regular bosonic states on the plane $\R^2$ on which the free Hamiltonian $-\sum_q \nabla_q^2/2$ is acting (i.e. our considered states are all exchange-symmetric and have finite expectation values w.r.t. this Hamiltonian). However, by coupling this Hamiltonian to the phonon Fock space in the form~\eqref{imp_ham_anyon_free}, we are effectively considering a semi-infinite ladder $\{\cH \tilde{F}^n \ket{n}\}$, $n=0,1,2,\ldots$, of even-integer bundles of composite bosons. The factor $\tilde{F}^n$ ensures that the winding number of the phase under simple exchange increases by $n$, and equivalently that the vorticity attached to each particle is $2n$. In Eq.~\eqref{imp_ham_anyon_free} we have introduced the possibility of hopping from one such bundle, winding number or vorticity, to the next higher one by means of the interaction term $\tilde{F}\ha^\dagger$ and hence the symmetric (thus staying within the family of bosonic bundles) attachment of a minimal number of vortices to each particle (cf. the arguments leading to Eq.~\eqref{beta_for_gauge}), as well as the corresponding hopping to the next lower bundle using the term $\tilde{F}^{-1}\ha$ and thus the detachment of a minimal number of vortices. We then have the interpretation of Eq.~\eqref{imp_ham_anyon_free} that we are introducing an energy gap $\omega$ between each level of the bundles (which in this context may be interpreted or defined as the energy cost of creating the corresponding number $N(N-1)$ of vortices), and enabling the hopping between consecutive bundles on the ladder by a non-zero amplitude $\lambda\omega$. In the simultaneous limit of both large energy gap $\omega$ and large hopping amplitude, while keeping their ratio $\lambda$ fixed, what then emerges according to Eq.~\eqref{free_anyons} is the fractional bundle labeled by the fraction of vorticity per particle $\alpha = 2\lambda^2$. The phonon state \eqref{coherent_state} attaches a Poisson-distributed sequence of weights on each integer bundle on the ladder,
resulting in the superposition \eqref{N_anyon_energy} of bundles. Furthermore, according to the alternative form \eqref{simple_expression_2}, it may be equivalently understood as a \emph{linear} (modulo weights) deformation between the two lowest bosonic bundles at $\alpha=0$ resp. $\alpha=2$. When $\lambda=1$ or $\alpha=2$ one effectively achieves a complete transmutation into the next integer level by means of the unitary equivalence of $(|\tilde{F}|/\tilde{F})\hat{H}(\tilde{F}/|\tilde{F}|) = F^* \hat{H} F$ to $\hat{H}$. We anticipate that this geometric perspective on the statistics transmutation could be extended, for instance to the setting of higher-rank vector bundles and non-Abelian anyons.

\section{Experimental realization} \label{exp_real}

We now investigate a possible experimental configuration of the quantum impurity model~\eqref{imp_ham}. Let us consider $N$ identical impurities, say bosons, with mass $M$, immersed into a weakly interacting many-particle bath, whose collective excitations are given by phonons with a gapped dispersion $\omega(k)$. This creates the necessary energy gap between the phonon states in the limit of $M \to \infty$, which allows us to consider the problem in the adiabatic limit. Furthermore, we consider impurities confined to two dimensions and leave out the direct interaction between impurities, as the latter is irrelevant for our discussion. In analogy to the Fr\"{o}hlich-Bogoliubov theory, the Hamiltonian of such a model is given by
\bal
\label{imp_ham_exp}
\hat H_\text{exp} & = -\frac{1}{2 M}\sum_{q=1}^N \, \vec{\nabla}_q^2   + \sum_{\vec{k}}  \omega(k) \, \hat b^\dagger_{\vec{k}} \hat b_{\vec{k}}  \\
\nonumber & + \sum_{\vec{k}} \left( V(\vec{k}, \vec{x}) \hat b^\dagger_{\vec{k}} +  V^*(\vec{k}, \vec{x}) \hat b_{\vec{k}} \right)
\eal
with $\sum_{\vec{k}} = \int d^2 k /(2\pi)^2$. Here $\hat b^\dagger_{\vec{k}}$ and $\hat b_{\vec{k}}$ are the creation and annihilation operators for a phonon with the wave vector $\vec{k}$ and frequency $\omega (k)$. They obey the commutation relation $[\hat b_{\vec{k}}, \hat b^\dagger_{\vec{k}'}] = (2\pi)^2 \delta(\vec{k}-\vec{k}')$. The last term in Eq.~\eqref{imp_ham_exp} describes the impurity-phonon interaction with the coupling $V(\vec{k}, \vec{x})$, which depends on the coordinates of impurities $\vec{x} = \{ \bx_1,\ldots,\bx_N \}$.

As a first step, we decompose the creation and annihilation operators in polar coordinates,
\be
\hat b^\dagger_{\vec{k}} = \sqrt{\frac{2 \pi}{k}} \sum_{\mu= - \infty}^\infty i^\mu e^{-i \mu \varphi_k} \hat b^\dagger_{k \mu} \, ,
\ee 
with $ \left[ \hat b_{k \mu}, \hat b^\dagger_{k' \mu'}\right]  = \delta(k - k')\delta_{\mu \mu'} $. The Hamiltonian~(\ref{imp_ham_exp}) is given by
\bal
\label{imp_ham_exp_1}
\hat H_\text{exp} & = -\frac{1}{2 M}\sum_{q=1}^N \, \vec{\nabla}_q^2   + \sum_{k, \mu} \omega(k) \, \hat b^\dagger_{k \mu} \hat b_{k \mu} \\
\nonumber & + \sum_{k, \mu} \lambda_\mu (k,\vec{x}) \left[ e^{-i \beta_\mu (k,\vec{x})}  \hat b^\dagger_{k \mu} +  e^{i \beta_\mu (k,\vec{x})}  \hat b_{k \mu} \right]
\eal
where $\sum_k = \int_0^\infty dk$, and 
\be
V_\mu (k,\vec{x}) = \sqrt{\frac{k}{(2\pi)^3}} \, \int d \varphi_k V(\vec{k},\vec{x})i^\mu \exp(-i\mu \varphi_k)\, ,
\ee
which has been further decomposed into 
$V_\mu (k,\vec{x}) = \lambda_\mu (k,\vec{x}) \exp(-i \beta_\mu (k,\vec{x}))$. 
The Hamiltonian~\eqref{imp_ham_exp_1} is of the form of the general model Hamiltonian~\eqref{imp_ham}. As a result, according to Eq.~\eqref{emergent_gauge_field}, the emergent gauge field can be written as
\be
\label{exp_gauge}
\vec{G}_q = - \sum_{k, \mu} \, \left( \lambda_\mu (k,\vec{x}) / \omega(k)\right)^2 \vec{\nabla}_q \beta_\mu (k,\vec{x}) \, .
\ee
If the bosonic impurities interact with the phonons in such a way that the emergent gauge field~\eqref{exp_gauge} matches with the statistics gauge field~\eqref{anyon_gauge_field_0}, they turn into anyons in the limit of $M \to \infty$. The eventual experimental realization of the model, therefore, reduces to the feasibility of such an impurity-phonon interaction. In general, the gauge field~\eqref{exp_gauge} is non-zero when the integral $\sum_{k} \, \left( \lambda_\mu (k,\vec{x}) / \omega(k)\right)^2 \vec{\nabla}_q \beta_\mu (k,\vec{x})$ is not an odd function under $\mu$, which is a manifestation of breaking time reversal symmetry. This can be achieved by applying a magnetic field or rotation to the system. In principle, such an interaction is feasible with the state-of-art techniques in ultracold atomic physics, for instance in a rotating Bose gas. Below we present a simple and intuitive realization within a well-known problem -- the Fr\"{o}hlich polaron.

\subsection{Fr\"{o}hlich Polarons as Anyons}

\begin{figure}
  \centering
  \includegraphics[width=\linewidth]{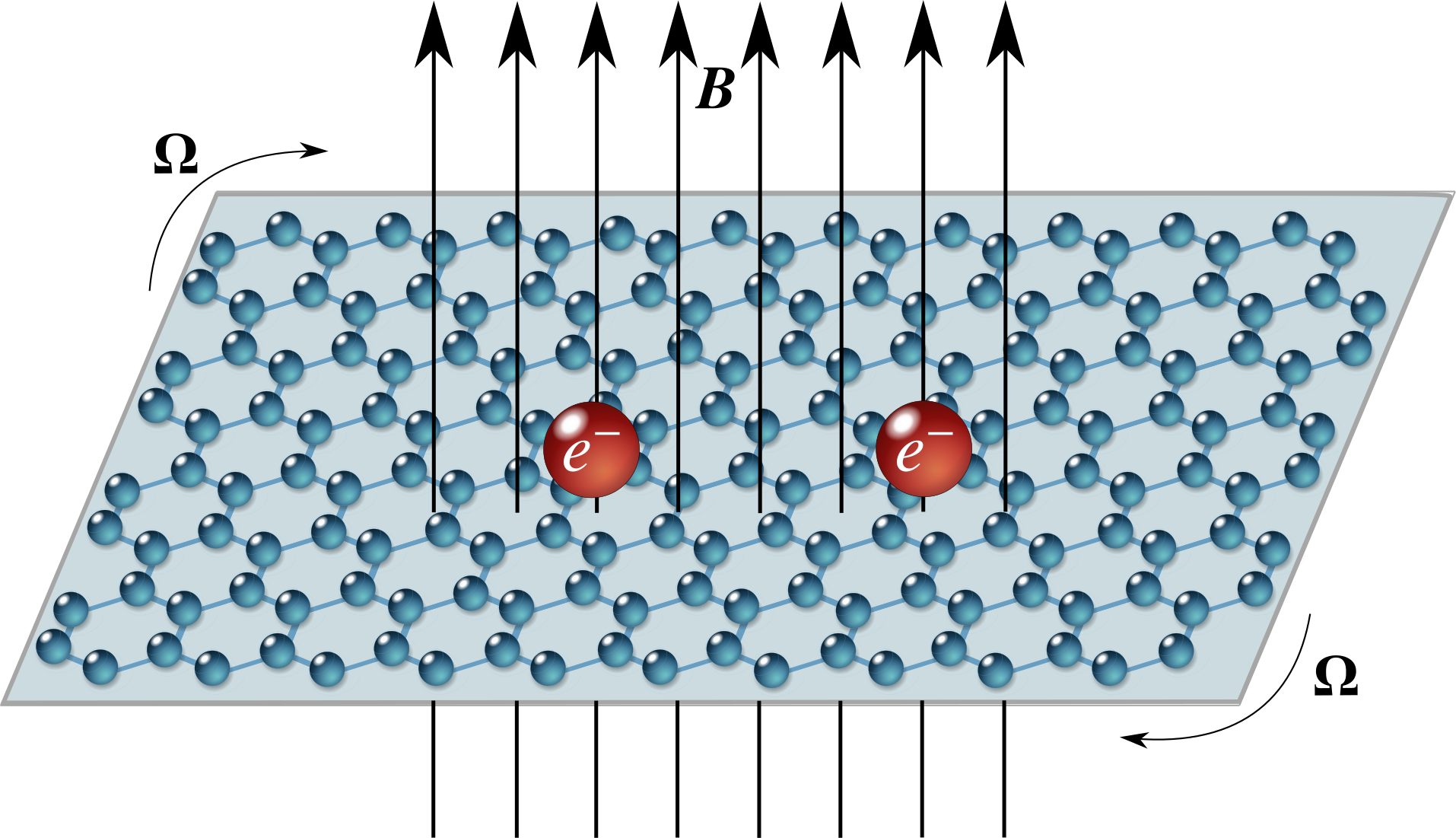}
 \caption{Experimental proposal where the Fr\"{o}hlich polarons turn into anyons. Heavy electrons with mass $M$ immersed in a 2D material are subjected to the magnetic field $\vec{B}$. If the electron-bath system is rotated at the cyclotron frequency $\Omega = B/(2M)$, the polarons become anyons. The setup can also be extended to a 2D Bose gas.}
 \label{fig_exp}
\end{figure} 

Let us consider two electrons confined in a plane interacting with longitudinal optical phonons, $\omega (k) = \omega_0$. The corresponding Fr\"{o}hlich Hamiltonian for two impurities is given by Eq.~\eqref{imp_ham_exp} with the following coupling~\cite{frohlich1954electrons}
\be
V(k, \vec{x}_1, \vec{x}_2) = V(k) \left( e^{-i \vec{k}\cdot \vec{x}_1 } +e^{-i \vec{k}\cdot \vec{x}_2 } \right) \, ,
\ee
where $V(k)$ is the Fourier component of the impurity-phonon interaction in real space. 
We further apply a magnetic field to the impurities along the $z$-direction. Then, the Fr\"{o}hlich Hamiltonian is given by
\bal
\label{imp_ham_exp_F_0}
\nonumber \hat H_\text{F}  &  = - \frac{\left(\vec{\nabla}_1 - i \vec{a}_1 \right)^2 }{2M} - \frac{\left(\vec{\nabla}_2 - i \vec{a}_2 \right)^2}{2M}  \\
& + \sum_{\vec{k}}  \omega_0 \, \hat b^\dagger_{\vec{k}} \hat b_{\vec{k}}  + \sum_{\vec{k}} V(k) \left( e^{-i \vec{k}\cdot \vec{x}_1 } +e^{-i \vec{k}\cdot \vec{x}_2 } \right) \hat b_{\vec{k}}^\dagger + \text{H.c.}
\eal
where $\vec{a}_i = B (-y_i,x_i)/2$ is the gauge field generating the magnetic field. Moreover, we rotate the impurity-bath system in the $x-y$ plane at the cyclotron frequency $\Omega = B/(2M)$. The experimental setup is depicted in Fig.~\ref{fig_exp}. The rotation of the system allows us to factor out the center-of-mass coordinates in the limit of $M\to \infty$. The Hamiltonian~\eqref{imp_ham_exp_F_0} can be written in the rotating coordinate system as 
\bal
\label{imp_ham_exp_F}
\hat H_\text{FR}  & = e^{-i t \Omega \hat J_z}\left( \hat H_\text{F} - i \frac{\del}{\del t}\right) e^{i t \Omega \hat J_z} +  i \frac{\del}{\del t} \\
\nonumber & = \frac{1}{2M} \left( - \vec{\nabla}_1^2 - \vec{\nabla}_2^2  + M^2 \Omega^2 (\vec{x}_1^2 + \vec{x}_2^2) \right)  + \Omega \hat \Lambda_z  \\
\nonumber & + \sum_{\vec{k}}  \omega_0 \, \hat b^\dagger_{\vec{k}} \hat b_{\vec{k}}  + \sum_{\vec{k}} V(k) \left( e^{-i \vec{k}\cdot \vec{x}_1 } +e^{-i \vec{k}\cdot \vec{x}_2 } \right) \hat b_{\vec{k}}^\dagger + \text{H.c.}
\eal
Here $\hat J_z = \hat L_{1\, z} + \hat L_{2\, z} + \hat \Lambda_z$ is the total angular momentum of the impurity-bath system along the $z$-direction, with $\hat L_{i\, z}$ being the angular momentum of the $i$-th impurity and $\hat \Lambda_z$ the collective angular momentum operator of the bath. Next we introduce relative and center-of-mass coordinates, $\vec{r}$ and $\vec{R}$, respectively, and then apply the unitary Lee-Low-Pines (LLP) transformation~\cite{LLP_53}, $ \hat T_\text{LLP} = \exp \left [ - i \hat{\vec{R}} \cdot \sum_{\vec{k}} \vec{k}\,  \hat b^\dagger_{\vec{k}} \hat b_{\vec{k}} /\sqrt{2}  \right] $. We decompose the creation and annihilation operators in polar coordinates, where the angular momentum operator simply reads $\hat\Lambda_z = \sum_{k, \mu} \mu \, \hat b^\dagger_{k \mu} \hat b_{k \mu}$. The transformed Hamiltonian can be rewritten as
\bal
\label{exp_two_imp}
\hat H'_\text{FR} &= - \frac{1}{2M} \vec{\nabla}_r^2  + \frac{1}{2}M \Omega^2 r^2 + \sum_{k, \mu} \omega_\mu  \hat b^\dagger_{k \mu} \hat b_{k \mu}  \\
\nonumber &+  \sum_{k, \mu} \lambda_\mu (k, r) \left[  e^{-i \mu \varphi }  \hat b^\dagger_{k \mu} + e^{i \mu \varphi }  \hat b_{k \mu} \right] + \hat h_R \, .
\eal
Here $\omega_\mu  = \omega_0 + \mu \Omega$ is the effective phonon dispersion relation, where the second term arises as a consequence of the rotation. The impurity-bath coupling strength is given by
\be
\label{coupling_in_polar}
\lambda_\mu (k, r) =  \sqrt{k/(2\pi)} \, V(k) J_\mu (k r /\sqrt{2}) \left[ 1+ (-1)^\mu \right] \, ,
\ee
which follows from the Jacobi-Anger expansion, $\exp[i \vec{k} \cdot \vec{x}] =  \sum_\mu i^\mu J_\mu (kr) \exp[i \mu (\varphi - \varphi_k)]$, with $J_\mu (kr)$ being the Bessel function of the first kind. 
The last term in Eq.~\eqref{exp_two_imp}, 
\be
\hat h_R = \frac{1}{2M}\left(\vec{\nabla}_R - i \sum_{\vec{k}} \vec{k}\,  \hat b^\dagger_{\vec{k}} \hat b_{\vec{k}} /\sqrt{2} \right)^2 + \frac{1}{2} M \Omega^2 R^2 \,,
\ee 
is the Hamiltonian for the center-of-mass motion that couples to the many-particle bath. We note that the coupling term, $\sum_{\vec{k}} \vec{\nabla}_R  \cdot \vec{k}\,  \hat b^\dagger_{\vec{k}} \hat b_{\vec{k}} /M$, is negligible in the limit of $M \to \infty$, as the momentum operator scales as $\sqrt{M}$. Therefore, the center-of-mass coordinate decouples in the transformed Hamiltonian. In a similar way, the contribution of the term $(\sum_{\vec{k}} \vec{k}\,  \hat b^\dagger_{\vec{k}} \hat b_{\vec{k}})^2/M$ to the fast Hamiltonian is also negligible in the limit of $M \to \infty$. Consequently, $\hat h_R $ will be omitted hereafter.

In realistic situations the maximum number of phonons $n_\text{max}$ interacting with impurities is finite. Furthermore, because of the finite size of the first Brillouin zone, we consider a natural cut-off for the phonon wave vector $\vec{k}_\text{max}$. This puts an upper limit for the $\mu$-summation as well as for the $k$-integral. The limitation on the $k$-integral can affect the small distance behavior of the impurities. Nevertheless, as the repulsive Coulomb interaction between two electrons prevents us from considering small distances, we will ignore this cut-off. The cut-off for the $\mu$-summation, on the other hand, is essential in order to have a spectrum bounded from below. Namely, the ground state energy of the fast Hamiltonian can be written as
\be
\varepsilon_\text{gs} = \min\left\{0, n_\text{max} (\omega_0 - \Omega \mu_\text{max})\right\} - \int_0^\infty dk \sum_{\mu = - \mu_\text{max}}^{\mu_\text{max}} \frac{\lambda_\mu (k,r)^2}{\omega_\mu} \, ,
\ee
and we consider the case $(\omega_0 - \Omega \mu_\text{max})>0$, where the ground state is given by the vacuum state $\hat S \hat U \ket{0}$.

It follows from Eq.~\eqref{exp_gauge} that the corresponding emergent gauge field for the ground state of the Hamiltonian~\eqref{exp_two_imp} is given by
\be
\label{gauge_field_exp_fro}
\vec{G} =  \frac{\alpha(r)}{r} \vec{e}_{\varphi} \, 
\ee
with 
\be
\label{general_alpha}
\alpha(r) = - \sum_{\mu = - \mu_\text{max}}^{\mu_\text{max}} \frac{(1 + (-1)^\mu)^2 \mu }{(\omega_0 + \mu \Omega)^2} \int_0^\infty \frac{dk \, k}{2 \pi}\, \left( V(k) J_\mu (k r / \sqrt{2})\right)^2 \, .
\ee
In general, the emergent gauge field does not necessarily correspond to a statistics gauge field, as the $r$-dependence of $\alpha(r)$ relies on the form of $V(k)$. Now we will have a closer look at the derivation of the Fr\"{o}hlich Hamiltonian, where $V(k)$ emerges as the Fourier component of the interaction between the impurity and surrounding many-particle bath in real space. 

In the regular Fr\"{o}hlich polaron, impurities are considered to be confined to two dimensions, but interact with a three dimensional bath, and $V(k)$ emerges as a consequence of the interaction between the electron and the polarization field~\cite{Sak_72}, which leads to $ V(k) = \sqrt{\sqrt{2}\pi \gamma_F/k}$ with $\gamma_F$ being the Fr\"{o}hlich coupling constant for the electron confined in two dimensions. This form of the coupling describes surface polarons~\cite{Sak_72,Wu_86,Peeters_87,devreese2012physics}. In this case, the statistics parameter~\eqref{general_alpha} scales as $1/r$. Nevertheless, if we apply a strong magnetic field, of the order of $\Omega \sim \omega_0$, the relative wave function of the impurities is localized in $r$-space, and hence the relative motion of two impurities is described by only the relative angle. In this case, we can omit the $r$-dependency of the statistics parameter, or consider a limited range, where $\alpha(r)$ is approximately constant. We note, however, that for such a strong magnetic field $(\omega_0 - \Omega \mu_\text{max})<0$, and hence the ground state will not be the vacuum state after the corresponding $\hat S$ and $\hat U$ transformations. The emergence of anyons in such a scenario is similar to the model investigated in Ref.~\cite{Yakaboylu_2018}, where the relative distance between impurities was assumed to be constant.

\begin{figure}
  \centering
  \includegraphics[width=0.8\linewidth]{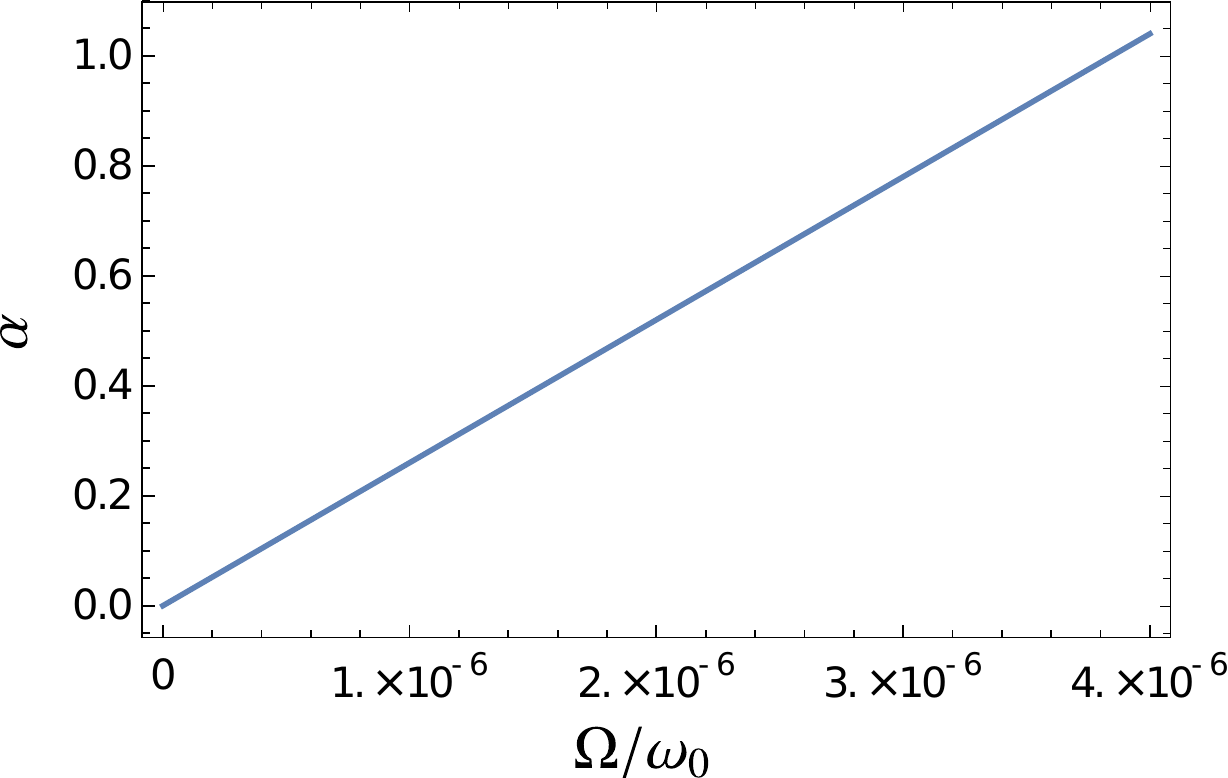}
 \caption{The relative statistics parameter~\eqref{alpha_exp} as a function of the dimensionless cyclotron frequency $\Omega/\omega_0$ for the impurities immersed in a 2D ionic crystal. We note that the effective scalar potential depends also on the statistics parameter. The applied parameters are $\gamma = 100 \, \omega_0^2$ and $\mu_\text{max} = 50$. For typical parameters of optical phonons ($\omega_0 \approx 10^{12} $ Hz) the cyclotron frequency is at the order of MHz.}
 \label{fig_mag}
\end{figure} 

\subsubsection{2D phonon bath}

Instead of a three-dimensional bath, we now consider a quasi-2D bath. Namely, we consider impurities confined to two dimensions, and also interacting with a 2D bath. The latter can be achieved by assuming that the confinement of a 3D bath in the $z$ direction is so strong that we can ignore the excitations in that direction, or we can consider a bath in the form of a single layer of atoms in a two-dimensional lattice, such as graphene. In this case, the polarization field behaves like a $1/r$ field, instead of the $1/r^2$ behavior of a 3D bath. This follows from the fact that in two-dimensional materials the Coulomb law scales as $\ln r$, which has already been observed and investigated in several experimental works~\cite{Kotov_09,Van_Tuan_18,Holten_18}. Then, the Fourier component is given by $V(k) = \sqrt{\pi \gamma/k^2}$ with some constant $\gamma$, which we call the 2D Fr\"{o}hlich coupling constant. As a result of this, the emergent gauge field yields $\vec{G} = \vec{e}_{\varphi} \alpha / r $ with the statistics parameter
\be
\label{alpha_exp}
\alpha  = 4 \gamma \omega_0 \Omega \sum_{\mu = 0, \text{even}}^{\mu_\text{max}} \mu (\omega_0^2 - \mu^2 \Omega^2)^{-2} \, ,
\ee
where we use the relation $\int_0^\infty dk J_\mu (k r)^2/k = 1/(2 |\mu|)$ for $\mu \neq 0$. We note that Eq.~\eqref{alpha_exp} reads as the relative statistics parameter for electrons. The absolute one is given by Eq.~\eqref{alpha_exp} minus one. Thus, the impurities, say electrons, immersed in a 2D many-particle bath in a magnetic field behave like anyons in the limit of $M\to \infty$, when the impurity-bath system is rotated at the cyclotron frequency. The adiabaticity condition~\eqref{transition} can be written as $ \left\langle[\vec{\nabla}_q,\vec{A}_q]_{+} \right\rangle/\omega_0 \propto \Omega / \omega_0 \ll 1 $ for the vacuum state, where we use the relations $\left\langle[\vec{\nabla}_q,\vec{A}_q]_{+} \right\rangle \propto \left\langle \del \hat{H}_\text{$2$-anyon} / \del \alpha  \right\rangle = \del E_\text{$2$-anyon} / \del \alpha \propto \Omega$. Therefore, the limit of $M\to \infty$ reads as $M \gg B/\omega_0$ in an experimental configuration. The absolute statistics parameter (\eqref{alpha_exp} minus one in the fermionic case) emerges as a function of the dispersion, cyclotron frequency, and the 2D Fr\"{o}hlich coupling constant. In Fig.~\ref{fig_mag} we show the statistics parameter as a function of the dimensionless cyclotron frequency $\Omega/\omega_0$.

The origin of the emergent anyons can be intuitively understood in terms of the relative angular momentum of impurities immersed in a bath. By using Eq.~\eqref{emergent_gauge_field} and bearing in mind that the $\hat S$ transformation~\eqref{s_trans} can be written as 
\be
\hat S = \exp\left( - i \varphi \sum_{k, \mu} \mu \, \hat b^\dagger_{k \mu} \hat b_{k \mu} \right)= \exp\left( - i \varphi \hat \Lambda_z \right) \, ,
\ee
the emergent gauge field is given by $ \vec{G} = - \bra{0} \hat U^{-1}\hat \Lambda_z \hat U \ket{0} \, \vec{e}_{\varphi}/r $. In other words, the statistics parameter is simply given by the expectation value of the angular momentum of the many-particle bath in the coherent state
\be
\label{alpha_exp_0}
\alpha = - \bk{\hat \Lambda_z}_\text{coherent state} \, ,
\ee
which can assume any number, as the coherent state is not an eigenstate of the angular momentum operator. Therefore, the relative angular momentum of the impurities is shifted by $\alpha$ and hence becomes nonintegral. 

This result is consistent with the conservation of the angular momentum of the impurity-bath system.  The total angular momentum of the impurity-bath system, which consists of the relative angular momentum of the impurities and collective angular momentum of the bath, assumes an integer value. However, this does not necessarily imply that each of them, separately, assumes an integer value. On the contrary, as we show in Eq.~\eqref{alpha_exp_0}, the relative angular momentum of the impurities is possibly nonintegral, and so is the collective angular momentum of the bath, in such a way that their sum is integer. Namely, the expectation value of the total angular momentum can be written as
\be
\label{tot_ang}
\bk{\hat J_z }_{\Psi^E}  = \bk{\hat L_z }_{\Psi^E} + \bk{\hat \Lambda_z }_{\Psi^E} \, ,
\ee
where $\ket{\Psi^E}$ is the total eigenstate of the impurity-bath system and $\hat L_z$ the relative angular momentum of the impurities. Then, it follows from Eq.~\eqref{total_state} that in the limit of $M \to \infty$ the first term is given by $\bk{\hat L_z }_{\Psi^E} = m - \bk{\hat \Lambda_z}_\text{coherent state}$, with $m$ being an integer. The second term in Eq.~\eqref{tot_ang}, on the other hand, reads $\bk{\hat \Lambda_z }_{\Psi^E} = \bk{\hat \Lambda_z}_\text{coherent state}$ so that the total angular momentum is integer at the end. This is the manifestation of anyons analogous to the picture of Wilczek's flux-tube-charged-particle composite. In this picture fractional values of the angular momentum stems from the fact that the photon field manifests itself as a classical field via the magnetic flux, even though the total angular momentum of the electron-photon system assumes a half-integer value, when the angular momentum of the magnetic flux is taken into account~\cite{Goldhaber_82,Jackiw_83}. Thereby, anyons can here be interpreted as impurities `orbiting around' a `magnetic flux' created by the many-particle bath through the coherent state.

Moreover, the manifestation of the statistics parameter in terms of the angular momentum of the many-particle bath, i.e., Eq.~\eqref{alpha_exp_0}, implies that the statistics parameter can be measured in experiment by detecting the phonon angular momentum. Recent works show that this is feasible; see for instance Refs.~\cite{Hamada_18,holanda2018detecting}. The relation~\eqref{alpha_exp_0} allows us also to propose a novel method to measure the statistics parameter. Namely, if we take the derivative of the Hamiltonian~\eqref{exp_two_imp} with respect to the cyclotron frequency $\Omega$, we obtain $\del \hat H'_\text{FR}/ \del \Omega = B \vec{r}^2 /2  + \hat \Lambda_z $. If we further use the Hellman-Feynman theorem in the limit of $M \to \infty$, where the total state, which is given by Eq.~\eqref{total_state}, is separable, the statistics parameter is given by
\be
\label{alpha_exp_1}
\alpha = \frac{B}{2} \bk{r^2}_0 - \frac{\del E_0}{\del \Omega} \, .
\ee
Here we take the expectation value of $r^2$ with respect to the ground state of the impurities such that $E_0$ is the anyonic ground state energy. We further note that there arises also the term $ \bk{\del \hat h_R / \del \Omega} = B \bk{R^2} /2$ in Eq.~\eqref{alpha_exp_1}. Nevertheless, as the center-of-mass coordinate decouples from the Hamiltonian $\hat H'_\text{FR}$, the expectation value simply assumes an integer number: $\bk{\del \hat h_R / \del \Omega} = n_{R_x} + n_{R_y} + 1$ with $n_{R_{x \, (y)}}$ being the energy level in the center-of-mass dimension $x \, (y)$. Consequently, we neglect its contribution to the statistics parameter. In Eq.~\eqref{alpha_exp_1} the second term defines the magnetization of the system $\mathcal{M}$, i.e., $\del E'_\text{FR}/\del \Omega =  - 2 M \mathcal{M} $, which is routinely measured in torque magnetometry setups to probe the 3D and 2D Fermi surfaces of different types of materials~\cite{Eisenstein_85,li2008phase,
sebastian2008multi,li2014two,tan2015unconventional}. The first term, on the other hand, can be measured with a standard time-of-flight measurement~\cite{Greiner_01}. We note that the relation~\eqref{alpha_exp_1} is reminiscent of the recently proposed method to observe anyonic statistics in the FQHE~\cite{UmuMacComCar-18}, where the statistics parameter is defined in terms of the mean square radius of the density distribution of atoms. Here, anyons can be observed in a much simpler condensed matter system by measuring the magnetization of the impurity-bath system and mean square distance of the impurities.

\subsubsection{2D weakly interacting Bose gas}

The proposed setup can be extended to different many-particle environments such as a two-dimensional weakly interacting Bose gas or a film of liquid helium. The Fr\"{o}hlich-Bogoliubov regime of these impurity problems is governed by the Hamiltonian~\eqref{imp_ham_exp}; see Refs.~\cite{Grusdt_2016,grusdt2017bose,ardila2019strong,Jackson_81} for the details on the validity of the Fr\"{o}hlich-Bogoliubov theory. In such environments, however, the dispersion of the corresponding excitations $\omega(k)$ is, in general, gapless. Nevertheless, realistic circumstances, such as finite size of the BEC, impose a natural low-momentum cut-off $k_\text{min}$ for the dispersion. This allows us to investigate these impurity problems within the adiabatic theorem as well. Furthermore, if the condition $\omega(k_\text{min}) - \mu_\text{max} \Omega > 0$ holds true,  then the ground state of the fast Hamiltonian is given by the vacuum state after the corresponding $\hat S$ and $\hat U$ transformations. This condition is satisfied for small values of the cyclotron frequency. Moreover, the cyclotron frequency should be less than the transverse trapping frequency. Otherwise, the atom density in the gas drops down and the healing length can become arbitrarily large, and the problem cannot be described within the Fr\"{o}hlich-Bogoliubov theory. Under these conditions Eq.~\eqref{general_alpha} remains valid by replacing $\omega_0$ with the dispersion $\omega(k)$:
\be
\label{r_dep_statistic_parameter}
\alpha(r) = - \sum_{\mu = - \mu_\text{max}}^{\mu_\text{max}}(1 + (-1)^\mu)^2 \mu \int_{k_\text{min}}^\infty \frac{dk \,  k}{2\pi}  \, \frac{\left( V(k) J_\mu (k r / \sqrt{2})\right)^2}{(\omega (k) + \mu \Omega)^2} \, .
\ee

\begin{figure}
  \centering
  \includegraphics[width=0.8\linewidth]{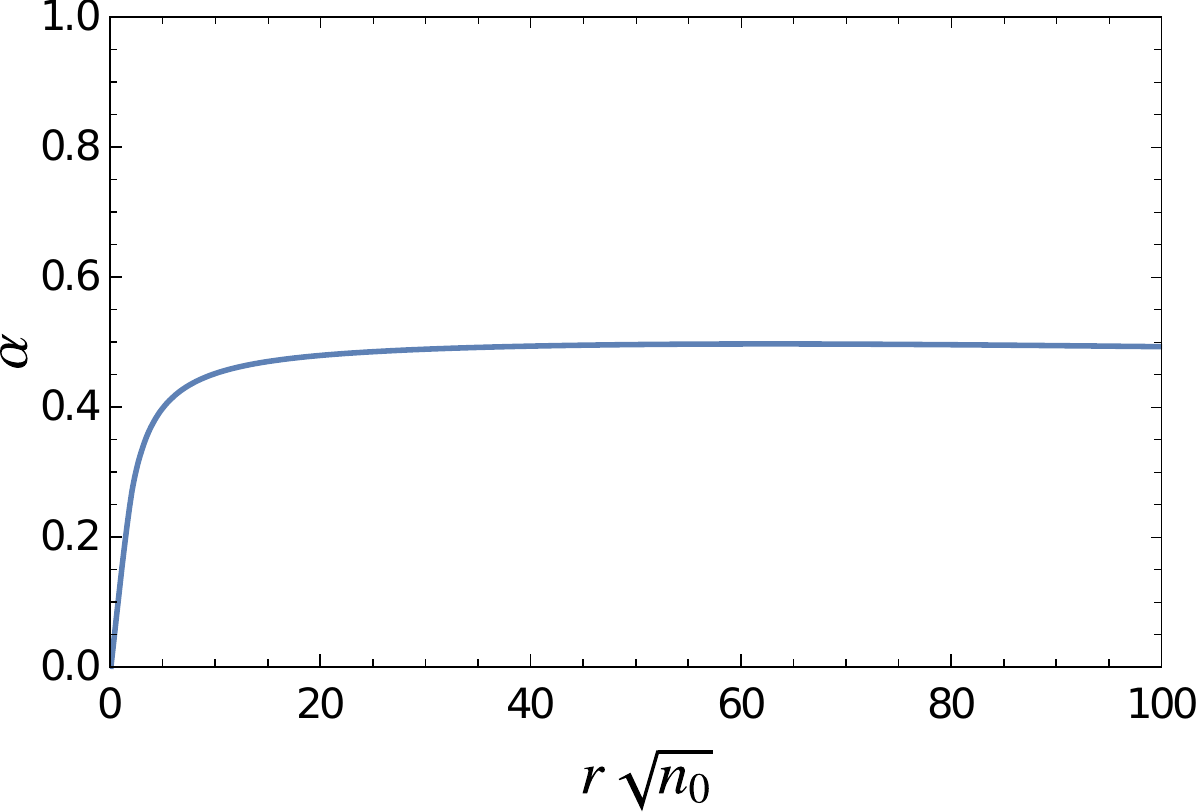}
 \caption{The relative statistics parameter of Eq.~\eqref{r_dep_statistic_parameter} as a function of the dimensionless distance $r \sqrt{n_0}$ for impurities immersed in a 2D Bose gas. The long distance behavior of the emergent gauge field corresponds to the statistics gauge field. The applied Bogoliubov parameters are $\beta_1 = 1000$, $\beta_2 = 1/30$, $\xi =0.1$, $B=1$, $\mu_\text{max} = 10$, and $k_\text{min} = 2 k_0$. The latter is found by solving the equation $\omega(k_0) - \mu_\text{max} \Omega = 0$.  For densities $n_0 > 100$ the parameter $\sqrt{n_0} \xi > 1$, which is accessible in the current experiments, see for instance Ref.~\cite{Ville_2018}.}
 \label{fig_ultracold}
\end{figure}

For instance, let us consider the impurities inside a two-dimensional Bose gas. The latter can be considered a weakly interacting gas if the condition $n_0 a_B^2 \ll 1$ is satisfied, where $n_0$ is the density of the Bose gas and $a_B$ the boson-boson scattering length parameterizing the contact boson-boson interaction. The Bogoliubov dispersion is given by
\be
\omega(k) = c k \sqrt{1+k^2 \xi^2/2} \, .
\ee
Here $c = \sqrt{g_{BB} n_0 / m_B}$ and $\xi = (2 m_B g_{BB} n_0)^{-1/2} $ are the speed of sound and the healing length of a weakly interacting Bose gas, respectively. $m_B$ is the boson mass and $g_{BB}$ is the boson-boson coupling constant given by $g_{BB} = 4 \pi / (m_B \ln(1/n_0 a_B^2))$. The coupling, on the other hand, is
\be
V(k) = \sqrt{n_0} (2\pi)^{-1} g_{IB} \left(\frac{\xi^2 k^2}{2 + \xi^2 k^2} \right)^{1/4}
\ee
with $g_{IB}$ being the impurity-boson coupling constant. We further define two dimensionless numbers $\beta_1 = g_{IB}^2 n_0 /(\pi c^2)$ and $\beta_2 = m_B /M$, which characterize the Bogoliubov-Fr\"{o}hlich Hamiltonian in 2D; see Ref.~\cite{Grusdt_2016} for details. In Fig.~\ref{fig_ultracold} we present $\alpha(r)$ as a function of the dimensionless distance $r \sqrt{n_0}$ in a parameter regime where the Fr\"{o}hlich-Bogoliubov theory is applicable. Apart from small distances, $\alpha(r)$ is approximately constant, and hence, the emergent gauge field behaves as the statistics gauge field. We finally note that the same formalism can also be extended beyond the Fr\"{o}hlich model, where two-phonon scattering processes as well as additional phonon-phonon interactions should be included~\cite{grusdt2017bose}. In this case, one can also investigate impurities strongly interacting with the bath.

Regarding the preparation of the system in order to observe the anyonic behavior of impurities in a 2D Bose gas, we first note that the impurity-bath interaction for a 2D Bose gas has not been achieved yet. Nevertheless, its realization is not out of reach as the sound propagation was already measured experimentally in a 2D Bose fluid~\cite{Ville_2018}. Rotation of the system, on the other hand, was already demonstrated for a cold atomic BEC~\cite{Engels_2003}. Furthermore, the adiabaticity condition $\Omega \ll \omega_0$ allows us to measure the statistics parameter by a time-of-flight measurement after sudden release of trapping potential and a rf spectroscopic measurement of impurity state, respectively, which have been already performed (see Ref.~\cite{Greiner_01} for the time-of-flight measurement and Refs.~\cite{Hu_2016,Jorgensen_16} for the rf spectroscopy of impurity states in 3D BEC systems). Finally, the anyonic behavior of impurities can also potentially be observed by direct interference measurements~\cite{Paredes_2001}. This is also feasible by imaging of the interference pattern of the impurity atoms after sudden release of the trapping potential.


\section{Discussions} \label{sec_conc}

In this paper, we have introduced a quantum impurity model where the surrounding many-particle bath manifests itself as the statistics gauge field with respect to the impurities, and the lowest energy spectra correspond to the anyonic spectra. In terms of the quasiparticle picture, anyons can here be identified as impurities which are first converted into composite bosons (fermions) and then dressed by a coherent state of phonons weighted according to the statistics parameter. Its magnitude in turn depends on the ratio between the phonon energy gap and the hopping amplitude in the adiabatic limit.

The introduced model reveals new numerical techniques for studying the $N$-anyon problem. Specifically, the analytical form of Eq.~\eqref{simple_expression_2} or Eq.~\eqref{simple_expression} provides new routes to obtaining the $N$-anyon spectrum. The direct approach is to calculate the matrices $\underline{\tilde{Z}}$ and $\underline{\tilde{Z}}^{-1}$ separately by explicitly evaluating interaction integrals in the impurity basis using Eq.~\eqref{Z_from_integrals}. One can also calculate $\underline{\tilde{Z}}$ algebraically by writing the harmonic-oscillator Hamiltonian in terms of the ladder operators. Afterwards the inverse matrix $\underline{\tilde{Z}}^{-1}$ can be evaluated from the pseudoinverse of $\underline{Z}$. The general procedure of the evaluation of $\underline{\tilde{Z}}$ and $\underline{\tilde{Z}}^{-1}$ numerically for more than three impurities will be the subject of future work. Moreover, some common techniques used in quantum impurity problems, such as the Diagrammatic Monte Carlo (DiagMC) and Density Matrix Renormalization Group (DMRG) technique, can be applied to the introduced impurity model for numerical studies of the N-anyon problem.

As an experimental proposal, we considered heavy electrons interacting with the excitations of a two-dimensional ionic crystal subject to a magnetic field. If the impurity-bath system is rotated at the cyclotron frequency, the impurities behave as anyons. We showed that the statistics parameter manifests itself through the expectation value of the angular momentum of the many-particle bath in the coherent state. This makes it possible to measure the statistics parameter in experiment in terms of the mean square distance of the impurities and the magnetization of the impurity-bath system. Furthermore, it has been shown that the proposed setup is applicable to other bosonic baths, such as a two-dimensional Bose gas. In this case the long distance behavior of the emergent gauge field resembles the statistics gauge field. A possible experimental measurement of the statistics parameter might be more feasible in such an environment due to recent advances in ultracold atomic physics.

To summarize, we have undertaken the first step towards realizing anyons by using quantum impurities. This new description of anyons promises to shed new light on the field of fractional statistics and related branches of physics, such as the FQHE. The formalism developed in this manuscript is based on the consideration of a  non-degenerate ground state of the fast Hamiltonian, which corresponds to a U(1) gauge field in the adiabatic limit, and hence Abelian anyons. If we further consider an impurity problem that exhibits some degeneracy in the ground state, then the emergent gauge field in the adiabatic limit could very well correspond to a non-Abelian gauge field. This would allow us to extend the model to realize non-Abelian anyons in terms of quantum impurities. In the context of impurity problems, this can potentially be achieved by considering internal degrees of freedom of phonons. Such an approach could then allow us to use quantum impurities as a platform for topological quantum computation.

\begin{acknowledgments}

We are grateful to M. Correggi, A. Deuchert, and P. Schmelcher for valuable discussions. We also thank the anonymous referees for helping to clarify a few important points in the experimental realization. 
A.G. acknowledges support by the European Unions Horizon 2020 research and innovation program under the Marie Sk\l{}odowska-Curie grant agreement No 754411. D.L. acknowledges financial support from the G\"oran Gustafsson Foundation (grant no. 1804) and LMU Munich. R.S., M.L., and N.R. gratefully acknowledge financial support by the European Research Council (ERC) under the European Union's Horizon 2020 research and innovation programme (grant agreements No 694227, No 801770, and No 758620, respectively).

\end{acknowledgments}

\section*{Appendix}

\subsection{Derivation of the emergent gauge field} \label{derivation_gauge_field}

In general, we consider a free $N$-boson system coupled to another system. The corresponding total Hamiltonian can be written as
\be
\label{tot_Ham} 
\hat H_\text{tot} =  -\frac{1}{2} \sum_{q=1}^N \vec{\nabla}_q^2 + \hat H_\text{fast} (\vec{x} ) \, ,
\ee
with $\vec{x} = \{ \bx_1,\ldots,\bx_N \}$ being the coordinates of bosons. Here the first term is the kinetic energy of bosons. Since it is often more convenient to study anyons in curvilinear coordinates, like polar coordinates as we discuss, we give the Laplacian in curvilinear coordinates in terms of the inverse metric tensor $g_q^{ij}$, with $g_q = \det g_q^{ij}$,
\be
\vec{\nabla}_q^2 = \frac{1}{\sqrt{g_q}} \sum_{i,j} \frac{\del}{\del x_q^i} \left(\sqrt{g_q} g_q^{i j} \frac{\del}{\del x_q^j} \right) \, .
\ee
The second term in Eq.~\eqref{tot_Ham} describes the Hamiltonian of the system that couples to bosons, and we assume that in general it is self-adjoint in a weighted space. The coordinates of bosons, $\vec{x}$, are regarded as parameters in the Hamiltonian $\hat H_\text{fast} ( \vec{x} )$, whose eigenvalue equation is given by
\be
\hat H_\text{fast} ( \vec{x} ) \hat S \ket{\psi_n( \vec{x} ) } = \varepsilon_n ( \vec{x} ) \hat S \ket{\psi_n( \vec{x} ) } \, ,
\ee
where $\hat S = \hat S(\vec{x})$ is a similarity transformation such that $\hat S^{-1} \hat H_\text{fast} \hat S$ is Hermitian, and $\braket{\psi_n( \vec{x} ) }{\psi_m( \vec{x} ) } = \delta_{n,m}$. Note that the identity operator in the weighted Hilbert space, where the Hamiltonian $\hat H_\text{fast}$ is defined, is given by $\hat S \sum_n \ket{\psi_n( \vec{x} ) } \bra{\psi_n( \vec{x} ) } \hat S^{-1} = \hat I $.

The total quantum state, which is defined via the eigenvalue equation $\hat H_\text{tot} \, \ket{\Psi^E(\vec{x} )} = E \, \ket{\Psi^E(\vec{x} )}$, can be expanded as
\be 
\ket{\Psi^E(\vec{x} )} = \sum_n \chi_n^E (\vec{x} ) \hat S \ket{\psi_n( \vec{x})} \, ,
\ee
where $\chi_n^E (\vec{x} ) = \bra{\psi_n( \vec{x} )} \hat S^{-1} \ket{\Psi^E(\vec{x} )} $. Then, the eigenvalue equation can be written as 
\be
\label{Sch_eqn_bosons}
\sum_m H_{n m}^\text{eff} \chi_m^E (\vec{x}) = E \, \chi_n^E (\vec{x}) \, ,
\ee
with the effective Hamiltonian 
\bal
\label{eff_ham}
\nonumber H_{n m}^\text{eff} & = - \frac{1}{2} \sum_{q=1}^N \sum_l \frac{1}{\sqrt{g_q}} \sum_{i,j} \left[ \delta_{nl} \frac{\del}{\del x^i_q} + \bra{\psi_n} \hat S^{-1} \frac{\del}{\del x^i_q} \hat S \ket{\psi_l} \right] \\
& \times \sqrt{g_q} g^{i j}_q \left[ \delta_{lm} \frac{\del}{\del x^j_q} + \bra{\psi_l} \hat S^{-1}\frac{\del}{\del x^j_q} \hat S\ket{\psi_m} \right] + \varepsilon_n \delta_{nm}\, ,
\eal
where $ \bra{\psi_n} \hat S^{-1} \del/\del x^i_q  \hat S \ket{\psi_m}$ is the emergent gauge field. We note that Eq.~\eqref{Sch_eqn_bosons} with the Hamiltonian~\eqref{eff_ham} is still exact. Now we assume that the spectrum of the Hamiltonian~$\hat H_\text{fast} (\vec{x})$ is discrete and non-degenerate at least in the $n$th level, and that the energy splittings between level $n$ and the other levels, $m\neq n$, are so large that 
\be
\label{adiab_condt}
\frac{\bk{H^\text{eff}_{nm}}}{\bk{H^\text{eff}_{mm}}- \bk{H^\text{eff}_{nn}}} \ll 1 \, .
\ee
Then, in this adiabatic limit the Schr\"{o}dinger equation~\eqref{Sch_eqn_bosons} simply reads
\bal
\nonumber & \left( - \frac{1}{2} \sum_{q=1}^N  \frac{1}{\sqrt{g_q}}\left[ \frac{\del}{\del x^i_q} + \bra{\psi_n}\hat S^{-1} \frac{\del}{\del x^i_q} \hat S \ket{\psi_n} \right] \sqrt{g_q} g^{i j}_q \right. \\
& \times \left.  \left[ \frac{\del}{\del x^j_q} + \bra{\psi_n} \hat S^{-1} \frac{\del}{\del x^j_q} \hat S \ket{\psi_n} \right] + W(\vec{x}) \right)  \chi_n^E (\vec{x}) = E \, \chi_n^E (\vec{x}) \, ,
\eal
where 
\bal
\label{scalar_pot}
W(\vec{x}) & = \varepsilon_n (\vec{x}) \\
\nonumber & - \frac{1}{2} \sum_{q=1}^N \sum_{l\neq n} \bra{\psi_n} \hat S^{-1} \frac{\del}{\del x^i_q} \hat S \ket{\psi_l} g^{ij}_q \bra{\psi_l} \hat S^{-1}\frac{\del}{\del x^j_q}\hat S \ket{\psi_n}\, 
\eal
is the emergent scalar potential.

\subsection{The three-impurity matrix element in hyperspherical coordinates}
\label{hyperspherical}

For the three-impurity problem the matrix elements of $\underline{\tilde{Z}}^{-1}$ in hyperspherical coordinates are given by
\be
\bra{\Phi^{\pm}_{n m \nu \mu}} 2 \rho^{-6} e^{6 i \psi} \left[ A(\theta) \cos(3\phi) + i  B(\theta) \sin(3\phi) \right]^{-2}  \ket{\Phi^{\pm}_{n' m' \nu' \mu'}} \, ,
\ee
where $A(\theta) =  \cos (\theta ) (2 - \cos (2 \theta )) $, $B(\theta) = \sin (\theta ) (2 + \cos (2 \theta ))$ with $-\pi/4 \le \theta \le \pi/4$. The $\rho$- and $\psi$-integrals can be evaluated analytically. The latter integral, similar to the two-impurity problem, excludes the diverging $\rho$-integrals. The $\phi$-integral, on the other hand, can be written as
\be
\label{con_int_phi}
I_w = \int_{-\pi/2}^{\pi/2} \frac{d \phi \, e^{i w \phi}}{\left[ A(\theta) \cos(3\phi) + i  B(\theta) \sin(3\phi) \right]^2} \, ,
\ee
with $w = 6 \times \text{integer}$. Here, we used the hyperspherical harmonics~\cite{kilpatrick1987set,khare1991perturbative}, which are given by
\bal
Y^{\pm}_{m \nu \mu} (\theta,\phi,\psi)& = \left(\left(\frac{1}{\sqrt{2}}-1 \right)\delta_{\nu 0} + 1\right)\\
\nonumber & \times \frac{1}{2\pi} \left(\bk{m,\nu, \mu} \pm (-1)^{\mu + \tilde{m}} \bk{m,-\nu, \mu} \right) \, ,
\eal
where 
\bal
\bk{m,\nu, \mu} & = \sqrt{\frac{\tilde{m}!(\tilde{m}+\alpha+\beta)!(m+1)}{2^{\alpha+\beta} (\tilde{m}+\alpha)!(\tilde{m}+\beta)! }} \Theta_{\tilde{m}}^{\alpha \beta} (\sin(2\theta)) e^{i \nu \phi} e^{i \mu \psi} \, , \\
\Theta_{\tilde{m}}^{\alpha \beta} (x) & = (1-x)^{\alpha/2} (1+x)^{\beta/2} P_{\tilde{m}}^{\alpha \beta} (x) \, ,
\eal
with the Jacobi polynomials $P_{\tilde{m}}^{\alpha \beta} (x)$ and the following numbers
\be
\tilde{m} = \frac{m- \max(|\mu|, |\nu|)}{2} \, , \alpha = \frac{|\nu + \mu|}{2} \, ,  \quad \text{and}  \quad \beta = \frac{|\nu - \mu|}{2} \, .
\ee

The integral~\eqref{con_int_phi} can be transformed into a rational function of a complex variable by the substitution of $z = \exp(2 i \phi)$:
\be
I_w = -\frac{2i}{(A+B)^2} \oint_{\mathcal{C}} \frac{d z\, z^{2+w/2}}{(z^3 + \kappa)^2} \, ,
\ee
where the contour $\mathcal{C}$ is the unit circle, and $\kappa = (A-B)/(A+B)$. This contour integral can be further written as
\be
I_w = \frac{2i}{(A+B)^2} \frac{\del}{\del \kappa }\oint_{\mathcal{C}} \frac{d z\, z^{2+w/2}}{(z^3 + \kappa)} \, ,
\ee
where the latter contour integral can be straightforwardly determined by using the residue theorem. For  $w\ge 0$ the integral is given by
\be
I_w = -\frac{4\pi \left(e^{i\pi w/2} + 2 \cos(\pi w/6) \right) }{3(A+B)^2} \frac{\del}{\del \kappa } \left( H(1-\kappa)\kappa^{w/6}  \right) \, ,
\ee
which yields
\be
\label{int_phi}
I_w  = \frac{2\pi e^{i\pi w/2}}{3} \left(\delta(\theta) - \frac{w H(\theta)}{(A+B)^2}\left(\frac{A-B}{A+B}\right)^{\frac{w}{6}-1} \right) \, ,
\ee
with $H(\theta)$ being the step function. The $w<0$ case, on the other hand, can be easily obtained by changing $w \to -w$ and $\theta \to - \theta$ in Eq.~\eqref{int_phi}. Finally, the remaining $\theta$-integral can be calculated numerically.

\subsection{Diagonalization of the impurity Hamiltonian} \label{app:diagonalization}

In this appendix we elaborate on why we expect to see that the spectrum of the full Hamiltonian operator converges to that of the lowest sector of the Fock space. 
Namely, we find that any eigenstate with low energy will have vanishing components outside the lowest sector in the adiabatic limit $\omega \to \infty$. However, a few technical assumptions enter because the various components of the full operator depend on the particle number.

Starting with an impurity-bath coupled Hamiltonian of the form \eqref{imp_ham_anyon_int},
\begin{align*}
	H_\omega &:= H_0 + \omega\, \ha^\dagger \ha + \lambda \omega (F \ha^\dagger + F^{-1} \ha) + \lambda^2 \omega 
\end{align*}
where $\omega \ge 0$ and $\lambda \in \R$ are parameters, 
and 
$$
	H_0 := \sum_{j=1}^N \left[ -\nabla_{\bx_j}^2 + V(\bx) \right] 
$$
acts in some $N$-body Hilbert space $\cH$,
let us define for arbitrary coupling $\gamma \in \R$
a deformed $N$-body operator
\be
\label{transmuted_ham}
	H_0^{\gamma\bF} := \sum_{j=1}^N \left[ -(\nabla_{\bx_j} + \gamma\bF_j)^2 + V(\bx) \right]. 
\ee
For generality we allow for 
a potential $V$ (which may depend on all the variables and thus also include interactions),
and also for $F$ to be any function of $\bx_1,\ldots,\bx_N$ such that
$$
	\bF_j := \nabla_{\bx_j} \log F = F^{-1} \nabla_{\bx_j} F
$$
in \eqref{transmuted_ham} is well defined, say smooth,
at least for non-coincident $\bx_j$ 
(we may then take an appropriate dense domain in $\cH$).

Using $\hS = F^\cN$, $\cN = \ha^\dagger \ha$, and $\hU = e^{-\lambda(\ha^\dagger-\ha)}$
in the expansion
$$
	e^X Y e^{-X} = Y + [X,Y] + \frac{1}{2!}[X,[X,Y]] + \frac{1}{3!}[X,[X,[X,Y]]] + \ldots
$$
we obtain the transformations
\begin{align*}
	\hS^{-1} \nabla_{\bx_j} \hS &= \nabla_{\bx_j} + \bF_j \cN, \\
	\hS^{-1} \ha^{(\dagger)} \hS &= F^{(-1)} \ha^{(\dagger)}, \\
	\hU^{-1} \ha^{(\dagger)} \hU &= \ha^{(\dagger)} - \lambda,
\end{align*}
and thus $H_\omega$ is similar to
\begin{align*}
	H_\omega' &:= \hU^{-1}\hS^{-1} H_\omega \hS \hU = H_0' + \omega \cN, \\
	H_0' &= \sum_{j=1}^N \Big[ -\nabla_{\bx_j}^2 + V(\bx) 
		-\bigl(\nabla_{\bx_j} \cdot \bF_j + 2\bF_j \cdot \nabla_{\bx_j}\bigr) \hU^{-1} \cN \hU \\
		&\qquad -\bF_j^2 \,\hU^{-1} \cN^2 \hU \Big].
\end{align*}
We compute for arbitrary $n=0,1,2,\ldots$
\begin{align*}
	\hU^{-1} \cN \hU|n\rangle &= \bigl( \cN - \lambda(\ha^\dagger+\ha) + \lambda^2 \bigr)|n\rangle \\
	&= (n+\lambda^2)|n\rangle - \lambda\sqrt{n+1}|n+1\rangle - \lambda\sqrt{n}|n-1\rangle,
\end{align*}
\begin{widetext}
\begin{align*}
	\hU^{-1} \cN^2 \hU|n\rangle &= \bigl( \cN^2 - \lambda\cN(\ha^\dagger+\ha) - \lambda(\ha^\dagger+\ha)\cN + 4\lambda^2\cN 
	 	+ \lambda^2((\ha^\dagger)^2+\ha^2) - 2\lambda^3(\ha^\dagger+\ha) + \lambda^2 + \lambda^4 \bigr)|n\rangle \\
	&= (n^2+4\lambda^2 n +\lambda^2+\lambda^4)|n\rangle 
		- \lambda(2n+1+2\lambda^2)\sqrt{n+1}|n+1\rangle 
		+ \lambda^2\sqrt{n+2}\sqrt{n+1}|n+2\rangle \\
	&\quad - \lambda(2n-1+2\lambda^2)\sqrt{n}  |n-1\rangle 
		+ \mu^2\sqrt{n}\sqrt{n-1}|n-2\rangle,
\end{align*}
and thus obtain the non-trivial operator matrix elements
\begin{align*}
	\langle n| H_\omega' |n\rangle &= \sum_{j=1}^N \Big[ -\nabla_{\bx_j}^2 + V(\bx) + \omega n 
		-\bigl(\nabla_{\bx_j} \cdot \bF_j + 2\bF_j \cdot \nabla_{\bx_j}\bigr) (n+\lambda^2) 
		-\bF_j^2 \bigl( (n+\lambda^2)^2 + \lambda^2(1+2n) \bigr)
		\Big] \\
	&= H_0^{(n+\lambda^2)\bF} + \omega n - \lambda^2 (1+2n) \bF^2,
\end{align*}
\begin{align*}
	\langle n+1| H_\omega' |n\rangle &= \sum_{j=1}^N \Big[ 
		-\bigl(\nabla_{\bx_j} \cdot \bF_j + 2\bF_j \cdot \nabla_{\bx_j}\bigr) (-\lambda\sqrt{n+1}) 
		-\bF_j^2 (-\lambda(2n+1+2\lambda^2)\sqrt{n+1})
		\Big] \\
	&= \lambda\sqrt{n+1}\bigl( H_0 - H_0^{\bF} + 2(n+\lambda^2)\bF^2 \bigr),
\end{align*}
\begin{align*}
	\langle n-1| H_\omega' |n\rangle &= \sum_{j=1}^N \left[ 
		-\bigl(\nabla_{\bx_j} \cdot \bF_j + 2\bF_j \cdot \nabla_{\bx_j}\bigr) (-\lambda\sqrt{n})
		-\bF_j^2 (-\lambda(2n-1+2\lambda^2)\sqrt{n})
		\right] \\
	&= \lambda\sqrt{n}\bigl( H_0 - H_0^{\bF} + 2(n-1+\lambda^2)\bF^2 \bigr),
\end{align*}
\begin{align*}
	\langle n+2| H_\omega' |n\rangle &= \sum_{j=1}^N \left[ 
		-\bF_j^2 \lambda^2\sqrt{n+2}\sqrt{n+1}
		\right] 
	= -\lambda^2\sqrt{n+2}\sqrt{n+1}\,\bF^2,
\end{align*}
\begin{align*}
	\langle n-2| H_\omega' |n\rangle &= \sum_{j=1}^N \left[ 
		-\bF_j^2 \lambda^2\sqrt{n}\sqrt{n-1}
		\right] 
	= -\lambda^2\sqrt{n}\sqrt{n-1}\,\bF^2.
\end{align*}
Hence, we have a symmetric pentadiagonal matrix of operators
$$
	\begin{bmatrix} \langle k| H_\omega' |n\rangle \end{bmatrix} =
	\begin{bmatrix}
	H_0^{\lambda^2\bF} - \lambda^2\bF^2 & \lambda(H_0 - H_0^\bF + 2\lambda^2\bF^2) & -\lambda^2\sqrt{2}\bF^2 & 0 & 0 & \ldots \\
	\lambda(H_0 - H_0^\bF + 2\lambda^2\bF^2) & H_0^{(1+\lambda^2)\bF} + \omega - 3\lambda^2\bF^2 & \lambda\sqrt{2}(H_0 - H_0^\bF + 2(1+\lambda^2)\bF^2) & -\lambda^2\sqrt{6}\bF^2 & 0 & \ldots \\
	-\lambda^2\sqrt{2}\bF^2 & \lambda\sqrt{2}(H_0 - H_0^\bF + 2(1+\lambda^2)\bF^2) & H_0^{(2+\lambda^2)\bF} + 2\omega - 5\lambda^2\bF^2 & \ldots \\ 
	0 & -\lambda^2\sqrt{6}\bF^2 & \lambda\sqrt{3}(H_0 - H_0^\bF + 2(2+\lambda^2)\bF^2) & \ldots \\
	0 & 0 & -\lambda^2 2\sqrt{3}\bF^2 & \ldots \\
	0 & 0 & 0 \\
	\vdots & \vdots & \vdots & \ddots
	\end{bmatrix}
$$

Note that for the choice \eqref{F_tilde}, $\bF_j = 2i\tilde\bA_j$, 
so $\bF^2 = 0$ and these expressions simplify significantly, 
while with the choice \eqref{F}, $\bF_j = 2i\bA_j$, $-\bF^2 = 4\bA^2 \ge 0$,
and $H_0^\bF = H_0^{2i\bA} = F^*H_0F$ 
is unitary equivalent to $H_0$.
Furthermore, one has the unitary equivalence of 
$H_0^{i\alpha\bA}$ to $H_0^{i(\alpha+2n)\bA}$ for any integer $n$ 
as well as the diamagnetic inequality 
$\langle H_0^{i\alpha\bA} \rangle_\Phi \ge \langle H_0 \rangle_{|\Phi|}$ 
for any $\alpha \in \R$ and $\Phi \in \cH$ \cite{LunSol-14}, while 
$H_0^{i\gamma\tilde\bA} = |\tilde F|^{-\gamma/2} H_0^{i\gamma\bA} |\tilde F|^{\gamma/2}$,
also implying isospectrality for $H_0^{i(\alpha+2n)\tilde\bA}$.
Hence, if $H_0 \ge -C$ and
we consider a truncated subspace of $N$-body states $\Phi_n \in \cH$ for which 
$\re\, \langle \pm H_0^{\gamma\bF} \rangle_{\Phi_n} \le C$
and $0 \le \langle -\bF^2 \rangle_{\Phi_n} \le C$ independent of $n$ and $\omega$,
then for any normalized $|\Psi\rangle = \sum_{n=0}^\infty \Phi_n |n\rangle$
with finite expectation $\langle \cN^{3/2} \rangle_\Psi$
(in the case $\bF^2 = 0$ it is sufficient that $\langle \cN \rangle_\Psi \le C$), 
\begin{align*}
	\re\, \langle \Psi| H_\omega' |\Psi\rangle 
	&\ge \sum_{n=0}^\infty \Big( (n\omega-C) \|\Phi_n\|^2 
	- 2|\lambda|\sqrt{n+1}\big|\langle\Phi_n|(H_0-H_0^\bF+2(n+\lambda^2)\bF^2)|\Phi_{n+1}\rangle\big|
	- 2\lambda^2\sqrt{n+1}\sqrt{n+2}\big|\langle\Phi_n|\bF^2|\Phi_{n+2}\rangle\big|
	\Big) \\
	&\ge -2C(1+|\lambda|+\sqrt{2}\lambda^2)\|\Phi_0\|^2 
	+ \sum_{n=1}^\infty \Big( n\omega - C
	- 8|\lambda|\sqrt{n+1}(1+n+\lambda^2) C
	- 4\lambda^2\sqrt{n+1}\sqrt{n+2}C
	\Big) \|\Phi_n\|^2,
\end{align*}
by the Cauchy-Schwarz inequality.
Keeping $N$ and $\lambda$ fixed, and
demanding that $\langle H_\omega' \rangle_\Psi$ stays bounded 
while $\omega^{-1} \langle \cN^{3/2} \rangle_\Psi \to 0$ as $\omega \to \infty$,
thus requires that
$1 - \|\Phi_0\|^2 = \sum_{n=1}^\infty \|\Phi_n\|^2 \le \langle \cN \rangle_\Psi \to 0$
by the above inequality.

Therefore,
in the limit as $\omega \to \infty$, 
the lowest part (as quantified by $C$) of the spectrum of $H_\omega'$, and equivalently $H_\omega$,
is described by that of
\be
	H_0^{\lambda^2\bF} - \lambda^2\bF^2 
	= \langle 0|H_\omega'|0\rangle
	= \langle 0|\hU^*\hS^{-1} H_0 \hS\hU|0\rangle 
	= e^{-\lambda^2} \sum_{n=0}^\infty \frac{\lambda^{2n}}{n!} F^{-n}H_0 F^n,
\ee
where in the last identity we used the coherent state~\eqref{coherent_state}.

\end{widetext}

\bibliography{anyon_ref.bib}

\begin{thebibliography}{123}%
\makeatletter
\providecommand \@ifxundefined [1]{%
 \@ifx{#1\undefined}
}%
\providecommand \@ifnum [1]{%
 \ifnum #1\expandafter \@firstoftwo
 \else \expandafter \@secondoftwo
 \fi
}%
\providecommand \@ifx [1]{%
 \ifx #1\expandafter \@firstoftwo
 \else \expandafter \@secondoftwo
 \fi
}%
\providecommand \natexlab [1]{#1}%
\providecommand \enquote  [1]{``#1''}%
\providecommand \bibnamefont  [1]{#1}%
\providecommand \bibfnamefont [1]{#1}%
\providecommand \citenamefont [1]{#1}%
\providecommand \href@noop [0]{\@secondoftwo}%
\providecommand \href [0]{\begingroup \@sanitize@url \@href}%
\providecommand \@href[1]{\@@startlink{#1}\@@href}%
\providecommand \@@href[1]{\endgroup#1\@@endlink}%
\providecommand \@sanitize@url [0]{\catcode `\\12\catcode `\$12\catcode
  `\&12\catcode `\#12\catcode `\^12\catcode `\_12\catcode `\%12\relax}%
\providecommand \@@startlink[1]{}%
\providecommand \@@endlink[0]{}%
\providecommand \url  [0]{\begingroup\@sanitize@url \@url }%
\providecommand \@url [1]{\endgroup\@href {#1}{\urlprefix }}%
\providecommand \urlprefix  [0]{URL }%
\providecommand \Eprint [0]{\href }%
\providecommand \doibase [0]{http://dx.doi.org/}%
\providecommand \selectlanguage [0]{\@gobble}%
\providecommand \bibinfo  [0]{\@secondoftwo}%
\providecommand \bibfield  [0]{\@secondoftwo}%
\providecommand \translation [1]{[#1]}%
\providecommand \BibitemOpen [0]{}%
\providecommand \bibitemStop [0]{}%
\providecommand \bibitemNoStop [0]{.\EOS\space}%
\providecommand \EOS [0]{\spacefactor3000\relax}%
\providecommand \BibitemShut  [1]{\csname bibitem#1\endcsname}%
\let\auto@bib@innerbib\@empty
\bibitem [{\citenamefont {Thouless}\ \emph {et~al.}(1982)\citenamefont
  {Thouless}, \citenamefont {Kohmoto}, \citenamefont {Nightingale},\ and\
  \citenamefont {den Nijs}}]{PhysRevLett.49.405}%
  \BibitemOpen
  \bibfield  {author} {\bibinfo {author} {\bibfnamefont {D.~J.}\ \bibnamefont
  {Thouless}}, \bibinfo {author} {\bibfnamefont {M.}~\bibnamefont {Kohmoto}},
  \bibinfo {author} {\bibfnamefont {M.~P.}\ \bibnamefont {Nightingale}}, \ and\
  \bibinfo {author} {\bibfnamefont {M.}~\bibnamefont {den Nijs}},\ }\href
  {\doibase 10.1103/PhysRevLett.49.405} {\bibfield  {journal} {\bibinfo
  {journal} {Phys. Rev. Lett.}\ }\textbf {\bibinfo {volume} {49}},\ \bibinfo
  {pages} {405} (\bibinfo {year} {1982})}\BibitemShut {NoStop}%
\bibitem [{\citenamefont {Kane}\ and\ \citenamefont
  {Mele}(2005)}]{PhysRevLett.95.146802}%
  \BibitemOpen
  \bibfield  {author} {\bibinfo {author} {\bibfnamefont {C.~L.}\ \bibnamefont
  {Kane}}\ and\ \bibinfo {author} {\bibfnamefont {E.~J.}\ \bibnamefont
  {Mele}},\ }\href {\doibase 10.1103/PhysRevLett.95.146802} {\bibfield
  {journal} {\bibinfo  {journal} {Phys. Rev. Lett.}\ }\textbf {\bibinfo
  {volume} {95}},\ \bibinfo {pages} {146802} (\bibinfo {year}
  {2005})}\BibitemShut {NoStop}%
\bibitem [{\citenamefont {Fu}\ \emph {et~al.}(2007)\citenamefont {Fu},
  \citenamefont {Kane},\ and\ \citenamefont {Mele}}]{PhysRevLett.98.106803}%
  \BibitemOpen
  \bibfield  {author} {\bibinfo {author} {\bibfnamefont {L.}~\bibnamefont
  {Fu}}, \bibinfo {author} {\bibfnamefont {C.~L.}\ \bibnamefont {Kane}}, \ and\
  \bibinfo {author} {\bibfnamefont {E.~J.}\ \bibnamefont {Mele}},\ }\href
  {\doibase 10.1103/PhysRevLett.98.106803} {\bibfield  {journal} {\bibinfo
  {journal} {Phys. Rev. Lett.}\ }\textbf {\bibinfo {volume} {98}},\ \bibinfo
  {pages} {106803} (\bibinfo {year} {2007})}\BibitemShut {NoStop}%
\bibitem [{\citenamefont {Haldane}(1988)}]{PhysRevLett.61.2015}%
  \BibitemOpen
  \bibfield  {author} {\bibinfo {author} {\bibfnamefont {F.~D.~M.}\
  \bibnamefont {Haldane}},\ }\href {\doibase 10.1103/PhysRevLett.61.2015}
  {\bibfield  {journal} {\bibinfo  {journal} {Phys. Rev. Lett.}\ }\textbf
  {\bibinfo {volume} {61}},\ \bibinfo {pages} {2015} (\bibinfo {year}
  {1988})}\BibitemShut {NoStop}%
\bibitem [{\citenamefont {Tsui}\ \emph {et~al.}(1982)\citenamefont {Tsui},
  \citenamefont {Stormer},\ and\ \citenamefont {Gossard}}]{Tsui_82}%
  \BibitemOpen
  \bibfield  {author} {\bibinfo {author} {\bibfnamefont {D.~C.}\ \bibnamefont
  {Tsui}}, \bibinfo {author} {\bibfnamefont {H.~L.}\ \bibnamefont {Stormer}}, \
  and\ \bibinfo {author} {\bibfnamefont {A.~C.}\ \bibnamefont {Gossard}},\
  }\href {\doibase 10.1103/PhysRevLett.48.1559} {\bibfield  {journal} {\bibinfo
   {journal} {Phys. Rev. Lett.}\ }\textbf {\bibinfo {volume} {48}},\ \bibinfo
  {pages} {1559} (\bibinfo {year} {1982})}\BibitemShut {NoStop}%
\bibitem [{\citenamefont {Laughlin}(1983)}]{Laughlin_83}%
  \BibitemOpen
  \bibfield  {author} {\bibinfo {author} {\bibfnamefont {R.~B.}\ \bibnamefont
  {Laughlin}},\ }\href {\doibase 10.1103/PhysRevLett.50.1395} {\bibfield
  {journal} {\bibinfo  {journal} {Phys. Rev. Lett.}\ }\textbf {\bibinfo
  {volume} {50}},\ \bibinfo {pages} {1395} (\bibinfo {year}
  {1983})}\BibitemShut {NoStop}%
\bibitem [{\citenamefont {Arovas}\ \emph {et~al.}(1984)\citenamefont {Arovas},
  \citenamefont {Schrieffer},\ and\ \citenamefont {Wilczek}}]{Arovas_84}%
  \BibitemOpen
  \bibfield  {author} {\bibinfo {author} {\bibfnamefont {D.}~\bibnamefont
  {Arovas}}, \bibinfo {author} {\bibfnamefont {J.~R.}\ \bibnamefont
  {Schrieffer}}, \ and\ \bibinfo {author} {\bibfnamefont {F.}~\bibnamefont
  {Wilczek}},\ }\href {\doibase 10.1103/PhysRevLett.53.722} {\bibfield
  {journal} {\bibinfo  {journal} {Phys. Rev. Lett.}\ }\textbf {\bibinfo
  {volume} {53}},\ \bibinfo {pages} {722} (\bibinfo {year} {1984})}\BibitemShut
  {NoStop}%
\bibitem [{\citenamefont {Lloyd}(2002)}]{lloyd2002quantum}%
  \BibitemOpen
  \bibfield  {author} {\bibinfo {author} {\bibfnamefont {S.}~\bibnamefont
  {Lloyd}},\ }\href@noop {} {\bibfield  {journal} {\bibinfo  {journal} {Quantum
  Information Processing}\ }\textbf {\bibinfo {volume} {1}},\ \bibinfo {pages}
  {13} (\bibinfo {year} {2002})}\BibitemShut {NoStop}%
\bibitem [{\citenamefont {Kitaev}(2003)}]{kitaev2003fault}%
  \BibitemOpen
  \bibfield  {author} {\bibinfo {author} {\bibfnamefont {A.~Y.}\ \bibnamefont
  {Kitaev}},\ }\href@noop {} {\bibfield  {journal} {\bibinfo  {journal} {Annals
  of Physics}\ }\textbf {\bibinfo {volume} {303}},\ \bibinfo {pages} {2}
  (\bibinfo {year} {2003})}\BibitemShut {NoStop}%
\bibitem [{\citenamefont {Freedman}\ \emph {et~al.}(2003)\citenamefont
  {Freedman}, \citenamefont {Kitaev}, \citenamefont {Larsen},\ and\
  \citenamefont {Wang}}]{freedman2003topological}%
  \BibitemOpen
  \bibfield  {author} {\bibinfo {author} {\bibfnamefont {M.}~\bibnamefont
  {Freedman}}, \bibinfo {author} {\bibfnamefont {A.}~\bibnamefont {Kitaev}},
  \bibinfo {author} {\bibfnamefont {M.}~\bibnamefont {Larsen}}, \ and\ \bibinfo
  {author} {\bibfnamefont {Z.}~\bibnamefont {Wang}},\ }\href@noop {} {\bibfield
   {journal} {\bibinfo  {journal} {Bulletin of the American Mathematical
  Society}\ }\textbf {\bibinfo {volume} {40}},\ \bibinfo {pages} {31} (\bibinfo
  {year} {2003})}\BibitemShut {NoStop}%
\bibitem [{\citenamefont {Nayak}\ \emph {et~al.}(2008)\citenamefont {Nayak},
  \citenamefont {Simon}, \citenamefont {Stern}, \citenamefont {Freedman},\ and\
  \citenamefont {Das~Sarma}}]{Nayak_08}%
  \BibitemOpen
  \bibfield  {author} {\bibinfo {author} {\bibfnamefont {C.}~\bibnamefont
  {Nayak}}, \bibinfo {author} {\bibfnamefont {S.~H.}\ \bibnamefont {Simon}},
  \bibinfo {author} {\bibfnamefont {A.}~\bibnamefont {Stern}}, \bibinfo
  {author} {\bibfnamefont {M.}~\bibnamefont {Freedman}}, \ and\ \bibinfo
  {author} {\bibfnamefont {S.}~\bibnamefont {Das~Sarma}},\ }\href {\doibase
  10.1103/RevModPhys.80.1083} {\bibfield  {journal} {\bibinfo  {journal} {Rev.
  Mod. Phys.}\ }\textbf {\bibinfo {volume} {80}},\ \bibinfo {pages} {1083}
  (\bibinfo {year} {2008})}\BibitemShut {NoStop}%
\bibitem [{\citenamefont {Bartolomei}\ \emph {et~al.}(2020)\citenamefont
  {Bartolomei}, \citenamefont {Kumar}, \citenamefont {Bisognin}, \citenamefont
  {Marguerite}, \citenamefont {Berroir}, \citenamefont {Bocquillon},
  \citenamefont {Pla{\c{c}}ais}, \citenamefont {Cavanna}, \citenamefont {Dong},
  \citenamefont {Gennser} \emph {et~al.}}]{bartolomei2020fractional}%
  \BibitemOpen
  \bibfield  {author} {\bibinfo {author} {\bibfnamefont {H.}~\bibnamefont
  {Bartolomei}}, \bibinfo {author} {\bibfnamefont {M.}~\bibnamefont {Kumar}},
  \bibinfo {author} {\bibfnamefont {R.}~\bibnamefont {Bisognin}}, \bibinfo
  {author} {\bibfnamefont {A.}~\bibnamefont {Marguerite}}, \bibinfo {author}
  {\bibfnamefont {J.-M.}\ \bibnamefont {Berroir}}, \bibinfo {author}
  {\bibfnamefont {E.}~\bibnamefont {Bocquillon}}, \bibinfo {author}
  {\bibfnamefont {B.}~\bibnamefont {Pla{\c{c}}ais}}, \bibinfo {author}
  {\bibfnamefont {A.}~\bibnamefont {Cavanna}}, \bibinfo {author} {\bibfnamefont
  {Q.}~\bibnamefont {Dong}}, \bibinfo {author} {\bibfnamefont {U.}~\bibnamefont
  {Gennser}},  \emph {et~al.},\ }\href@noop {} {\bibfield  {journal} {\bibinfo
  {journal} {Science}\ }\textbf {\bibinfo {volume} {368}},\ \bibinfo {pages}
  {173} (\bibinfo {year} {2020})}\BibitemShut {NoStop}%
\bibitem [{\citenamefont {Nakamura}\ \emph {et~al.}(2020)\citenamefont
  {Nakamura}, \citenamefont {Liang}, \citenamefont {Gardner},\ and\
  \citenamefont {Manfra}}]{nakamura2020direct}%
  \BibitemOpen
  \bibfield  {author} {\bibinfo {author} {\bibfnamefont {J.}~\bibnamefont
  {Nakamura}}, \bibinfo {author} {\bibfnamefont {S.}~\bibnamefont {Liang}},
  \bibinfo {author} {\bibfnamefont {G.}~\bibnamefont {Gardner}}, \ and\
  \bibinfo {author} {\bibfnamefont {M.}~\bibnamefont {Manfra}},\ }\href@noop {}
  {\bibfield  {journal} {\bibinfo  {journal} {Nature Physics}\ ,\ \bibinfo
  {pages} {1}} (\bibinfo {year} {2020})}\BibitemShut {NoStop}%
\bibitem [{\citenamefont {Girardeau}(1965)}]{Girardeau_1965}%
  \BibitemOpen
  \bibfield  {author} {\bibinfo {author} {\bibfnamefont {M.~D.}\ \bibnamefont
  {Girardeau}},\ }\href {\doibase 10.1103/PhysRev.139.B500} {\bibfield
  {journal} {\bibinfo  {journal} {Phys. Rev.}\ }\textbf {\bibinfo {volume}
  {139}},\ \bibinfo {pages} {B500} (\bibinfo {year} {1965})}\BibitemShut
  {NoStop}%
\bibitem [{\citenamefont {Souriau}(1970)}]{Souriau_1970}%
  \BibitemOpen
  \bibfield  {author} {\bibinfo {author} {\bibfnamefont {J.-M.}\ \bibnamefont
  {Souriau}},\ }\href
  {http://www.jmsouriau.com/structure_des_systemes_dynamiques.htm} {\emph
  {\bibinfo {title} {Structure des syst\`emes dynamiques}}},\ Ma{\^{i}}trises
  de math\'ematiques\ (\bibinfo  {publisher} {Dunod, Paris},\ \bibinfo {year}
  {1970})\ pp.\ \bibinfo {pages} {xxxii+414},\ \bibinfo {note} {english
  translation by R. H. Cushman and G. M. Tuynman, Progress in Mathematics, 149,
  Birkh\"auser Boston Inc., Boston, MA, 1997}\BibitemShut {NoStop}%
\bibitem [{\citenamefont {Streater}\ and\ \citenamefont
  {Wilde}(1970)}]{StrWil-70}%
  \BibitemOpen
  \bibfield  {author} {\bibinfo {author} {\bibfnamefont {R.~F.}\ \bibnamefont
  {Streater}}\ and\ \bibinfo {author} {\bibfnamefont {I.~F.}\ \bibnamefont
  {Wilde}},\ }\href {\doibase 10.1016/0550-3213(70)90445-1} {\bibfield
  {journal} {\bibinfo  {journal} {Nucl. Phys. B}\ }\textbf {\bibinfo {volume}
  {24}},\ \bibinfo {pages} {561} (\bibinfo {year} {1970})}\BibitemShut
  {NoStop}%
\bibitem [{\citenamefont {Laidlaw}\ and\ \citenamefont
  {DeWitt}(1971)}]{laidlaw1971feynman}%
  \BibitemOpen
  \bibfield  {author} {\bibinfo {author} {\bibfnamefont {M.~G.}\ \bibnamefont
  {Laidlaw}}\ and\ \bibinfo {author} {\bibfnamefont {C.~M.}\ \bibnamefont
  {DeWitt}},\ }\href@noop {} {\bibfield  {journal} {\bibinfo  {journal}
  {Physical Review D}\ }\textbf {\bibinfo {volume} {3}},\ \bibinfo {pages}
  {1375} (\bibinfo {year} {1971})}\BibitemShut {NoStop}%
\bibitem [{\citenamefont {Leinaas}\ and\ \citenamefont
  {Myrheim}(1977)}]{leinaas1977theory}%
  \BibitemOpen
  \bibfield  {author} {\bibinfo {author} {\bibfnamefont {J.~M.}\ \bibnamefont
  {Leinaas}}\ and\ \bibinfo {author} {\bibfnamefont {J.}~\bibnamefont
  {Myrheim}},\ }\href@noop {} {\bibfield  {journal} {\bibinfo  {journal} {Il
  Nuovo Cimento B (1971-1996)}\ }\textbf {\bibinfo {volume} {37}},\ \bibinfo
  {pages} {1} (\bibinfo {year} {1977})}\BibitemShut {NoStop}%
\bibitem [{\citenamefont {Goldin}\ \emph {et~al.}(1981)\citenamefont {Goldin},
  \citenamefont {Menikoff},\ and\ \citenamefont
  {Sharp}}]{goldin1981representations}%
  \BibitemOpen
  \bibfield  {author} {\bibinfo {author} {\bibfnamefont {G.~A.}\ \bibnamefont
  {Goldin}}, \bibinfo {author} {\bibfnamefont {R.}~\bibnamefont {Menikoff}}, \
  and\ \bibinfo {author} {\bibfnamefont {D.~H.}\ \bibnamefont {Sharp}},\
  }\href@noop {} {\bibfield  {journal} {\bibinfo  {journal} {Journal of
  Mathematical Physics}\ }\textbf {\bibinfo {volume} {22}},\ \bibinfo {pages}
  {1664} (\bibinfo {year} {1981})}\BibitemShut {NoStop}%
\bibitem [{\citenamefont {Wilczek}(1982{\natexlab{a}})}]{Wilczek_82}%
  \BibitemOpen
  \bibfield  {author} {\bibinfo {author} {\bibfnamefont {F.}~\bibnamefont
  {Wilczek}},\ }\href {\doibase 10.1103/PhysRevLett.48.1144} {\bibfield
  {journal} {\bibinfo  {journal} {Phys. Rev. Lett.}\ }\textbf {\bibinfo
  {volume} {48}},\ \bibinfo {pages} {1144} (\bibinfo {year}
  {1982}{\natexlab{a}})}\BibitemShut {NoStop}%
\bibitem [{\citenamefont {Wilczek}(1982{\natexlab{b}})}]{Wilczek_82b}%
  \BibitemOpen
  \bibfield  {author} {\bibinfo {author} {\bibfnamefont {F.}~\bibnamefont
  {Wilczek}},\ }\href {\doibase 10.1103/PhysRevLett.49.957} {\bibfield
  {journal} {\bibinfo  {journal} {Phys. Rev. Lett.}\ }\textbf {\bibinfo
  {volume} {49}},\ \bibinfo {pages} {957} (\bibinfo {year}
  {1982}{\natexlab{b}})}\BibitemShut {NoStop}%
\bibitem [{\citenamefont {Wu}(1984{\natexlab{a}})}]{Wu_84}%
  \BibitemOpen
  \bibfield  {author} {\bibinfo {author} {\bibfnamefont {Y.-S.}\ \bibnamefont
  {Wu}},\ }\href {\doibase 10.1103/PhysRevLett.52.2103} {\bibfield  {journal}
  {\bibinfo  {journal} {Phys. Rev. Lett.}\ }\textbf {\bibinfo {volume} {52}},\
  \bibinfo {pages} {2103} (\bibinfo {year} {1984}{\natexlab{a}})}\BibitemShut
  {NoStop}%
\bibitem [{\citenamefont {Fr{\"o}hlich}(1990)}]{Froehlich_1990}%
  \BibitemOpen
  \bibfield  {author} {\bibinfo {author} {\bibfnamefont {J.}~\bibnamefont
  {Fr{\"o}hlich}},\ }in\ \href@noop {} {\emph {\bibinfo {booktitle}
  {Proceedings of the {G}ibbs {S}ymposium ({N}ew {H}aven, {CT}, 1989)}}}\
  (\bibinfo  {publisher} {Amer. Math. Soc., Providence, RI},\ \bibinfo {year}
  {1990})\ pp.\ \bibinfo {pages} {89--142}\BibitemShut {NoStop}%
\bibitem [{\citenamefont {Cooper}\ and\ \citenamefont
  {Simon}(2015)}]{CooSim-15}%
  \BibitemOpen
  \bibfield  {author} {\bibinfo {author} {\bibfnamefont {N.~R.}\ \bibnamefont
  {Cooper}}\ and\ \bibinfo {author} {\bibfnamefont {S.~H.}\ \bibnamefont
  {Simon}},\ }\href {\doibase 10.1103/PhysRevLett.114.106802} {\bibfield
  {journal} {\bibinfo  {journal} {Phys. Rev. Lett.}\ }\textbf {\bibinfo
  {volume} {114}},\ \bibinfo {pages} {106802} (\bibinfo {year}
  {2015})}\BibitemShut {NoStop}%
\bibitem [{\citenamefont {Zhang}\ \emph {et~al.}(2014)\citenamefont {Zhang},
  \citenamefont {Sreejith}, \citenamefont {Gemelke},\ and\ \citenamefont
  {Jain}}]{ZhaSreGemJai-14}%
  \BibitemOpen
  \bibfield  {author} {\bibinfo {author} {\bibfnamefont {Y.}~\bibnamefont
  {Zhang}}, \bibinfo {author} {\bibfnamefont {G.~J.}\ \bibnamefont {Sreejith}},
  \bibinfo {author} {\bibfnamefont {N.~D.}\ \bibnamefont {Gemelke}}, \ and\
  \bibinfo {author} {\bibfnamefont {J.~K.}\ \bibnamefont {Jain}},\ }\href
  {\doibase 10.1103/PhysRevLett.113.160404} {\bibfield  {journal} {\bibinfo
  {journal} {Phys. Rev. Lett.}\ }\textbf {\bibinfo {volume} {113}},\ \bibinfo
  {pages} {160404} (\bibinfo {year} {2014})}\BibitemShut {NoStop}%
\bibitem [{\citenamefont {Zhang}\ \emph {et~al.}(2015)\citenamefont {Zhang},
  \citenamefont {Sreejith},\ and\ \citenamefont {Jain}}]{ZhaSreJai-15}%
  \BibitemOpen
  \bibfield  {author} {\bibinfo {author} {\bibfnamefont {Y.}~\bibnamefont
  {Zhang}}, \bibinfo {author} {\bibfnamefont {G.~J.}\ \bibnamefont {Sreejith}},
  \ and\ \bibinfo {author} {\bibfnamefont {J.~K.}\ \bibnamefont {Jain}},\
  }\href {\doibase 10.1103/PhysRevB.92.075116} {\bibfield  {journal} {\bibinfo
  {journal} {Phys. Rev. B.}\ }\textbf {\bibinfo {volume} {92}},\ \bibinfo
  {pages} {075116} (\bibinfo {year} {2015})}\BibitemShut {NoStop}%
\bibitem [{\citenamefont {Morampudi}\ \emph {et~al.}(2017)\citenamefont
  {Morampudi}, \citenamefont {Turner}, \citenamefont {Pollmann},\ and\
  \citenamefont {Wilczek}}]{MorTurPolWil-17}%
  \BibitemOpen
  \bibfield  {author} {\bibinfo {author} {\bibfnamefont {S.~C.}\ \bibnamefont
  {Morampudi}}, \bibinfo {author} {\bibfnamefont {A.~M.}\ \bibnamefont
  {Turner}}, \bibinfo {author} {\bibfnamefont {F.}~\bibnamefont {Pollmann}}, \
  and\ \bibinfo {author} {\bibfnamefont {F.}~\bibnamefont {Wilczek}},\ }\href
  {\doibase 10.1103/PhysRevLett.118.227201} {\bibfield  {journal} {\bibinfo
  {journal} {Phys. Rev. Lett.}\ }\textbf {\bibinfo {volume} {118}},\ \bibinfo
  {pages} {227201} (\bibinfo {year} {2017})}\BibitemShut {NoStop}%
\bibitem [{\citenamefont {Umucal\ifmmode\imath\else\i\fi{}lar}\ \emph
  {et~al.}(2018)\citenamefont {Umucal\ifmmode\imath\else\i\fi{}lar},
  \citenamefont {Macaluso}, \citenamefont {Comparin},\ and\ \citenamefont
  {Carusotto}}]{UmuMacComCar-18}%
  \BibitemOpen
  \bibfield  {author} {\bibinfo {author} {\bibfnamefont {R.~O.}\ \bibnamefont
  {Umucal\ifmmode\imath\else\i\fi{}lar}}, \bibinfo {author} {\bibfnamefont
  {E.}~\bibnamefont {Macaluso}}, \bibinfo {author} {\bibfnamefont
  {T.}~\bibnamefont {Comparin}}, \ and\ \bibinfo {author} {\bibfnamefont
  {I.}~\bibnamefont {Carusotto}},\ }\href {\doibase
  10.1103/PhysRevLett.120.230403} {\bibfield  {journal} {\bibinfo  {journal}
  {Phys. Rev. Lett.}\ }\textbf {\bibinfo {volume} {120}},\ \bibinfo {pages}
  {230403} (\bibinfo {year} {2018})}\BibitemShut {NoStop}%
\bibitem [{\citenamefont {Yakaboylu}\ and\ \citenamefont
  {Lemeshko}(2018)}]{Yakaboylu_2018}%
  \BibitemOpen
  \bibfield  {author} {\bibinfo {author} {\bibfnamefont {E.}~\bibnamefont
  {Yakaboylu}}\ and\ \bibinfo {author} {\bibfnamefont {M.}~\bibnamefont
  {Lemeshko}},\ }\href {\doibase 10.1103/PhysRevB.98.045402} {\bibfield
  {journal} {\bibinfo  {journal} {Phys. Rev. B}\ }\textbf {\bibinfo {volume}
  {98}},\ \bibinfo {pages} {045402} (\bibinfo {year} {2018})}\BibitemShut
  {NoStop}%
\bibitem [{\citenamefont {Correggi}\ \emph {et~al.}(2019)\citenamefont
  {Correggi}, \citenamefont {Duboscq}, \citenamefont {Lundholm},\ and\
  \citenamefont {Rougerie}}]{CorDubLunRou-19}%
  \BibitemOpen
  \bibfield  {author} {\bibinfo {author} {\bibfnamefont {M.}~\bibnamefont
  {Correggi}}, \bibinfo {author} {\bibfnamefont {R.}~\bibnamefont {Duboscq}},
  \bibinfo {author} {\bibfnamefont {D.}~\bibnamefont {Lundholm}}, \ and\
  \bibinfo {author} {\bibfnamefont {N.}~\bibnamefont {Rougerie}},\ }\href
  {\doibase 10.1209/0295-5075/126/20005} {\bibfield  {journal} {\bibinfo
  {journal} {EPL (Europhysics Letters)}\ }\textbf {\bibinfo {volume} {126}},\
  \bibinfo {pages} {20005} (\bibinfo {year} {2019})}\BibitemShut {NoStop}%
\bibitem [{\citenamefont {Lundholm}\ and\ \citenamefont
  {Rougerie}(2016)}]{Lundholm_2016}%
  \BibitemOpen
  \bibfield  {author} {\bibinfo {author} {\bibfnamefont {D.}~\bibnamefont
  {Lundholm}}\ and\ \bibinfo {author} {\bibfnamefont {N.}~\bibnamefont
  {Rougerie}},\ }\href {\doibase 10.1103/PhysRevLett.116.170401} {\bibfield
  {journal} {\bibinfo  {journal} {Phys. Rev. Lett.}\ }\textbf {\bibinfo
  {volume} {116}},\ \bibinfo {pages} {170401} (\bibinfo {year}
  {2016})}\BibitemShut {NoStop}%
\bibitem [{\citenamefont {Moody}\ \emph {et~al.}(1986)\citenamefont {Moody},
  \citenamefont {Shapere},\ and\ \citenamefont {Wilczek}}]{Moody_86}%
  \BibitemOpen
  \bibfield  {author} {\bibinfo {author} {\bibfnamefont {J.}~\bibnamefont
  {Moody}}, \bibinfo {author} {\bibfnamefont {A.}~\bibnamefont {Shapere}}, \
  and\ \bibinfo {author} {\bibfnamefont {F.}~\bibnamefont {Wilczek}},\ }\href
  {\doibase 10.1103/PhysRevLett.56.893} {\bibfield  {journal} {\bibinfo
  {journal} {Phys. Rev. Lett.}\ }\textbf {\bibinfo {volume} {56}},\ \bibinfo
  {pages} {893} (\bibinfo {year} {1986})}\BibitemShut {NoStop}%
\bibitem [{\citenamefont {Yakaboylu}\ \emph {et~al.}(2017)\citenamefont
  {Yakaboylu}, \citenamefont {Deuchert},\ and\ \citenamefont
  {Lemeshko}}]{Yakaboylu17}%
  \BibitemOpen
  \bibfield  {author} {\bibinfo {author} {\bibfnamefont {E.}~\bibnamefont
  {Yakaboylu}}, \bibinfo {author} {\bibfnamefont {A.}~\bibnamefont {Deuchert}},
  \ and\ \bibinfo {author} {\bibfnamefont {M.}~\bibnamefont {Lemeshko}},\
  }\href {\doibase 10.1103/PhysRevLett.119.235301} {\bibfield  {journal}
  {\bibinfo  {journal} {Phys. Rev. Lett.}\ }\textbf {\bibinfo {volume} {119}},\
  \bibinfo {pages} {235301} (\bibinfo {year} {2017})}\BibitemShut {NoStop}%
\bibitem [{\citenamefont {Landau}(1933)}]{landau1933uber}%
  \BibitemOpen
  \bibfield  {author} {\bibinfo {author} {\bibfnamefont {L.}~\bibnamefont
  {Landau}},\ }\href@noop {} {\bibfield  {journal} {\bibinfo  {journal} {Phys.
  Z. Sowjetunion}\ }\textbf {\bibinfo {volume} {3}},\ \bibinfo {pages} {644}
  (\bibinfo {year} {1933})}\BibitemShut {NoStop}%
\bibitem [{\citenamefont {Pekar}(1946)}]{pekar1946local}%
  \BibitemOpen
  \bibfield  {author} {\bibinfo {author} {\bibfnamefont {S.}~\bibnamefont
  {Pekar}},\ }\href@noop {} {\bibfield  {journal} {\bibinfo  {journal} {Zhurnal
  Eksperimentalnoi I Teoreticheskoi Fiziki}\ }\textbf {\bibinfo {volume}
  {16}},\ \bibinfo {pages} {341} (\bibinfo {year} {1946})}\BibitemShut
  {NoStop}%
\bibitem [{\citenamefont {Fr{\"o}hlich}(1954)}]{frohlich1954electrons}%
  \BibitemOpen
  \bibfield  {author} {\bibinfo {author} {\bibfnamefont {H.}~\bibnamefont
  {Fr{\"o}hlich}},\ }\href@noop {} {\bibfield  {journal} {\bibinfo  {journal}
  {Advances in Physics}\ }\textbf {\bibinfo {volume} {3}},\ \bibinfo {pages}
  {325} (\bibinfo {year} {1954})}\BibitemShut {NoStop}%
\bibitem [{\citenamefont {Schmidt}\ and\ \citenamefont
  {Lemeshko}(2015)}]{Lemeshko_2015}%
  \BibitemOpen
  \bibfield  {author} {\bibinfo {author} {\bibfnamefont {R.}~\bibnamefont
  {Schmidt}}\ and\ \bibinfo {author} {\bibfnamefont {M.}~\bibnamefont
  {Lemeshko}},\ }\href {\doibase 10.1103/PhysRevLett.114.203001} {\bibfield
  {journal} {\bibinfo  {journal} {Phys. Rev. Lett.}\ }\textbf {\bibinfo
  {volume} {114}},\ \bibinfo {pages} {203001} (\bibinfo {year}
  {2015})}\BibitemShut {NoStop}%
\bibitem [{\citenamefont {Duer}(2018)}]{duer2018probing}%
  \BibitemOpen
  \bibfield  {author} {\bibinfo {author} {\bibfnamefont {M.}~\bibnamefont
  {Duer}},\ }\href@noop {} {\bibfield  {journal} {\bibinfo  {journal} {Nature
  (London)}\ }\textbf {\bibinfo {volume} {560}} (\bibinfo {year}
  {2018})}\BibitemShut {NoStop}%
\bibitem [{\citenamefont {Kondo}(1964)}]{kondo1964resistance}%
  \BibitemOpen
  \bibfield  {author} {\bibinfo {author} {\bibfnamefont {J.}~\bibnamefont
  {Kondo}},\ }\href@noop {} {\bibfield  {journal} {\bibinfo  {journal}
  {Progress of theoretical physics}\ }\textbf {\bibinfo {volume} {32}},\
  \bibinfo {pages} {37} (\bibinfo {year} {1964})}\BibitemShut {NoStop}%
\bibitem [{\citenamefont {Anderson}(1967)}]{Anderson_67}%
  \BibitemOpen
  \bibfield  {author} {\bibinfo {author} {\bibfnamefont {P.~W.}\ \bibnamefont
  {Anderson}},\ }\href {\doibase 10.1103/PhysRevLett.18.1049} {\bibfield
  {journal} {\bibinfo  {journal} {Phys. Rev. Lett.}\ }\textbf {\bibinfo
  {volume} {18}},\ \bibinfo {pages} {1049} (\bibinfo {year}
  {1967})}\BibitemShut {NoStop}%
\bibitem [{\citenamefont {Schirotzek}\ \emph {et~al.}(2009)\citenamefont
  {Schirotzek}, \citenamefont {Wu}, \citenamefont {Sommer},\ and\ \citenamefont
  {Zwierlein}}]{Schirotzek_09}%
  \BibitemOpen
  \bibfield  {author} {\bibinfo {author} {\bibfnamefont {A.}~\bibnamefont
  {Schirotzek}}, \bibinfo {author} {\bibfnamefont {C.-H.}\ \bibnamefont {Wu}},
  \bibinfo {author} {\bibfnamefont {A.}~\bibnamefont {Sommer}}, \ and\ \bibinfo
  {author} {\bibfnamefont {M.~W.}\ \bibnamefont {Zwierlein}},\ }\href {\doibase
  10.1103/PhysRevLett.102.230402} {\bibfield  {journal} {\bibinfo  {journal}
  {Phys. Rev. Lett.}\ }\textbf {\bibinfo {volume} {102}},\ \bibinfo {pages}
  {230402} (\bibinfo {year} {2009})}\BibitemShut {NoStop}%
\bibitem [{\citenamefont {J\o{}rgensen}\ \emph {et~al.}(2016)\citenamefont
  {J\o{}rgensen}, \citenamefont {Wacker}, \citenamefont {Skalmstang},
  \citenamefont {Parish}, \citenamefont {Levinsen}, \citenamefont
  {Christensen}, \citenamefont {Bruun},\ and\ \citenamefont
  {Arlt}}]{Jorgensen_16}%
  \BibitemOpen
  \bibfield  {author} {\bibinfo {author} {\bibfnamefont {N.~B.}\ \bibnamefont
  {J\o{}rgensen}}, \bibinfo {author} {\bibfnamefont {L.}~\bibnamefont
  {Wacker}}, \bibinfo {author} {\bibfnamefont {K.~T.}\ \bibnamefont
  {Skalmstang}}, \bibinfo {author} {\bibfnamefont {M.~M.}\ \bibnamefont
  {Parish}}, \bibinfo {author} {\bibfnamefont {J.}~\bibnamefont {Levinsen}},
  \bibinfo {author} {\bibfnamefont {R.~S.}\ \bibnamefont {Christensen}},
  \bibinfo {author} {\bibfnamefont {G.~M.}\ \bibnamefont {Bruun}}, \ and\
  \bibinfo {author} {\bibfnamefont {J.~J.}\ \bibnamefont {Arlt}},\ }\href
  {\doibase 10.1103/PhysRevLett.117.055302} {\bibfield  {journal} {\bibinfo
  {journal} {Phys. Rev. Lett.}\ }\textbf {\bibinfo {volume} {117}},\ \bibinfo
  {pages} {055302} (\bibinfo {year} {2016})}\BibitemShut {NoStop}%
\bibitem [{\citenamefont {Cetina}\ \emph {et~al.}(2016)\citenamefont {Cetina},
  \citenamefont {Jag}, \citenamefont {Lous}, \citenamefont {Fritsche},
  \citenamefont {Walraven}, \citenamefont {Grimm}, \citenamefont {Levinsen},
  \citenamefont {Parish}, \citenamefont {Schmidt}, \citenamefont {Knap},\ and\
  \citenamefont {Demler}}]{Cetina96}%
  \BibitemOpen
  \bibfield  {author} {\bibinfo {author} {\bibfnamefont {M.}~\bibnamefont
  {Cetina}}, \bibinfo {author} {\bibfnamefont {M.}~\bibnamefont {Jag}},
  \bibinfo {author} {\bibfnamefont {R.~S.}\ \bibnamefont {Lous}}, \bibinfo
  {author} {\bibfnamefont {I.}~\bibnamefont {Fritsche}}, \bibinfo {author}
  {\bibfnamefont {J.~T.~M.}\ \bibnamefont {Walraven}}, \bibinfo {author}
  {\bibfnamefont {R.}~\bibnamefont {Grimm}}, \bibinfo {author} {\bibfnamefont
  {J.}~\bibnamefont {Levinsen}}, \bibinfo {author} {\bibfnamefont {M.~M.}\
  \bibnamefont {Parish}}, \bibinfo {author} {\bibfnamefont {R.}~\bibnamefont
  {Schmidt}}, \bibinfo {author} {\bibfnamefont {M.}~\bibnamefont {Knap}}, \
  and\ \bibinfo {author} {\bibfnamefont {E.}~\bibnamefont {Demler}},\ }\href
  {\doibase 10.1126/science.aaf5134} {\bibfield  {journal} {\bibinfo  {journal}
  {Science}\ }\textbf {\bibinfo {volume} {354}},\ \bibinfo {pages} {96}
  (\bibinfo {year} {2016})}\BibitemShut {NoStop}%
\bibitem [{\citenamefont {Fukuhara}\ \emph {et~al.}(2013)\citenamefont
  {Fukuhara}, \citenamefont {Kantian}, \citenamefont {Endres}, \citenamefont
  {Cheneau}, \citenamefont {Schau{\ss}}, \citenamefont {Hild}, \citenamefont
  {Bellem}, \citenamefont {Schollw{\"o}ck}, \citenamefont {Giamarchi},
  \citenamefont {Gross} \emph {et~al.}}]{fukuhara2013quantum}%
  \BibitemOpen
  \bibfield  {author} {\bibinfo {author} {\bibfnamefont {T.}~\bibnamefont
  {Fukuhara}}, \bibinfo {author} {\bibfnamefont {A.}~\bibnamefont {Kantian}},
  \bibinfo {author} {\bibfnamefont {M.}~\bibnamefont {Endres}}, \bibinfo
  {author} {\bibfnamefont {M.}~\bibnamefont {Cheneau}}, \bibinfo {author}
  {\bibfnamefont {P.}~\bibnamefont {Schau{\ss}}}, \bibinfo {author}
  {\bibfnamefont {S.}~\bibnamefont {Hild}}, \bibinfo {author} {\bibfnamefont
  {D.}~\bibnamefont {Bellem}}, \bibinfo {author} {\bibfnamefont
  {U.}~\bibnamefont {Schollw{\"o}ck}}, \bibinfo {author} {\bibfnamefont
  {T.}~\bibnamefont {Giamarchi}}, \bibinfo {author} {\bibfnamefont
  {C.}~\bibnamefont {Gross}},  \emph {et~al.},\ }\href@noop {} {\bibfield
  {journal} {\bibinfo  {journal} {Nature Physics}\ }\textbf {\bibinfo {volume}
  {9}},\ \bibinfo {pages} {235} (\bibinfo {year} {2013})}\BibitemShut {NoStop}%
\bibitem [{\citenamefont {Koschorreck}\ \emph {et~al.}(2012)\citenamefont
  {Koschorreck}, \citenamefont {Pertot}, \citenamefont {Vogt}, \citenamefont
  {Fr{\"o}hlich}, \citenamefont {Feld},\ and\ \citenamefont
  {K{\"o}hl}}]{koschorreck2012attractive}%
  \BibitemOpen
  \bibfield  {author} {\bibinfo {author} {\bibfnamefont {M.}~\bibnamefont
  {Koschorreck}}, \bibinfo {author} {\bibfnamefont {D.}~\bibnamefont {Pertot}},
  \bibinfo {author} {\bibfnamefont {E.}~\bibnamefont {Vogt}}, \bibinfo {author}
  {\bibfnamefont {B.}~\bibnamefont {Fr{\"o}hlich}}, \bibinfo {author}
  {\bibfnamefont {M.}~\bibnamefont {Feld}}, \ and\ \bibinfo {author}
  {\bibfnamefont {M.}~\bibnamefont {K{\"o}hl}},\ }\href@noop {} {\bibfield
  {journal} {\bibinfo  {journal} {Nature}\ }\textbf {\bibinfo {volume} {485}},\
  \bibinfo {pages} {619} (\bibinfo {year} {2012})}\BibitemShut {NoStop}%
\bibitem [{\citenamefont {Kohstall}\ \emph {et~al.}(2012)\citenamefont
  {Kohstall}, \citenamefont {Zaccanti}, \citenamefont {Jag}, \citenamefont
  {Trenkwalder}, \citenamefont {Massignan}, \citenamefont {Bruun},
  \citenamefont {Schreck},\ and\ \citenamefont
  {Grimm}}]{kohstall2012metastability}%
  \BibitemOpen
  \bibfield  {author} {\bibinfo {author} {\bibfnamefont {C.}~\bibnamefont
  {Kohstall}}, \bibinfo {author} {\bibfnamefont {M.}~\bibnamefont {Zaccanti}},
  \bibinfo {author} {\bibfnamefont {M.}~\bibnamefont {Jag}}, \bibinfo {author}
  {\bibfnamefont {A.}~\bibnamefont {Trenkwalder}}, \bibinfo {author}
  {\bibfnamefont {P.}~\bibnamefont {Massignan}}, \bibinfo {author}
  {\bibfnamefont {G.~M.}\ \bibnamefont {Bruun}}, \bibinfo {author}
  {\bibfnamefont {F.}~\bibnamefont {Schreck}}, \ and\ \bibinfo {author}
  {\bibfnamefont {R.}~\bibnamefont {Grimm}},\ }\href@noop {} {\bibfield
  {journal} {\bibinfo  {journal} {Nature}\ }\textbf {\bibinfo {volume} {485}},\
  \bibinfo {pages} {615} (\bibinfo {year} {2012})}\BibitemShut {NoStop}%
\bibitem [{\citenamefont {Camargo}\ \emph {et~al.}(2018)\citenamefont
  {Camargo}, \citenamefont {Schmidt}, \citenamefont {Whalen}, \citenamefont
  {Ding}, \citenamefont {Woehl}, \citenamefont {Yoshida}, \citenamefont
  {Burgd\"orfer}, \citenamefont {Dunning}, \citenamefont {Sadeghpour},
  \citenamefont {Demler},\ and\ \citenamefont {Killian}}]{Camargo_18}%
  \BibitemOpen
  \bibfield  {author} {\bibinfo {author} {\bibfnamefont {F.}~\bibnamefont
  {Camargo}}, \bibinfo {author} {\bibfnamefont {R.}~\bibnamefont {Schmidt}},
  \bibinfo {author} {\bibfnamefont {J.~D.}\ \bibnamefont {Whalen}}, \bibinfo
  {author} {\bibfnamefont {R.}~\bibnamefont {Ding}}, \bibinfo {author}
  {\bibfnamefont {G.}~\bibnamefont {Woehl}}, \bibinfo {author} {\bibfnamefont
  {S.}~\bibnamefont {Yoshida}}, \bibinfo {author} {\bibfnamefont
  {J.}~\bibnamefont {Burgd\"orfer}}, \bibinfo {author} {\bibfnamefont {F.~B.}\
  \bibnamefont {Dunning}}, \bibinfo {author} {\bibfnamefont {H.~R.}\
  \bibnamefont {Sadeghpour}}, \bibinfo {author} {\bibfnamefont
  {E.}~\bibnamefont {Demler}}, \ and\ \bibinfo {author} {\bibfnamefont {T.~C.}\
  \bibnamefont {Killian}},\ }\href {\doibase 10.1103/PhysRevLett.120.083401}
  {\bibfield  {journal} {\bibinfo  {journal} {Phys. Rev. Lett.}\ }\textbf
  {\bibinfo {volume} {120}},\ \bibinfo {pages} {083401} (\bibinfo {year}
  {2018})}\BibitemShut {NoStop}%
\bibitem [{\citenamefont {Schmidt}\ and\ \citenamefont
  {Lemeshko}(2016)}]{PhysRevX.6.011012}%
  \BibitemOpen
  \bibfield  {author} {\bibinfo {author} {\bibfnamefont {R.}~\bibnamefont
  {Schmidt}}\ and\ \bibinfo {author} {\bibfnamefont {M.}~\bibnamefont
  {Lemeshko}},\ }\href {\doibase 10.1103/PhysRevX.6.011012} {\bibfield
  {journal} {\bibinfo  {journal} {Phys. Rev. X}\ }\textbf {\bibinfo {volume}
  {6}},\ \bibinfo {pages} {011012} (\bibinfo {year} {2016})}\BibitemShut
  {NoStop}%
\bibitem [{\citenamefont {Yakaboylu}\ \emph {et~al.}(2018)\citenamefont
  {Yakaboylu}, \citenamefont {Shkolnikov},\ and\ \citenamefont
  {Lemeshko}}]{Yakaboylu_2018_quantum}%
  \BibitemOpen
  \bibfield  {author} {\bibinfo {author} {\bibfnamefont {E.}~\bibnamefont
  {Yakaboylu}}, \bibinfo {author} {\bibfnamefont {M.}~\bibnamefont
  {Shkolnikov}}, \ and\ \bibinfo {author} {\bibfnamefont {M.}~\bibnamefont
  {Lemeshko}},\ }\href {\doibase 10.1103/PhysRevLett.121.255302} {\bibfield
  {journal} {\bibinfo  {journal} {Phys. Rev. Lett.}\ }\textbf {\bibinfo
  {volume} {121}},\ \bibinfo {pages} {255302} (\bibinfo {year}
  {2018})}\BibitemShut {NoStop}%
\bibitem [{\citenamefont {Koepsell}\ \emph {et~al.}(2019)\citenamefont
  {Koepsell}, \citenamefont {Vijayan}, \citenamefont {Sompet}, \citenamefont
  {Grusdt}, \citenamefont {Hilker}, \citenamefont {Demler}, \citenamefont
  {Salomon}, \citenamefont {Bloch},\ and\ \citenamefont
  {Gross}}]{koepsell2019imaging}%
  \BibitemOpen
  \bibfield  {author} {\bibinfo {author} {\bibfnamefont {J.}~\bibnamefont
  {Koepsell}}, \bibinfo {author} {\bibfnamefont {J.}~\bibnamefont {Vijayan}},
  \bibinfo {author} {\bibfnamefont {P.}~\bibnamefont {Sompet}}, \bibinfo
  {author} {\bibfnamefont {F.}~\bibnamefont {Grusdt}}, \bibinfo {author}
  {\bibfnamefont {T.~A.}\ \bibnamefont {Hilker}}, \bibinfo {author}
  {\bibfnamefont {E.}~\bibnamefont {Demler}}, \bibinfo {author} {\bibfnamefont
  {G.}~\bibnamefont {Salomon}}, \bibinfo {author} {\bibfnamefont
  {I.}~\bibnamefont {Bloch}}, \ and\ \bibinfo {author} {\bibfnamefont
  {C.}~\bibnamefont {Gross}},\ }\href@noop {} {\bibfield  {journal} {\bibinfo
  {journal} {Nature}\ }\textbf {\bibinfo {volume} {572}},\ \bibinfo {pages}
  {358} (\bibinfo {year} {2019})}\BibitemShut {NoStop}%
\bibitem [{\citenamefont {Jackiw}(1990)}]{Jackiw-90}%
  \BibitemOpen
  \bibfield  {author} {\bibinfo {author} {\bibfnamefont {R.}~\bibnamefont
  {Jackiw}},\ }\href {\doibase 10.1016/0920-5632(90)90648-E} {\bibfield
  {journal} {\bibinfo  {journal} {Nuclear Phys. B Proc. Suppl.}\ }\textbf
  {\bibinfo {volume} {18A}},\ \bibinfo {pages} {107} (\bibinfo {year}
  {1990})},\ \bibinfo {note} {{Integrability and quantization (Jaca,
  1989)}}\BibitemShut {NoStop}%
\bibitem [{\citenamefont {Iengo}\ and\ \citenamefont
  {Lechner}(1992)}]{iengo1992anyon}%
  \BibitemOpen
  \bibfield  {author} {\bibinfo {author} {\bibfnamefont {R.}~\bibnamefont
  {Iengo}}\ and\ \bibinfo {author} {\bibfnamefont {K.}~\bibnamefont
  {Lechner}},\ }\href@noop {} {\bibfield  {journal} {\bibinfo  {journal}
  {Physics Reports}\ }\textbf {\bibinfo {volume} {213}},\ \bibinfo {pages}
  {179} (\bibinfo {year} {1992})}\BibitemShut {NoStop}%
\bibitem [{\citenamefont {Dunne}(1999)}]{dunne1999aspects}%
  \BibitemOpen
  \bibfield  {author} {\bibinfo {author} {\bibfnamefont {G.~V.}\ \bibnamefont
  {Dunne}},\ }in\ \href@noop {} {\emph {\bibinfo {booktitle} {Aspects
  topologiques de la physique en basse dimension. Topological aspects of low
  dimensional systems}}}\ (\bibinfo  {publisher} {Springer},\ \bibinfo {year}
  {1999})\ pp.\ \bibinfo {pages} {177--263}\BibitemShut {NoStop}%
\bibitem [{\citenamefont {de~Veigy}\ and\ \citenamefont
  {Ouvry}(1993)}]{de1993topological}%
  \BibitemOpen
  \bibfield  {author} {\bibinfo {author} {\bibfnamefont {A.~D.}\ \bibnamefont
  {de~Veigy}}\ and\ \bibinfo {author} {\bibfnamefont {S.}~\bibnamefont
  {Ouvry}},\ }\href@noop {} {\bibfield  {journal} {\bibinfo  {journal} {Physics
  Letters B}\ }\textbf {\bibinfo {volume} {307}},\ \bibinfo {pages} {91}
  (\bibinfo {year} {1993})}\BibitemShut {NoStop}%
\bibitem [{Note1()}]{Note1}%
  \BibitemOpen
  \bibinfo {note} {We note that in Chern-Simons theory, the flux is given by
  $\Phi = 1/\kappa $ so that the statistics parameter reads $\alpha = \Phi
  /(2\pi )$, whereas in the flux-tube-charged-particle composite picture, i.e.,
  in Maxwell theory, the statistics parameter is given by $\alpha = 2 \xi
  /(2\pi )$ with $\xi $ being the flux of the each composite. In other words,
  the statistics phase is half of the flux in the former case, whereas in the
  latter case it is equivalent to the flux. This can be intuitively understood
  as follows. Interchanging two particles in Chern-Simons theory gives only the
  phase from the charges moving around the fluxes, but no contribution from the
  fluxes moving around the charges, whereas in Maxwell theory interchanging two
  composites gives the sum of these two phases~\cite
  {iengo1992anyon}.}\BibitemShut {Stop}%
\bibitem [{\citenamefont {Arovas}\ \emph {et~al.}(1985)\citenamefont {Arovas},
  \citenamefont {Schrieffer}, \citenamefont {Wilczek},\ and\ \citenamefont
  {Zee}}]{AroSchWilZee-85}%
  \BibitemOpen
  \bibfield  {author} {\bibinfo {author} {\bibfnamefont {D.~P.}\ \bibnamefont
  {Arovas}}, \bibinfo {author} {\bibfnamefont {R.}~\bibnamefont {Schrieffer}},
  \bibinfo {author} {\bibfnamefont {F.}~\bibnamefont {Wilczek}}, \ and\
  \bibinfo {author} {\bibfnamefont {A.}~\bibnamefont {Zee}},\ }\href {\doibase
  10.1016/0550-3213(85)90252-4} {\bibfield  {journal} {\bibinfo  {journal}
  {Nuclear Physics B}\ }\textbf {\bibinfo {volume} {251}},\ \bibinfo {pages}
  {117 } (\bibinfo {year} {1985})}\BibitemShut {NoStop}%
\bibitem [{\citenamefont {Vercin}(1991)}]{vercin1991two}%
  \BibitemOpen
  \bibfield  {author} {\bibinfo {author} {\bibfnamefont {A.}~\bibnamefont
  {Vercin}},\ }\href@noop {} {\bibfield  {journal} {\bibinfo  {journal}
  {Physics Letters B}\ }\textbf {\bibinfo {volume} {260}},\ \bibinfo {pages}
  {120} (\bibinfo {year} {1991})}\BibitemShut {NoStop}%
\bibitem [{\citenamefont {Myrheim}\ \emph {et~al.}(1992)\citenamefont
  {Myrheim}, \citenamefont {Halvorsen},\ and\ \citenamefont
  {Vercin}}]{myrheim1992two}%
  \BibitemOpen
  \bibfield  {author} {\bibinfo {author} {\bibfnamefont {J.}~\bibnamefont
  {Myrheim}}, \bibinfo {author} {\bibfnamefont {E.}~\bibnamefont {Halvorsen}},
  \ and\ \bibinfo {author} {\bibfnamefont {A.}~\bibnamefont {Vercin}},\
  }\href@noop {} {\bibfield  {journal} {\bibinfo  {journal} {Physics Letters
  B}\ }\textbf {\bibinfo {volume} {278}},\ \bibinfo {pages} {171} (\bibinfo
  {year} {1992})}\BibitemShut {NoStop}%
\bibitem [{\citenamefont {Murthy}\ \emph {et~al.}(1991)\citenamefont {Murthy},
  \citenamefont {Law}, \citenamefont {Brack},\ and\ \citenamefont
  {Bhaduri}}]{Murthy_91}%
  \BibitemOpen
  \bibfield  {author} {\bibinfo {author} {\bibfnamefont {M.~V.~N.}\
  \bibnamefont {Murthy}}, \bibinfo {author} {\bibfnamefont {J.}~\bibnamefont
  {Law}}, \bibinfo {author} {\bibfnamefont {M.}~\bibnamefont {Brack}}, \ and\
  \bibinfo {author} {\bibfnamefont {R.~K.}\ \bibnamefont {Bhaduri}},\ }\href
  {\doibase 10.1103/PhysRevLett.67.1817} {\bibfield  {journal} {\bibinfo
  {journal} {Phys. Rev. Lett.}\ }\textbf {\bibinfo {volume} {67}},\ \bibinfo
  {pages} {1817} (\bibinfo {year} {1991})}\BibitemShut {NoStop}%
\bibitem [{\citenamefont {Sporre}\ \emph {et~al.}(1991)\citenamefont {Sporre},
  \citenamefont {Verbaarschot},\ and\ \citenamefont {Zahed}}]{Sporre_91}%
  \BibitemOpen
  \bibfield  {author} {\bibinfo {author} {\bibfnamefont {M.}~\bibnamefont
  {Sporre}}, \bibinfo {author} {\bibfnamefont {J.~J.~M.}\ \bibnamefont
  {Verbaarschot}}, \ and\ \bibinfo {author} {\bibfnamefont {I.}~\bibnamefont
  {Zahed}},\ }\href {\doibase 10.1103/PhysRevLett.67.1813} {\bibfield
  {journal} {\bibinfo  {journal} {Phys. Rev. Lett.}\ }\textbf {\bibinfo
  {volume} {67}},\ \bibinfo {pages} {1813} (\bibinfo {year}
  {1991})}\BibitemShut {NoStop}%
\bibitem [{\citenamefont {Sporre}\ \emph {et~al.}(1992)\citenamefont {Sporre},
  \citenamefont {Verbaarschot},\ and\ \citenamefont {Zahed}}]{SpoVerZah-92}%
  \BibitemOpen
  \bibfield  {author} {\bibinfo {author} {\bibfnamefont {M.}~\bibnamefont
  {Sporre}}, \bibinfo {author} {\bibfnamefont {J.~J.~M.}\ \bibnamefont
  {Verbaarschot}}, \ and\ \bibinfo {author} {\bibfnamefont {I.}~\bibnamefont
  {Zahed}},\ }\href {\doibase 10.1103/PhysRevB.46.5738} {\bibfield  {journal}
  {\bibinfo  {journal} {Phys. Rev. B}\ }\textbf {\bibinfo {volume} {46}},\
  \bibinfo {pages} {5738} (\bibinfo {year} {1992})}\BibitemShut {NoStop}%
\bibitem [{\citenamefont {Sporre}\ \emph {et~al.}(1993)\citenamefont {Sporre},
  \citenamefont {Verbaarschot},\ and\ \citenamefont {Zahed}}]{SpoVerZah-93}%
  \BibitemOpen
  \bibfield  {author} {\bibinfo {author} {\bibfnamefont {M.}~\bibnamefont
  {Sporre}}, \bibinfo {author} {\bibfnamefont {J.}~\bibnamefont
  {Verbaarschot}}, \ and\ \bibinfo {author} {\bibfnamefont {I.}~\bibnamefont
  {Zahed}},\ }\href {\doibase 10.1016/0550-3213(93)90357-U} {\bibfield
  {journal} {\bibinfo  {journal} {Nuclear Physics B}\ }\textbf {\bibinfo
  {volume} {389}},\ \bibinfo {pages} {645} (\bibinfo {year}
  {1993})}\BibitemShut {NoStop}%
\bibitem [{\citenamefont {Wu}(1984{\natexlab{b}})}]{Wu_84_anyon}%
  \BibitemOpen
  \bibfield  {author} {\bibinfo {author} {\bibfnamefont {Y.-S.}\ \bibnamefont
  {Wu}},\ }\href {\doibase 10.1103/PhysRevLett.53.111} {\bibfield  {journal}
  {\bibinfo  {journal} {Phys. Rev. Lett.}\ }\textbf {\bibinfo {volume} {53}},\
  \bibinfo {pages} {111} (\bibinfo {year} {1984}{\natexlab{b}})}\BibitemShut
  {NoStop}%
\bibitem [{\citenamefont {Chou}(1991{\natexlab{a}})}]{Chou-91a}%
  \BibitemOpen
  \bibfield  {author} {\bibinfo {author} {\bibfnamefont {C.}~\bibnamefont
  {Chou}},\ }\href {\doibase 10.1103/PhysRevD.44.2533} {\bibfield  {journal}
  {\bibinfo  {journal} {Phys. Rev. D}\ }\textbf {\bibinfo {volume} {44}},\
  \bibinfo {pages} {2533} (\bibinfo {year} {1991}{\natexlab{a}})}\BibitemShut
  {NoStop}%
\bibitem [{\citenamefont {Chou}(1991{\natexlab{b}})}]{Chou-91b}%
  \BibitemOpen
  \bibfield  {author} {\bibinfo {author} {\bibfnamefont {C.}~\bibnamefont
  {Chou}},\ }\href {\doibase 10.1016/0375-9601(91)90477-P} {\bibfield
  {journal} {\bibinfo  {journal} {Phys. Lett. A}\ }\textbf {\bibinfo {volume}
  {155}},\ \bibinfo {pages} {245 } (\bibinfo {year}
  {1991}{\natexlab{b}})}\BibitemShut {NoStop}%
\bibitem [{\citenamefont {Myrheim}(1999)}]{Myrheim-99}%
  \BibitemOpen
  \bibfield  {author} {\bibinfo {author} {\bibfnamefont {J.}~\bibnamefont
  {Myrheim}},\ }in\ \href {\doibase 10.1007/3-540-46637-1_4} {\emph {\bibinfo
  {booktitle} {Topological aspects of low dimensional systems}}},\ \bibinfo
  {series} {Les Houches - Ecole d'Ete de Physique Theorique}, Vol.~\bibinfo
  {volume} {69},\ \bibinfo {editor} {edited by\ \bibinfo {editor}
  {\bibfnamefont {A.}~\bibnamefont {Comtet}}, \bibinfo {editor} {\bibfnamefont
  {T.}~\bibnamefont {Jolic{\oe}ur}}, \bibinfo {editor} {\bibfnamefont
  {S.}~\bibnamefont {Ouvry}}, \ and\ \bibinfo {editor} {\bibfnamefont
  {F.}~\bibnamefont {David}}}\ (\bibinfo  {publisher} {(Springer-Verlag,
  Berlin, Germany)},\ \bibinfo {year} {1999})\ pp.\ \bibinfo {pages}
  {265--413}\BibitemShut {NoStop}%
\bibitem [{\citenamefont {Khare}(2005)}]{Khare-05}%
  \BibitemOpen
  \bibfield  {author} {\bibinfo {author} {\bibfnamefont {A.}~\bibnamefont
  {Khare}},\ }\href@noop {} {\emph {\bibinfo {title} {{Fractional Statistics
  and Quantum Theory}}}},\ \bibinfo {edition} {2nd}\ ed.\ (\bibinfo
  {publisher} {World Scientific, Singapore},\ \bibinfo {year}
  {2005})\BibitemShut {NoStop}%
\bibitem [{\citenamefont {Chitra}\ and\ \citenamefont {Sen}(1992)}]{ChiSen-92}%
  \BibitemOpen
  \bibfield  {author} {\bibinfo {author} {\bibfnamefont {R.}~\bibnamefont
  {Chitra}}\ and\ \bibinfo {author} {\bibfnamefont {D.}~\bibnamefont {Sen}},\
  }\href {\doibase 10.1103/PhysRevB.46.10923} {\bibfield  {journal} {\bibinfo
  {journal} {Phys. Rev. B}\ }\textbf {\bibinfo {volume} {46}},\ \bibinfo
  {pages} {10923} (\bibinfo {year} {1992})}\BibitemShut {NoStop}%
\bibitem [{\citenamefont {Lundholm}\ and\ \citenamefont
  {Solovej}(2013{\natexlab{a}})}]{LunSol-13a}%
  \BibitemOpen
  \bibfield  {author} {\bibinfo {author} {\bibfnamefont {D.}~\bibnamefont
  {Lundholm}}\ and\ \bibinfo {author} {\bibfnamefont {J.~P.}\ \bibnamefont
  {Solovej}},\ }\href {\doibase 10.1007/s00220-013-1748-4} {\bibfield
  {journal} {\bibinfo  {journal} {Comm. Math. Phys.}\ }\textbf {\bibinfo
  {volume} {322}},\ \bibinfo {pages} {883} (\bibinfo {year}
  {2013}{\natexlab{a}})}\BibitemShut {NoStop}%
\bibitem [{\citenamefont {Lundholm}\ and\ \citenamefont
  {Solovej}(2013{\natexlab{b}})}]{LunSol-13b}%
  \BibitemOpen
  \bibfield  {author} {\bibinfo {author} {\bibfnamefont {D.}~\bibnamefont
  {Lundholm}}\ and\ \bibinfo {author} {\bibfnamefont {J.~P.}\ \bibnamefont
  {Solovej}},\ }\href {\doibase 10.1103/PhysRevA.88.062106} {\bibfield
  {journal} {\bibinfo  {journal} {Phys. Rev. A}\ }\textbf {\bibinfo {volume}
  {88}},\ \bibinfo {pages} {062106} (\bibinfo {year}
  {2013}{\natexlab{b}})}\BibitemShut {NoStop}%
\bibitem [{\citenamefont {Lundholm}\ and\ \citenamefont
  {Solovej}(2014)}]{LunSol-14}%
  \BibitemOpen
  \bibfield  {author} {\bibinfo {author} {\bibfnamefont {D.}~\bibnamefont
  {Lundholm}}\ and\ \bibinfo {author} {\bibfnamefont {J.~P.}\ \bibnamefont
  {Solovej}},\ }\href {\doibase 10.1007/s00023-013-0273-5} {\bibfield
  {journal} {\bibinfo  {journal} {Ann. Henri Poincar\'e}\ }\textbf {\bibinfo
  {volume} {15}},\ \bibinfo {pages} {1061} (\bibinfo {year}
  {2014})}\BibitemShut {NoStop}%
\bibitem [{\citenamefont {Lundholm}(2017)}]{Lundholm_2017}%
  \BibitemOpen
  \bibfield  {author} {\bibinfo {author} {\bibfnamefont {D.}~\bibnamefont
  {Lundholm}},\ }\href {\doibase 10.1103/PhysRevA.96.012116} {\bibfield
  {journal} {\bibinfo  {journal} {Phys. Rev. A}\ }\textbf {\bibinfo {volume}
  {96}},\ \bibinfo {pages} {012116} (\bibinfo {year} {2017})}\BibitemShut
  {NoStop}%
\bibitem [{\citenamefont {Lundholm}\ and\ \citenamefont
  {Seiringer}(2018)}]{LunSei-17}%
  \BibitemOpen
  \bibfield  {author} {\bibinfo {author} {\bibfnamefont {D.}~\bibnamefont
  {Lundholm}}\ and\ \bibinfo {author} {\bibfnamefont {R.}~\bibnamefont
  {Seiringer}},\ }\href {\doibase 10.1007/s11005-018-1091-y} {\bibfield
  {journal} {\bibinfo  {journal} {Lett. Math. Phys.}\ }\textbf {\bibinfo
  {volume} {108}},\ \bibinfo {pages} {2523} (\bibinfo {year}
  {2018})}\BibitemShut {NoStop}%
\bibitem [{\citenamefont {Trugenberger}(1992{\natexlab{a}})}]{Trugenberger-92}%
  \BibitemOpen
  \bibfield  {author} {\bibinfo {author} {\bibfnamefont {C.}~\bibnamefont
  {Trugenberger}},\ }\href {\doibase 10.1103/PhysRevD.45.3807} {\bibfield
  {journal} {\bibinfo  {journal} {Phys. Rev. D}\ }\textbf {\bibinfo {volume}
  {45}},\ \bibinfo {pages} {3807} (\bibinfo {year}
  {1992}{\natexlab{a}})}\BibitemShut {NoStop}%
\bibitem [{\citenamefont
  {Trugenberger}(1992{\natexlab{b}})}]{Trugenberger-92b}%
  \BibitemOpen
  \bibfield  {author} {\bibinfo {author} {\bibfnamefont {C.}~\bibnamefont
  {Trugenberger}},\ }\href {\doibase 10.1016/0370-2693(92)91965-C} {\bibfield
  {journal} {\bibinfo  {journal} {Phys. Lett. B}\ }\textbf {\bibinfo {volume}
  {288}},\ \bibinfo {pages} {121} (\bibinfo {year}
  {1992}{\natexlab{b}})}\BibitemShut {NoStop}%
\bibitem [{\citenamefont {Choi}\ \emph {et~al.}(1992)\citenamefont {Choi},
  \citenamefont {Lee},\ and\ \citenamefont {Lee}}]{ChoLeeLee-92}%
  \BibitemOpen
  \bibfield  {author} {\bibinfo {author} {\bibfnamefont {M.~Y.}\ \bibnamefont
  {Choi}}, \bibinfo {author} {\bibfnamefont {C.}~\bibnamefont {Lee}}, \ and\
  \bibinfo {author} {\bibfnamefont {J.}~\bibnamefont {Lee}},\ }\href {\doibase
  10.1103/PhysRevB.46.1489} {\bibfield  {journal} {\bibinfo  {journal} {Phys.
  Rev. B}\ }\textbf {\bibinfo {volume} {46}},\ \bibinfo {pages} {1489}
  (\bibinfo {year} {1992})}\BibitemShut {NoStop}%
\bibitem [{\citenamefont {Larson}\ and\ \citenamefont
  {Lundholm}(2018)}]{LarLun-16}%
  \BibitemOpen
  \bibfield  {author} {\bibinfo {author} {\bibfnamefont {S.}~\bibnamefont
  {Larson}}\ and\ \bibinfo {author} {\bibfnamefont {D.}~\bibnamefont
  {Lundholm}},\ }\href {\doibase 10.1007/s00205-017-1161-9} {\bibfield
  {journal} {\bibinfo  {journal} {Arch. Ration. Mech. Anal.}\ }\textbf
  {\bibinfo {volume} {227}},\ \bibinfo {pages} {309} (\bibinfo {year}
  {2018})}\BibitemShut {NoStop}%
\bibitem [{\citenamefont {Lundholm}\ and\ \citenamefont
  {Rougerie}(2015)}]{LunRou-15}%
  \BibitemOpen
  \bibfield  {author} {\bibinfo {author} {\bibfnamefont {D.}~\bibnamefont
  {Lundholm}}\ and\ \bibinfo {author} {\bibfnamefont {N.}~\bibnamefont
  {Rougerie}},\ }\href {\doibase 10.1007/s10955-015-1382-y} {\bibfield
  {journal} {\bibinfo  {journal} {J. Stat. Phys.}\ }\textbf {\bibinfo {volume}
  {161}},\ \bibinfo {pages} {1236} (\bibinfo {year} {2015})}\BibitemShut
  {NoStop}%
\bibitem [{\citenamefont {Correggi}\ \emph {et~al.}(2017)\citenamefont
  {Correggi}, \citenamefont {Lundholm},\ and\ \citenamefont
  {Rougerie}}]{CorLunRou-16}%
  \BibitemOpen
  \bibfield  {author} {\bibinfo {author} {\bibfnamefont {M.}~\bibnamefont
  {Correggi}}, \bibinfo {author} {\bibfnamefont {D.}~\bibnamefont {Lundholm}},
  \ and\ \bibinfo {author} {\bibfnamefont {N.}~\bibnamefont {Rougerie}},\
  }\href {\doibase 10.2140/apde.2017.10.1169} {\bibfield  {journal} {\bibinfo
  {journal} {Analysis \& PDE}\ }\textbf {\bibinfo {volume} {10}},\ \bibinfo
  {pages} {1169} (\bibinfo {year} {2017})}\BibitemShut {NoStop}%
\bibitem [{\citenamefont {Correggi}\ \emph {et~al.}(2018)\citenamefont
  {Correggi}, \citenamefont {Lundholm},\ and\ \citenamefont
  {Rougerie}}]{CorLunRou-proc-17}%
  \BibitemOpen
  \bibfield  {author} {\bibinfo {author} {\bibfnamefont {M.}~\bibnamefont
  {Correggi}}, \bibinfo {author} {\bibfnamefont {D.}~\bibnamefont {Lundholm}},
  \ and\ \bibinfo {author} {\bibfnamefont {N.}~\bibnamefont {Rougerie}},\ }in\
  \href {\doibase 10.1090/conm/717} {\emph {\bibinfo {booktitle} {Proceedings
  of QMath13, Atlanta, October 8--11, 2016, Mathematical problems in quantum
  physics}}},\ \bibinfo {series} {Contemp. Math.}, Vol.\ \bibinfo {volume}
  {717},\ \bibinfo {editor} {edited by\ \bibinfo {editor} {\bibfnamefont
  {F.}~\bibnamefont {Bonetto}}, \bibinfo {editor} {\bibfnamefont
  {D.}~\bibnamefont {Borthwick}}, \bibinfo {editor} {\bibfnamefont
  {E.}~\bibnamefont {Harrell}}, \ and\ \bibinfo {editor} {\bibfnamefont
  {M.}~\bibnamefont {Loss}}}\ (\bibinfo {year} {2018})\ pp.\ \bibinfo {pages}
  {77--92}\BibitemShut {NoStop}%
\bibitem [{\citenamefont {Girardot}(2019)}]{Girardot-19}%
  \BibitemOpen
  \bibfield  {author} {\bibinfo {author} {\bibfnamefont {T.}~\bibnamefont
  {Girardot}},\ }\href@noop {} {\enquote {\bibinfo {title} {Average field
  approximation for almost bosonic anyons in a magnetic field},}\ } (\bibinfo
  {year} {2019}),\ \Eprint {http://arxiv.org/abs/1910.09310} {arXiv:1910.09310
  [math.AP]} \BibitemShut {NoStop}%
\bibitem [{\citenamefont {Grundberg}\ \emph {et~al.}(1991)\citenamefont
  {Grundberg}, \citenamefont {Hansson}, \citenamefont {Karlhede},\ and\
  \citenamefont {Leinaas}}]{Grundberg-etal-91}%
  \BibitemOpen
  \bibfield  {author} {\bibinfo {author} {\bibfnamefont {J.}~\bibnamefont
  {Grundberg}}, \bibinfo {author} {\bibfnamefont {T.}~\bibnamefont {Hansson}},
  \bibinfo {author} {\bibfnamefont {A.}~\bibnamefont {Karlhede}}, \ and\
  \bibinfo {author} {\bibfnamefont {J.}~\bibnamefont {Leinaas}},\ }\href
  {\doibase 10.1142/S0217984991000642} {\bibfield  {journal} {\bibinfo
  {journal} {Mod. Phys. Lett. B}\ }\textbf {\bibinfo {volume} {05}},\ \bibinfo
  {pages} {539} (\bibinfo {year} {1991})}\BibitemShut {NoStop}%
\bibitem [{\citenamefont {Manuel}\ and\ \citenamefont
  {Tarrach}(1991)}]{ManTar-91}%
  \BibitemOpen
  \bibfield  {author} {\bibinfo {author} {\bibfnamefont {C.}~\bibnamefont
  {Manuel}}\ and\ \bibinfo {author} {\bibfnamefont {R.}~\bibnamefont
  {Tarrach}},\ }\href {\doibase 10.1016/0370-2693(91)90807-3} {\bibfield
  {journal} {\bibinfo  {journal} {Phys. Lett. B}\ }\textbf {\bibinfo {volume}
  {268}},\ \bibinfo {pages} {222} (\bibinfo {year} {1991})}\BibitemShut
  {NoStop}%
\bibitem [{\citenamefont {Bourdeau}\ and\ \citenamefont
  {Sorkin}(1992)}]{BorSor-92}%
  \BibitemOpen
  \bibfield  {author} {\bibinfo {author} {\bibfnamefont {M.}~\bibnamefont
  {Bourdeau}}\ and\ \bibinfo {author} {\bibfnamefont {R.~D.}\ \bibnamefont
  {Sorkin}},\ }\href {\doibase 10.1103/PhysRevD.45.687} {\bibfield  {journal}
  {\bibinfo  {journal} {Phys. Rev. D}\ }\textbf {\bibinfo {volume} {45}},\
  \bibinfo {pages} {687} (\bibinfo {year} {1992})}\BibitemShut {NoStop}%
\bibitem [{\citenamefont {Murthy}\ \emph {et~al.}(1992)\citenamefont {Murthy},
  \citenamefont {Law}, \citenamefont {Bhaduri},\ and\ \citenamefont
  {Date}}]{MurLawBhaDat_92}%
  \BibitemOpen
  \bibfield  {author} {\bibinfo {author} {\bibfnamefont {M.~V.~N.}\
  \bibnamefont {Murthy}}, \bibinfo {author} {\bibfnamefont {J.}~\bibnamefont
  {Law}}, \bibinfo {author} {\bibfnamefont {R.~K.}\ \bibnamefont {Bhaduri}}, \
  and\ \bibinfo {author} {\bibfnamefont {G.}~\bibnamefont {Date}},\ }\href
  {\doibase 10.1088/0305-4470/25/23/013} {\bibfield  {journal} {\bibinfo
  {journal} {J. Phys. A: Math. Gen.}\ }\textbf {\bibinfo {volume} {25}},\
  \bibinfo {pages} {6163} (\bibinfo {year} {1992})}\BibitemShut {NoStop}%
\bibitem [{\citenamefont {Correggi}\ and\ \citenamefont
  {Oddis}(2018)}]{CorOdd-18}%
  \BibitemOpen
  \bibfield  {author} {\bibinfo {author} {\bibfnamefont {M.}~\bibnamefont
  {Correggi}}\ and\ \bibinfo {author} {\bibfnamefont {L.}~\bibnamefont
  {Oddis}},\ }\href
  {http://www1.mat.uniroma1.it/ricerca/rendiconti/39_2_(2018)_277-292.html}
  {\bibfield  {journal} {\bibinfo  {journal} {Rend. Mat. Appl.}\ }\textbf
  {\bibinfo {volume} {39}},\ \bibinfo {pages} {277} (\bibinfo {year}
  {2018})}\BibitemShut {NoStop}%
\bibitem [{\citenamefont {Ouvry}(2007)}]{Ouvry-07}%
  \BibitemOpen
  \bibfield  {author} {\bibinfo {author} {\bibfnamefont {S.}~\bibnamefont
  {Ouvry}},\ }\href {\doibase 10.1007/978-3-7643-8799-0_3} {\bibfield
  {journal} {\bibinfo  {journal} {S\'eminaire Poincar\'e}\ }\textbf {\bibinfo
  {volume} {11}},\ \bibinfo {pages} {77} (\bibinfo {year} {2007})}\BibitemShut
  {NoStop}%
\bibitem [{\citenamefont {McCabe}\ and\ \citenamefont
  {Ouvry}(1991)}]{mccabe1991perturbative}%
  \BibitemOpen
  \bibfield  {author} {\bibinfo {author} {\bibfnamefont {J.}~\bibnamefont
  {McCabe}}\ and\ \bibinfo {author} {\bibfnamefont {S.~p.}\ \bibnamefont
  {Ouvry}},\ }\href@noop {} {\bibfield  {journal} {\bibinfo  {journal} {Physics
  Letters B}\ }\textbf {\bibinfo {volume} {260}},\ \bibinfo {pages} {113}
  (\bibinfo {year} {1991})}\BibitemShut {NoStop}%
\bibitem [{\citenamefont {Amelino-Camelia}(1994)}]{amelino1994perturbative}%
  \BibitemOpen
  \bibfield  {author} {\bibinfo {author} {\bibfnamefont {G.}~\bibnamefont
  {Amelino-Camelia}},\ }\href@noop {} {\bibfield  {journal} {\bibinfo
  {journal} {Physics Letters B}\ }\textbf {\bibinfo {volume} {326}},\ \bibinfo
  {pages} {282} (\bibinfo {year} {1994})}\BibitemShut {NoStop}%
\bibitem [{\citenamefont {Comtet}\ \emph {et~al.}(1995)\citenamefont {Comtet},
  \citenamefont {Mashkevich},\ and\ \citenamefont
  {Ouvry}}]{comtet1995magnetic}%
  \BibitemOpen
  \bibfield  {author} {\bibinfo {author} {\bibfnamefont {A.}~\bibnamefont
  {Comtet}}, \bibinfo {author} {\bibfnamefont {S.}~\bibnamefont {Mashkevich}},
  \ and\ \bibinfo {author} {\bibfnamefont {S.}~\bibnamefont {Ouvry}},\
  }\href@noop {} {\bibfield  {journal} {\bibinfo  {journal} {Physical Review
  D}\ }\textbf {\bibinfo {volume} {52}},\ \bibinfo {pages} {2594} (\bibinfo
  {year} {1995})}\BibitemShut {NoStop}%
\bibitem [{\citenamefont {Mashkevich}(1996)}]{Mashkevich_1996}%
  \BibitemOpen
  \bibfield  {author} {\bibinfo {author} {\bibfnamefont {S.}~\bibnamefont
  {Mashkevich}},\ }\href {\doibase 10.1103/PhysRevD.54.6537} {\bibfield
  {journal} {\bibinfo  {journal} {Phys. Rev. D}\ }\textbf {\bibinfo {volume}
  {54}},\ \bibinfo {pages} {6537} (\bibinfo {year} {1996})}\BibitemShut
  {NoStop}%
\bibitem [{\citenamefont {Bogolyubov}(1947)}]{bogolyubov1947theory}%
  \BibitemOpen
  \bibfield  {author} {\bibinfo {author} {\bibfnamefont {N.~N.}\ \bibnamefont
  {Bogolyubov}},\ }\href@noop {} {\bibfield  {journal} {\bibinfo  {journal}
  {Izv. Akad. Nauk Ser. Fiz.}\ }\textbf {\bibinfo {volume} {11}},\ \bibinfo
  {pages} {23} (\bibinfo {year} {1947})}\BibitemShut {NoStop}%
\bibitem [{\citenamefont {Pitaevskii}\ and\ \citenamefont
  {Stringari}(2016)}]{Pitaevskii2016}%
  \BibitemOpen
  \bibfield  {author} {\bibinfo {author} {\bibfnamefont {L.~P.}\ \bibnamefont
  {Pitaevskii}}\ and\ \bibinfo {author} {\bibfnamefont {S.}~\bibnamefont
  {Stringari}},\ }\href@noop {} {\emph {\bibinfo {title} {Bose-Einstein
  Condensation and Superfluidity}}}\ (\bibinfo  {publisher} {Oxford University
  Press},\ \bibinfo {year} {2016})\BibitemShut {NoStop}%
\bibitem [{\citenamefont {Jain}(2007)}]{jain2007composite}%
  \BibitemOpen
  \bibfield  {author} {\bibinfo {author} {\bibfnamefont {J.~K.}\ \bibnamefont
  {Jain}},\ }\href@noop {} {\emph {\bibinfo {title} {Composite fermions}}}\
  (\bibinfo  {publisher} {Cambridge University Press},\ \bibinfo {year}
  {2007})\BibitemShut {NoStop}%
\bibitem [{\citenamefont {Kilpatrick}\ and\ \citenamefont
  {Larsen}(1987)}]{kilpatrick1987set}%
  \BibitemOpen
  \bibfield  {author} {\bibinfo {author} {\bibfnamefont {J.}~\bibnamefont
  {Kilpatrick}}\ and\ \bibinfo {author} {\bibfnamefont {S.}~\bibnamefont
  {Larsen}},\ }\href@noop {} {\bibfield  {journal} {\bibinfo  {journal}
  {Few-body systems}\ }\textbf {\bibinfo {volume} {3}},\ \bibinfo {pages} {75}
  (\bibinfo {year} {1987})}\BibitemShut {NoStop}%
\bibitem [{\citenamefont {Khare}\ and\ \citenamefont
  {McCabe}(1991)}]{khare1991perturbative}%
  \BibitemOpen
  \bibfield  {author} {\bibinfo {author} {\bibfnamefont {A.}~\bibnamefont
  {Khare}}\ and\ \bibinfo {author} {\bibfnamefont {J.}~\bibnamefont {McCabe}},\
  }\href@noop {} {\bibfield  {journal} {\bibinfo  {journal} {Physics Letters
  B}\ }\textbf {\bibinfo {volume} {269}},\ \bibinfo {pages} {330} (\bibinfo
  {year} {1991})}\BibitemShut {NoStop}%
\bibitem [{Note2()}]{Note2}%
  \BibitemOpen
  \bibinfo {note} {In order to uniquely label the bundles and identify them as
  composite bosons one needs to select some sub-bundle where the exact winding
  comes out algebraically (and not just via the phase which is periodic over
  the even integers), such as some subspace of holomorphic functions or some
  finite range of allowed angular momenta. Note in particular that the space
  $\protect \mathcal {H}\protect \tilde {F}^n$ contains also functions on the
  form $|\protect \tilde {F}|^{2n}$ for which the winding is zero, however by
  further restricting $\protect \mathcal {H}$ it is possible to ensure zero
  overlap between such function spaces and thus the one-to-one labeling by $n$
  via the holomorphic factor $\protect \tilde {F}^n$. This happens for example
  when one switches on a strong external magnetic field, and indeed the bosonic
  Laughlin factor $\protect \tilde {F}$ represents the smallest symmetric
  attachment of an integer number of vortices to every particle.}\BibitemShut
  {Stop}%
\bibitem [{\citenamefont {Lee}\ \emph {et~al.}(1953)\citenamefont {Lee},
  \citenamefont {Low},\ and\ \citenamefont {Pines}}]{LLP_53}%
  \BibitemOpen
  \bibfield  {author} {\bibinfo {author} {\bibfnamefont {T.~D.}\ \bibnamefont
  {Lee}}, \bibinfo {author} {\bibfnamefont {F.~E.}\ \bibnamefont {Low}}, \ and\
  \bibinfo {author} {\bibfnamefont {D.}~\bibnamefont {Pines}},\ }\href
  {\doibase 10.1103/PhysRev.90.297} {\bibfield  {journal} {\bibinfo  {journal}
  {Phys. Rev.}\ }\textbf {\bibinfo {volume} {90}},\ \bibinfo {pages} {297}
  (\bibinfo {year} {1953})}\BibitemShut {NoStop}%
\bibitem [{\citenamefont {Sak}(1972)}]{Sak_72}%
  \BibitemOpen
  \bibfield  {author} {\bibinfo {author} {\bibfnamefont {J.}~\bibnamefont
  {Sak}},\ }\href {\doibase 10.1103/PhysRevB.6.3981} {\bibfield  {journal}
  {\bibinfo  {journal} {Phys. Rev. B}\ }\textbf {\bibinfo {volume} {6}},\
  \bibinfo {pages} {3981} (\bibinfo {year} {1972})}\BibitemShut {NoStop}%
\bibitem [{\citenamefont {Wu}\ \emph {et~al.}(1986)\citenamefont {Wu},
  \citenamefont {Peeters},\ and\ \citenamefont {Devreese}}]{Wu_86}%
  \BibitemOpen
  \bibfield  {author} {\bibinfo {author} {\bibfnamefont {X.}~\bibnamefont
  {Wu}}, \bibinfo {author} {\bibfnamefont {F.~M.}\ \bibnamefont {Peeters}}, \
  and\ \bibinfo {author} {\bibfnamefont {J.~T.}\ \bibnamefont {Devreese}},\
  }\href {\doibase 10.1103/PhysRevB.34.2621} {\bibfield  {journal} {\bibinfo
  {journal} {Phys. Rev. B}\ }\textbf {\bibinfo {volume} {34}},\ \bibinfo
  {pages} {2621} (\bibinfo {year} {1986})}\BibitemShut {NoStop}%
\bibitem [{\citenamefont {Peeters}\ and\ \citenamefont
  {Devreese}(1987)}]{Peeters_87}%
  \BibitemOpen
  \bibfield  {author} {\bibinfo {author} {\bibfnamefont {F.~M.}\ \bibnamefont
  {Peeters}}\ and\ \bibinfo {author} {\bibfnamefont {J.~T.}\ \bibnamefont
  {Devreese}},\ }\href {\doibase 10.1103/PhysRevB.36.4442} {\bibfield
  {journal} {\bibinfo  {journal} {Phys. Rev. B}\ }\textbf {\bibinfo {volume}
  {36}},\ \bibinfo {pages} {4442} (\bibinfo {year} {1987})}\BibitemShut
  {NoStop}%
\bibitem [{\citenamefont {Devreese}\ and\ \citenamefont
  {Peeters}(2012)}]{devreese2012physics}%
  \BibitemOpen
  \bibfield  {author} {\bibinfo {author} {\bibfnamefont {J.~T.}\ \bibnamefont
  {Devreese}}\ and\ \bibinfo {author} {\bibfnamefont {F.~M.}\ \bibnamefont
  {Peeters}},\ }\href@noop {} {\emph {\bibinfo {title} {The physics of the
  two-dimensional electron gas}}},\ Vol.\ \bibinfo {volume} {157}\ (\bibinfo
  {publisher} {Springer Science \& Business Media},\ \bibinfo {year}
  {2012})\BibitemShut {NoStop}%
\bibitem [{\citenamefont {Kotov}\ \emph {et~al.}(2009)\citenamefont {Kotov},
  \citenamefont {Uchoa},\ and\ \citenamefont {Castro~Neto}}]{Kotov_09}%
  \BibitemOpen
  \bibfield  {author} {\bibinfo {author} {\bibfnamefont {V.~N.}\ \bibnamefont
  {Kotov}}, \bibinfo {author} {\bibfnamefont {B.}~\bibnamefont {Uchoa}}, \ and\
  \bibinfo {author} {\bibfnamefont {A.~H.}\ \bibnamefont {Castro~Neto}},\
  }\href {\doibase 10.1103/PhysRevB.80.165424} {\bibfield  {journal} {\bibinfo
  {journal} {Phys. Rev. B}\ }\textbf {\bibinfo {volume} {80}},\ \bibinfo
  {pages} {165424} (\bibinfo {year} {2009})}\BibitemShut {NoStop}%
\bibitem [{\citenamefont {Van~Tuan}\ \emph {et~al.}(2018)\citenamefont
  {Van~Tuan}, \citenamefont {Yang},\ and\ \citenamefont {Dery}}]{Van_Tuan_18}%
  \BibitemOpen
  \bibfield  {author} {\bibinfo {author} {\bibfnamefont {D.}~\bibnamefont
  {Van~Tuan}}, \bibinfo {author} {\bibfnamefont {M.}~\bibnamefont {Yang}}, \
  and\ \bibinfo {author} {\bibfnamefont {H.}~\bibnamefont {Dery}},\ }\href
  {\doibase 10.1103/PhysRevB.98.125308} {\bibfield  {journal} {\bibinfo
  {journal} {Phys. Rev. B}\ }\textbf {\bibinfo {volume} {98}},\ \bibinfo
  {pages} {125308} (\bibinfo {year} {2018})}\BibitemShut {NoStop}%
\bibitem [{\citenamefont {Holten}\ \emph {et~al.}(2018)\citenamefont {Holten},
  \citenamefont {Bayha}, \citenamefont {Klein}, \citenamefont {Murthy},
  \citenamefont {Preiss},\ and\ \citenamefont {Jochim}}]{Holten_18}%
  \BibitemOpen
  \bibfield  {author} {\bibinfo {author} {\bibfnamefont {M.}~\bibnamefont
  {Holten}}, \bibinfo {author} {\bibfnamefont {L.}~\bibnamefont {Bayha}},
  \bibinfo {author} {\bibfnamefont {A.~C.}\ \bibnamefont {Klein}}, \bibinfo
  {author} {\bibfnamefont {P.~A.}\ \bibnamefont {Murthy}}, \bibinfo {author}
  {\bibfnamefont {P.~M.}\ \bibnamefont {Preiss}}, \ and\ \bibinfo {author}
  {\bibfnamefont {S.}~\bibnamefont {Jochim}},\ }\href {\doibase
  10.1103/PhysRevLett.121.120401} {\bibfield  {journal} {\bibinfo  {journal}
  {Phys. Rev. Lett.}\ }\textbf {\bibinfo {volume} {121}},\ \bibinfo {pages}
  {120401} (\bibinfo {year} {2018})}\BibitemShut {NoStop}%
\bibitem [{\citenamefont {Goldhaber}(1982)}]{Goldhaber_82}%
  \BibitemOpen
  \bibfield  {author} {\bibinfo {author} {\bibfnamefont {A.~S.}\ \bibnamefont
  {Goldhaber}},\ }\href {\doibase 10.1103/PhysRevLett.49.905} {\bibfield
  {journal} {\bibinfo  {journal} {Phys. Rev. Lett.}\ }\textbf {\bibinfo
  {volume} {49}},\ \bibinfo {pages} {905} (\bibinfo {year} {1982})}\BibitemShut
  {NoStop}%
\bibitem [{\citenamefont {Jackiw}\ and\ \citenamefont
  {Redlich}(1983)}]{Jackiw_83}%
  \BibitemOpen
  \bibfield  {author} {\bibinfo {author} {\bibfnamefont {R.}~\bibnamefont
  {Jackiw}}\ and\ \bibinfo {author} {\bibfnamefont {A.~N.}\ \bibnamefont
  {Redlich}},\ }\href {\doibase 10.1103/PhysRevLett.50.555} {\bibfield
  {journal} {\bibinfo  {journal} {Phys. Rev. Lett.}\ }\textbf {\bibinfo
  {volume} {50}},\ \bibinfo {pages} {555} (\bibinfo {year} {1983})}\BibitemShut
  {NoStop}%
\bibitem [{\citenamefont {Hamada}\ \emph {et~al.}(2018)\citenamefont {Hamada},
  \citenamefont {Minamitani}, \citenamefont {Hirayama},\ and\ \citenamefont
  {Murakami}}]{Hamada_18}%
  \BibitemOpen
  \bibfield  {author} {\bibinfo {author} {\bibfnamefont {M.}~\bibnamefont
  {Hamada}}, \bibinfo {author} {\bibfnamefont {E.}~\bibnamefont {Minamitani}},
  \bibinfo {author} {\bibfnamefont {M.}~\bibnamefont {Hirayama}}, \ and\
  \bibinfo {author} {\bibfnamefont {S.}~\bibnamefont {Murakami}},\ }\href
  {\doibase 10.1103/PhysRevLett.121.175301} {\bibfield  {journal} {\bibinfo
  {journal} {Phys. Rev. Lett.}\ }\textbf {\bibinfo {volume} {121}},\ \bibinfo
  {pages} {175301} (\bibinfo {year} {2018})}\BibitemShut {NoStop}%
\bibitem [{\citenamefont {Holanda}\ \emph {et~al.}(2018)\citenamefont
  {Holanda}, \citenamefont {Maior}, \citenamefont {Azevedo},\ and\
  \citenamefont {Rezende}}]{holanda2018detecting}%
  \BibitemOpen
  \bibfield  {author} {\bibinfo {author} {\bibfnamefont {J.}~\bibnamefont
  {Holanda}}, \bibinfo {author} {\bibfnamefont {D.}~\bibnamefont {Maior}},
  \bibinfo {author} {\bibfnamefont {A.}~\bibnamefont {Azevedo}}, \ and\
  \bibinfo {author} {\bibfnamefont {S.}~\bibnamefont {Rezende}},\ }\href@noop
  {} {\bibfield  {journal} {\bibinfo  {journal} {Nature Physics}\ }\textbf
  {\bibinfo {volume} {14}},\ \bibinfo {pages} {500} (\bibinfo {year}
  {2018})}\BibitemShut {NoStop}%
\bibitem [{\citenamefont {Eisenstein}\ \emph {et~al.}(1985)\citenamefont
  {Eisenstein}, \citenamefont {Stormer}, \citenamefont {Narayanamurti},
  \citenamefont {Cho}, \citenamefont {Gossard},\ and\ \citenamefont
  {Tu}}]{Eisenstein_85}%
  \BibitemOpen
  \bibfield  {author} {\bibinfo {author} {\bibfnamefont {J.~P.}\ \bibnamefont
  {Eisenstein}}, \bibinfo {author} {\bibfnamefont {H.~L.}\ \bibnamefont
  {Stormer}}, \bibinfo {author} {\bibfnamefont {V.}~\bibnamefont
  {Narayanamurti}}, \bibinfo {author} {\bibfnamefont {A.~Y.}\ \bibnamefont
  {Cho}}, \bibinfo {author} {\bibfnamefont {A.~C.}\ \bibnamefont {Gossard}}, \
  and\ \bibinfo {author} {\bibfnamefont {C.~W.}\ \bibnamefont {Tu}},\ }\href
  {\doibase 10.1103/PhysRevLett.55.875} {\bibfield  {journal} {\bibinfo
  {journal} {Phys. Rev. Lett.}\ }\textbf {\bibinfo {volume} {55}},\ \bibinfo
  {pages} {875} (\bibinfo {year} {1985})}\BibitemShut {NoStop}%
\bibitem [{\citenamefont {Li}\ \emph {et~al.}(2008)\citenamefont {Li},
  \citenamefont {Checkelsky}, \citenamefont {Hor}, \citenamefont {Uher},
  \citenamefont {Hebard}, \citenamefont {Cava},\ and\ \citenamefont
  {Ong}}]{li2008phase}%
  \BibitemOpen
  \bibfield  {author} {\bibinfo {author} {\bibfnamefont {L.}~\bibnamefont
  {Li}}, \bibinfo {author} {\bibfnamefont {J.~G.}\ \bibnamefont {Checkelsky}},
  \bibinfo {author} {\bibfnamefont {Y.~S.}\ \bibnamefont {Hor}}, \bibinfo
  {author} {\bibfnamefont {C.}~\bibnamefont {Uher}}, \bibinfo {author}
  {\bibfnamefont {A.~F.}\ \bibnamefont {Hebard}}, \bibinfo {author}
  {\bibfnamefont {R.~J.}\ \bibnamefont {Cava}}, \ and\ \bibinfo {author}
  {\bibfnamefont {N.~P.}\ \bibnamefont {Ong}},\ }\href@noop {} {\bibfield
  {journal} {\bibinfo  {journal} {Science}\ }\textbf {\bibinfo {volume}
  {321}},\ \bibinfo {pages} {547} (\bibinfo {year} {2008})}\BibitemShut
  {NoStop}%
\bibitem [{\citenamefont {Sebastian}\ \emph {et~al.}(2008)\citenamefont
  {Sebastian}, \citenamefont {Harrison}, \citenamefont {Palm}, \citenamefont
  {Murphy}, \citenamefont {Mielke}, \citenamefont {Liang}, \citenamefont
  {Bonn}, \citenamefont {Hardy},\ and\ \citenamefont
  {Lonzarich}}]{sebastian2008multi}%
  \BibitemOpen
  \bibfield  {author} {\bibinfo {author} {\bibfnamefont {S.~E.}\ \bibnamefont
  {Sebastian}}, \bibinfo {author} {\bibfnamefont {N.}~\bibnamefont {Harrison}},
  \bibinfo {author} {\bibfnamefont {E.}~\bibnamefont {Palm}}, \bibinfo {author}
  {\bibfnamefont {T.}~\bibnamefont {Murphy}}, \bibinfo {author} {\bibfnamefont
  {C.}~\bibnamefont {Mielke}}, \bibinfo {author} {\bibfnamefont
  {R.}~\bibnamefont {Liang}}, \bibinfo {author} {\bibfnamefont
  {D.}~\bibnamefont {Bonn}}, \bibinfo {author} {\bibfnamefont {W.}~\bibnamefont
  {Hardy}}, \ and\ \bibinfo {author} {\bibfnamefont {G.}~\bibnamefont
  {Lonzarich}},\ }\href@noop {} {\bibfield  {journal} {\bibinfo  {journal}
  {Nature}\ }\textbf {\bibinfo {volume} {454}},\ \bibinfo {pages} {200}
  (\bibinfo {year} {2008})}\BibitemShut {NoStop}%
\bibitem [{\citenamefont {Li}\ \emph {et~al.}(2014)\citenamefont {Li},
  \citenamefont {Xiang}, \citenamefont {Yu}, \citenamefont {Asaba},
  \citenamefont {Lawson}, \citenamefont {Cai}, \citenamefont {Tinsman},
  \citenamefont {Berkley}, \citenamefont {Wolgast}, \citenamefont {Eo} \emph
  {et~al.}}]{li2014two}%
  \BibitemOpen
  \bibfield  {author} {\bibinfo {author} {\bibfnamefont {G.}~\bibnamefont
  {Li}}, \bibinfo {author} {\bibfnamefont {Z.}~\bibnamefont {Xiang}}, \bibinfo
  {author} {\bibfnamefont {F.}~\bibnamefont {Yu}}, \bibinfo {author}
  {\bibfnamefont {T.}~\bibnamefont {Asaba}}, \bibinfo {author} {\bibfnamefont
  {B.}~\bibnamefont {Lawson}}, \bibinfo {author} {\bibfnamefont
  {P.}~\bibnamefont {Cai}}, \bibinfo {author} {\bibfnamefont {C.}~\bibnamefont
  {Tinsman}}, \bibinfo {author} {\bibfnamefont {A.}~\bibnamefont {Berkley}},
  \bibinfo {author} {\bibfnamefont {S.}~\bibnamefont {Wolgast}}, \bibinfo
  {author} {\bibfnamefont {Y.~S.}\ \bibnamefont {Eo}},  \emph {et~al.},\
  }\href@noop {} {\bibfield  {journal} {\bibinfo  {journal} {Science}\ }\textbf
  {\bibinfo {volume} {346}},\ \bibinfo {pages} {1208} (\bibinfo {year}
  {2014})}\BibitemShut {NoStop}%
\bibitem [{\citenamefont {Tan}\ \emph {et~al.}(2015)\citenamefont {Tan},
  \citenamefont {Hsu}, \citenamefont {Zeng}, \citenamefont {Hatnean},
  \citenamefont {Harrison}, \citenamefont {Zhu}, \citenamefont {Hartstein},
  \citenamefont {Kiourlappou}, \citenamefont {Srivastava}, \citenamefont
  {Johannes} \emph {et~al.}}]{tan2015unconventional}%
  \BibitemOpen
  \bibfield  {author} {\bibinfo {author} {\bibfnamefont {B.}~\bibnamefont
  {Tan}}, \bibinfo {author} {\bibfnamefont {Y.-T.}\ \bibnamefont {Hsu}},
  \bibinfo {author} {\bibfnamefont {B.}~\bibnamefont {Zeng}}, \bibinfo {author}
  {\bibfnamefont {M.~C.}\ \bibnamefont {Hatnean}}, \bibinfo {author}
  {\bibfnamefont {N.}~\bibnamefont {Harrison}}, \bibinfo {author}
  {\bibfnamefont {Z.}~\bibnamefont {Zhu}}, \bibinfo {author} {\bibfnamefont
  {M.}~\bibnamefont {Hartstein}}, \bibinfo {author} {\bibfnamefont
  {M.}~\bibnamefont {Kiourlappou}}, \bibinfo {author} {\bibfnamefont
  {A.}~\bibnamefont {Srivastava}}, \bibinfo {author} {\bibfnamefont
  {M.}~\bibnamefont {Johannes}},  \emph {et~al.},\ }\href@noop {} {\bibfield
  {journal} {\bibinfo  {journal} {Science}\ }\textbf {\bibinfo {volume}
  {349}},\ \bibinfo {pages} {287} (\bibinfo {year} {2015})}\BibitemShut
  {NoStop}%
\bibitem [{\citenamefont {Greiner}\ \emph {et~al.}(2001)\citenamefont
  {Greiner}, \citenamefont {Bloch}, \citenamefont {Mandel}, \citenamefont
  {H\"ansch},\ and\ \citenamefont {Esslinger}}]{Greiner_01}%
  \BibitemOpen
  \bibfield  {author} {\bibinfo {author} {\bibfnamefont {M.}~\bibnamefont
  {Greiner}}, \bibinfo {author} {\bibfnamefont {I.}~\bibnamefont {Bloch}},
  \bibinfo {author} {\bibfnamefont {O.}~\bibnamefont {Mandel}}, \bibinfo
  {author} {\bibfnamefont {T.~W.}\ \bibnamefont {H\"ansch}}, \ and\ \bibinfo
  {author} {\bibfnamefont {T.}~\bibnamefont {Esslinger}},\ }\href {\doibase
  10.1103/PhysRevLett.87.160405} {\bibfield  {journal} {\bibinfo  {journal}
  {Phys. Rev. Lett.}\ }\textbf {\bibinfo {volume} {87}},\ \bibinfo {pages}
  {160405} (\bibinfo {year} {2001})}\BibitemShut {NoStop}%
\bibitem [{\citenamefont {Grusdt}\ and\ \citenamefont
  {Fleischhauer}(2016)}]{Grusdt_2016}%
  \BibitemOpen
  \bibfield  {author} {\bibinfo {author} {\bibfnamefont {F.}~\bibnamefont
  {Grusdt}}\ and\ \bibinfo {author} {\bibfnamefont {M.}~\bibnamefont
  {Fleischhauer}},\ }\href {\doibase 10.1103/PhysRevLett.116.053602} {\bibfield
   {journal} {\bibinfo  {journal} {Phys. Rev. Lett.}\ }\textbf {\bibinfo
  {volume} {116}},\ \bibinfo {pages} {053602} (\bibinfo {year}
  {2016})}\BibitemShut {NoStop}%
\bibitem [{\citenamefont {Grusdt}\ \emph {et~al.}(2017)\citenamefont {Grusdt},
  \citenamefont {Astrakharchik},\ and\ \citenamefont
  {Demler}}]{grusdt2017bose}%
  \BibitemOpen
  \bibfield  {author} {\bibinfo {author} {\bibfnamefont {F.}~\bibnamefont
  {Grusdt}}, \bibinfo {author} {\bibfnamefont {G.~E.}\ \bibnamefont
  {Astrakharchik}}, \ and\ \bibinfo {author} {\bibfnamefont {E.}~\bibnamefont
  {Demler}},\ }\href@noop {} {\bibfield  {journal} {\bibinfo  {journal} {New
  Journal of Physics}\ }\textbf {\bibinfo {volume} {19}},\ \bibinfo {pages}
  {103035} (\bibinfo {year} {2017})}\BibitemShut {NoStop}%
\bibitem [{\citenamefont {Ardila}\ \emph {et~al.}(2019)\citenamefont {Ardila},
  \citenamefont {Astrakharchik},\ and\ \citenamefont
  {Giorgini}}]{ardila2019strong}%
  \BibitemOpen
  \bibfield  {author} {\bibinfo {author} {\bibfnamefont {L.}~\bibnamefont
  {Ardila}}, \bibinfo {author} {\bibfnamefont {G.}~\bibnamefont
  {Astrakharchik}}, \ and\ \bibinfo {author} {\bibfnamefont {S.}~\bibnamefont
  {Giorgini}},\ }\href@noop {} {\bibfield  {journal} {\bibinfo  {journal}
  {arXiv preprint arXiv:1907.01533}\ } (\bibinfo {year} {2019})}\BibitemShut
  {NoStop}%
\bibitem [{\citenamefont {Jackson}\ and\ \citenamefont
  {Platzman}(1981)}]{Jackson_81}%
  \BibitemOpen
  \bibfield  {author} {\bibinfo {author} {\bibfnamefont {S.~A.}\ \bibnamefont
  {Jackson}}\ and\ \bibinfo {author} {\bibfnamefont {P.~M.}\ \bibnamefont
  {Platzman}},\ }\href {\doibase 10.1103/PhysRevB.24.499} {\bibfield  {journal}
  {\bibinfo  {journal} {Phys. Rev. B}\ }\textbf {\bibinfo {volume} {24}},\
  \bibinfo {pages} {499} (\bibinfo {year} {1981})}\BibitemShut {NoStop}%
\bibitem [{\citenamefont {Ville}\ \emph {et~al.}(2018)\citenamefont {Ville},
  \citenamefont {Saint-Jalm}, \citenamefont {Le~Cerf}, \citenamefont
  {Aidelsburger}, \citenamefont {Nascimb\`ene}, \citenamefont {Dalibard},\ and\
  \citenamefont {Beugnon}}]{Ville_2018}%
  \BibitemOpen
  \bibfield  {author} {\bibinfo {author} {\bibfnamefont {J.~L.}\ \bibnamefont
  {Ville}}, \bibinfo {author} {\bibfnamefont {R.}~\bibnamefont {Saint-Jalm}},
  \bibinfo {author} {\bibfnamefont {E.}~\bibnamefont {Le~Cerf}}, \bibinfo
  {author} {\bibfnamefont {M.}~\bibnamefont {Aidelsburger}}, \bibinfo {author}
  {\bibfnamefont {S.}~\bibnamefont {Nascimb\`ene}}, \bibinfo {author}
  {\bibfnamefont {J.}~\bibnamefont {Dalibard}}, \ and\ \bibinfo {author}
  {\bibfnamefont {J.}~\bibnamefont {Beugnon}},\ }\href {\doibase
  10.1103/PhysRevLett.121.145301} {\bibfield  {journal} {\bibinfo  {journal}
  {Phys. Rev. Lett.}\ }\textbf {\bibinfo {volume} {121}},\ \bibinfo {pages}
  {145301} (\bibinfo {year} {2018})}\BibitemShut {NoStop}%
\bibitem [{\citenamefont {Engels}\ \emph {et~al.}(2003)\citenamefont {Engels},
  \citenamefont {Coddington}, \citenamefont {Haljan}, \citenamefont
  {Schweikhard},\ and\ \citenamefont {Cornell}}]{Engels_2003}%
  \BibitemOpen
  \bibfield  {author} {\bibinfo {author} {\bibfnamefont {P.}~\bibnamefont
  {Engels}}, \bibinfo {author} {\bibfnamefont {I.}~\bibnamefont {Coddington}},
  \bibinfo {author} {\bibfnamefont {P.~C.}\ \bibnamefont {Haljan}}, \bibinfo
  {author} {\bibfnamefont {V.}~\bibnamefont {Schweikhard}}, \ and\ \bibinfo
  {author} {\bibfnamefont {E.~A.}\ \bibnamefont {Cornell}},\ }\href {\doibase
  10.1103/PhysRevLett.90.170405} {\bibfield  {journal} {\bibinfo  {journal}
  {Phys. Rev. Lett.}\ }\textbf {\bibinfo {volume} {90}},\ \bibinfo {pages}
  {170405} (\bibinfo {year} {2003})}\BibitemShut {NoStop}%
\bibitem [{\citenamefont {Hu}\ \emph {et~al.}(2016)\citenamefont {Hu},
  \citenamefont {Van~de Graaff}, \citenamefont {Kedar}, \citenamefont {Corson},
  \citenamefont {Cornell},\ and\ \citenamefont {Jin}}]{Hu_2016}%
  \BibitemOpen
  \bibfield  {author} {\bibinfo {author} {\bibfnamefont {M.-G.}\ \bibnamefont
  {Hu}}, \bibinfo {author} {\bibfnamefont {M.~J.}\ \bibnamefont {Van~de
  Graaff}}, \bibinfo {author} {\bibfnamefont {D.}~\bibnamefont {Kedar}},
  \bibinfo {author} {\bibfnamefont {J.~P.}\ \bibnamefont {Corson}}, \bibinfo
  {author} {\bibfnamefont {E.~A.}\ \bibnamefont {Cornell}}, \ and\ \bibinfo
  {author} {\bibfnamefont {D.~S.}\ \bibnamefont {Jin}},\ }\href {\doibase
  10.1103/PhysRevLett.117.055301} {\bibfield  {journal} {\bibinfo  {journal}
  {Phys. Rev. Lett.}\ }\textbf {\bibinfo {volume} {117}},\ \bibinfo {pages}
  {055301} (\bibinfo {year} {2016})}\BibitemShut {NoStop}%
\bibitem [{\citenamefont {Paredes}\ \emph {et~al.}(2001)\citenamefont
  {Paredes}, \citenamefont {Fedichev}, \citenamefont {Cirac},\ and\
  \citenamefont {Zoller}}]{Paredes_2001}%
  \BibitemOpen
  \bibfield  {author} {\bibinfo {author} {\bibfnamefont {B.}~\bibnamefont
  {Paredes}}, \bibinfo {author} {\bibfnamefont {P.}~\bibnamefont {Fedichev}},
  \bibinfo {author} {\bibfnamefont {J.~I.}\ \bibnamefont {Cirac}}, \ and\
  \bibinfo {author} {\bibfnamefont {P.}~\bibnamefont {Zoller}},\ }\href
  {\doibase 10.1103/PhysRevLett.87.010402} {\bibfield  {journal} {\bibinfo
  {journal} {Phys. Rev. Lett.}\ }\textbf {\bibinfo {volume} {87}},\ \bibinfo
  {pages} {010402} (\bibinfo {year} {2001})}\BibitemShut {NoStop}%
\end{thebibliography}%

\end{document}